\documentclass[showpacs,superscriptaddress,a4paper,floatfix,nofootinbib]{revtex4-2}
\usepackage{graphicx}
\usepackage{amsmath}
\usepackage{amssymb}
\usepackage{dcolumn}
\usepackage[version=4]{mhchem}
\usepackage{latexsym}
\usepackage{rotating}
\usepackage{epstopdf}
\usepackage[usenames,dvipsnames]{xcolor}
\usepackage{float}
\usepackage{soul}
\usepackage{epsfig}
\usepackage{psfrag}
\usepackage{natbib}
\usepackage{bm}
\usepackage{eucal}
\usepackage{mathrsfs}
\usepackage{braket}
\usepackage{enumerate}
\usepackage{enumitem}
\usepackage{longtable}
\usepackage{threeparttable}
\usepackage{bm}
\usepackage{amsfonts}
\usepackage{dcolumn}% Align table columns on decimal point
\usepackage{bm}
\usepackage{changes}

\usepackage{fancybox}
\usepackage{eepic}
\usepackage{times}
\usepackage{latexsym}
\usepackage{pifont}

\usepackage{scrextend}

\def\kk{\boldsymbol{k}}

\newcommand{\qsgw}{QS$GW$}
\newcommand{\qsgwl}{QS$G\hat{W}$}
\newcommand{\epsi}{$\epsilon_{\infty}$}
\newcommand{\qub}{School of Mathematics and Physics, Queen's University Belfast, Belfast BT7 1NN, Northern Ireland, United Kingdom}
\newcommand{\etsf}{European Theoretical Spectroscopy Facilities (ETSF)}
\newcommand{\kings}{King's College London, London WC2R 2LS, United Kingdom}

\begin{document}
%\article{article}{Title}
\title{\qsgwl: Quasiparticle Self consistent $GW$ with ladder diagrams in $W$}
%\title{Quasiparticle Self consistent $GW$ plus ladder diagrams for optical properties}
\author{Brian Cunningham}
\affiliation{\qub}
\author{Myrta Gr\"uning}
\affiliation{\qub}
\affiliation{\etsf}

\author{Dimitar Pashov}
\affiliation{\kings}

\author{Mark van Schilfgaarde}
\affiliation{National Renewable Energy Laboratories, Golden, CO 80401, USA}

%\ead{bcunningham12@qub.ac.uk}
\begin{abstract}

We present an extension of the quasiparticle self-consistent $GW$ approximation (QS\emph{GW}) [Phys. Rev. B, {\bf 76}
  165106 (2007)] to include vertex corrections in the screened Coulomb interaction $W$.  This is achieved by solving the
Bethe-Salpeter equation for the polarization matrix at all $k$-points in the Brillouin zone. We refer to this method as
\qsgwl.  QS\emph{GW} yields a reasonable and consistent description of the electronic structure and optical response,
but systematic errors in several properties appear, notably a tendency to overestimate insulating bandgaps, blue-shift
plasmon peaks in the imaginary part of the dielectric function, and underestimate the dielectric constant \epsi.  A
primary objective of this paper is to assess to what extent including ladder diagrams in $W$ ameliorates systematic errors
for insulators in the \qsgw\ approximation.  For benchmarking we consider about 40 well understood semiconductors, and
also examine a variety of less well characterized nonmagnetic systems, six antiferromagnetic oxides, and the ferrimagnet
{Fe\textsubscript{3}O\textsubscript{4}}.  We find ladders ameliorate shortcomings in \qsgw\ to a remarkable degree in
both the one-body Green's function and the dielectric function for a wide range of insulators.  New discrepancies with
experiment appear, and a key aim of this paper is to establish to what extent the errors are systematic and can be
traced to diagrams missing from the theory.  One key finding of this work is to establish a relation between the bandgap
and the dielectric constant \epsi.  Good description of both properties together provides a much more robust benchmark
than either alone.  We show how this information can be used to improve our understanding of the one-particle spectral
properties in materials systems such as SrTiO\textsubscript{3} and FeO.

\end{abstract}
\maketitle

%%%%%%%%%%%%%%%%%%%%%%%%%%%%%%%%%%%%%%%%%%%%%%%%
%
%
%
%%%%%%%%%%%%%%%%%%%%%%%%%%%%%%%%%%%%%%%%%%%%%%%%
\section{Introduction}\label{ss:intro}

The one-particle Green's function $G(\mathbf{r},\mathbf{r'},\omega)$ provides essential information about material
properties.  Besides having value in its own right, determining both ground state properties (total energy, charge and
magnetic densities), and excitation energies, it is the starting point for transport and other two-particle properties,
e.g. spin and charge response functions, and superconductivity.

As a consequence, knowledge of $G$ is of the first importance, and a vast amount of effort has been dedicated to finding
prescriptions to yield $G$ both efficiently and with high-fidelity \emph{ab initio} (without recourse to models or
adjustable parameters).  Density-functional theory \cite{PhysRev.136.B864} (DFT), where the electron density \emph{n}
replaces $G$ as the fundamental variable, is an alternative, and indeed it is far more popular because of its efficiency
and good scaling with system size.  DFT is a ground-state theory, but it generates an auxilliary one-body $H_{0}$, with
fictitious eigenvalues and eigenfunctions.  $H_{0}$ often provides a reasonable approximation to excitations of the real
system, but it is often unsatisfactory, e.g. its notorious tendency to underestimate splitting between occupied and
unoccupied levels.  Wave function methods, widely used in quantum chemistry, but less so in materials physics, use the
single-particle orbitals $\psi_i$ as the fundamental variable.  They can provide high-fidelity solutions to the
many-body Schrodinger equation.  As Walter Kohn noted in his Nobel prize lecture~\cite{Kohn99}, wave function methods
contain more information than is needed or useful, but nevertheless require concomitant effort needed to compute
observables.  For that reason they are expensive and scale poorly with system size.  Also, spectral properties are not
readily computed.

Green's function (GF) methods lie between the two: $G$ has more information than $n$ but less than the wave functions.
As with DFT, $G$-based methods create an effective one-body potential $\Sigma(\mathbf{r},\mathbf{r'},\omega)$, but
differ in that $\Sigma$ is nonlocal and energy-dependent.  They are computationally more intensive than DFT; however
they can be made to scale reasonably well with system size, and because of their better fidelity it is likely they will
ultimately outphase DFT methods for many functional materials, particularly when excitations are involved.  Thus, GF
theories might be called the ``Goldilocks'' approach.  GF methods possess a key advantage in another respect: dynamical
screening becomes the predominant many-body effect for systems involving many atoms.  Hedin's equations~\cite{hedin} can
be expanded diagramatically in powers of the screened coulomb interaction \emph{W}, and encapsulate this phenomenon in a
natural way, even in the lowest order (\emph{GW}).  The traditional target applications are also different: quantum
chemical methods focus mostly on ground state properties while GF methods focus on spectral properties, especially
two-particle spectra.  GF methods do not yet possess the fidelity of wave function methods, and to what extent their
fidelity can eventually approach them remains a key open question.\footnote{As regards total energy, not examined here,
GF methods are immature~\cite{Miyake02} but they have advanced significantly in recent years.  A particularly noteworthy
example is the study of the ``S66'' test set, where the authors achieved quantum-chemical accuracy by adding only
singles and second-order screened exchange to \emph{GW}~\cite{Paier12,Ren13}.}

One key aim of this paper is to provide a partial answer to this question.  The quasiparticle self-consistent \emph{GW}
approximation (QS\emph{GW}) provides an effective way to implement \emph{GW} theory without relying on a lower level
approximation as a starting point. This makes discrepancies with experimental data much more uniform, and it is
essential to distinguish errors intrinsic to the theory itself, from accidents as a result of the starting point.  We
assess in some detail the extent to which discrepancies QS\emph{GW} displays with experiment can be mitigated by adding
the low-order diagram (ladder diagram) to the random-phase approximation (RPA)\footnote{May also be referred to as the
independent-particle approximation} for the bare polarizability.  As noted, the \emph{GW} approximation is the lowest
order diagram in the many-body perturbation theory (MBPT) of Hedin~\cite{hedin}, and while \emph{GW} shows significant
improvement over DFT (including functionals designed to surmount the well-gap underestimate~\cite{Das19}), it has well known problems.  First, it is a perturbation theory, which typically starts from
some reference noninteracting $G_{0}$ and generates a correction to it.  The most common choice of $G_{0}$ is one based
in DFT, but many kinds of choices have been made to improve on the final result.  This situation is unsatisfactory in
two respects:

\begin{itemize}[leftmargin=*]

\item $G_{0}$ can be (and often is) tuned to improve agreement with experiments.  $G_{0}$ plays the role of a free
  parameter, and in this sense the theory isn't really \emph{ab initio} any more.

\item The errors inherent in low-order MBPT, e.g. \emph{GW}, can be masked by the arbitrariness in $G_{0}$.  Sometimes
  qualitiatively wrong conclusions can be drawn, or good agreement with experiment found, but for the wrong reason.  This is a
  quite common, albeit not well appreciated, difficulty with the theory; see for example Ref. \cite{Acharya21a}.

\end{itemize}

By employing the \emph{GW} approximation in the QS\emph{GW} form, we can circumvent these difficulties.  QS\emph{GW} is
a procedure where $G_{0}$ is determined self-consistently.  Self-consistency is used not to minimize the total energy,
but instead some measure (norm) of the difference between $G^{-1}_{0}$ and $G^{-1}$ \cite{QSGW_paper}.  With a
definition for optimal construction for $G_{0}$, it surmounts the ambiguities from arbitrariness in the
starting-point~\cite{notea}.  It provides a good and systematic $G_{0}$ so that discrepancies with experiment that appear
tend to be similarly systematic, making it possible to associate these discrepancies with diagrams missing from the
theory.

There is no unique definition of the norm, but one intuitively appealing definition leads to a static (quasiparticlized)
self-energy $\Sigma^0$ generated from the dynamical one as
\begin{equation}
\label{eq:veff}
{\Sigma^0}(\mathbf{r},\mathbf{r'}) = \sum_{ij} \psi_i(\mathbf{r})\ {\Sigma_{ij}^{0}}\  \psi_j^{*}(\mathbf{r'});~{\Sigma_{ij}^{0}} = \frac{1}{2}\left\{ {{\rm Re}[\Sigma({\varepsilon_i})]_{ij}+{\rm Re}[\Sigma({\varepsilon_j})]_{ij}} \right\}
\end{equation}
$i$ and $j$ are eigenstates of the one-particle hamiltonian.  Ismail-Beigi showed this construction satisfies a
variational principle, not for the total energy but its gradient \cite{Beigi17}.  One other important consequence of
Eq.~(\ref{eq:veff}) is that at self-consistency the poles of $G$ and the poles of $G_{0}$ coincide: thus in contrast to
DFT, the energy bands of $H_{0}$ generated by QS\emph{GW} correspond to true excitations of the system.

Results generated by QS\emph{GW} may sometimes worsen agreement with experiment over other forms of \emph{GW}.  For example,
$\varepsilon_\infty$ generated from a Kohn-Sham band structure is often better than the QS\emph{GW} one.  We will argue
that this stems from a fortutious cancellation of errors; see \S\ref{ss:consistency}.  Fortutitous error cancellation in
QS\emph{GW} is much less pronounced, and as a result, discrepancies with experiment are better exposed, and moreover
they are much more uniform.  Several of the most salient discrepancies are connected to the inadequate description of the
dielectric polarizability.  This forms the primary motivation for the present work: to make a detailed assessment of how
the simplest extension to the RPA polarizability improves both $G$ and the dielectric response.  In this work, the
excitonic contributions are taken into account by including ladder diagrams into the screened Coulomb interaction $W$
through use of the Bethe-Salpeter equation (BSE)~\cite{PhysRevLett.91.056402,PhysRevLett.91.256402} for the
polarization.  A high-fidelity $G$ is essential for a good description of any response function, including the magnetic
one, as shown for NiO \cite{KotaniSW08} and for yttrium iron garnet~\cite{Jerome20}; and the particle-particle
correlation function that governs superconductivity (see e.g.  Ref.~\cite{Acharyalafeas}).

Other works have considered the effect of vertex corrections to the dielectric screening on the band gap.  For example,
Refs.~\onlinecite{PhysRevB.81.085213,PhysRevLett.99.246403,PhysRevLett.94.186402} included ladder diagrams through an
effective nonlocal static kernel constructed within time-dependent density functional theory to mimic the BSE.  More
recently, Kutepov proposed several schemes for the self-consistent solution of Hedin's equations including vertex
corrections~\cite{PhysRevB.94.155101}. In particular, for selected semiconductors and
insulators~\cite{PhysRevB.95.195120}, he included the vertex correction for the dielectric screening at the BSE level
together with a so-called first-order approximation for the vertex in the self-energy, $\Sigma=iGW\Gamma$.  In all the
cases, an improvement over L\qsgw\ was observed (see \S\ref{ss:lqsgw} for a comparison between L\qsgw\ and \qsgw).  With
respect to these previous works, we include vertex corrections to the dielectric screening only at the BSE level, but
introducing the usual static approximation for the BSE kernel, which was lifted in
Refs~\onlinecite{PhysRevB.95.195120,PhysRevB.94.155101}.\footnote{An analogy in the
quantum-chemical literature is the fully self-consistent framework corresponding to ``multireference'' starting points
while the quasiparticlized form corresponds to an optimized single-reference.}  Here we omit the first-order vertex for
$\Sigma$, in keeping with our present objective --- to find the best single-Slater determinant construction.  The
QS\emph{GW} philosophy incorporates this vertex in an approximate way, via a Ward identity in $\Gamma$ that goes as
$1/Z$ in the $q{\rightarrow}0,\omega{\rightarrow}0$ limit, cancelling the $Z$ factor that is the predominant difference
between the quasiparticlized $G_0$ and the interacting $G$~\cite{QSGW_PRL}.  Adding this vertex explicitly jeapordizes
this cancellation.  It can be surmounted via a fully self-consistent $G$, as Kutepov did, but the cost is considerable.
Ladder diagrams can be included in \emph{W} while retaining $O(N^3)$ scaling~\cite{Ljungberg15}, but there is no obvious
analog to adding the vertex in $\Sigma$.  We show here that including the vertex in \emph{W} is more important: including it without
the vertex in $\Sigma$ usually yields quite satisfactory results.  Kutepov noted the interacting $G$ with both vertices
performs better in CuCl: more generally it seems to matter when the highest occupied states are flat and nearly
dispersionless (see point 2, \S\ref{ss:fidelity} and discussion around Table~\ref{tab:dvbmgaps}).  Finding a way to
surmount this shortcoming in a quasiparticle framework is a work in progress.

Another readily identifiable source of error is the contribution lattice vibrations make to $\Sigma$.  It can be a few
tenths of an eV in diamond and in polar compounds with light elements, so we include that contribution here in an
approximate way, by obtaining the reduction in the gap by an independent method (\S\ref{ss:frolich}) and using a hybrid self
energy to reproduce this shift (\S\ref{ss:hybrid}).

%%%%%%%%%%%%%%%%%%%%%%%%%%%%%%%%%%%%%%%
%
%METHOD
%
%%%%%%%%%%%%%%%%%%%%%%%%%%%%%%%%%%%%%%%%%%%%%
\section{Theory and numerical implementation}

Starting from the Hedin equations (\S\ref{ss:g0w0}), we outline how the original \qsgw\ approximation
(\S\ref{ss:qsgw}) is modified to include excitonic contributions (\S\ref{ss:bse}). The resulting approach, which
is referred to as \qsgwl\ (where the substitution $W{\rightarrow}\hat{W}$ implies that vertex corrections are included
in $W$), was numerically implemented within an all-electron framework using a linear muffin-tin orbital basis set
(\S\ref{ss:lmto}) in the Questaal package\cite{questaal_web}.  Later sections show applications to a broad range of
materials.

\subsection{The GWA from the Hedin's equations}\label{ss:g0w0}

The approach described in this work, the standard GWA and the QS\emph{GW}, are all derived from the many-body
perturbative approach developed by Hedin\cite{hedin}. In this method, the following set of closed coupled
equations\cite{PhysRev.139.A796,schwinger,schwinger2} are to be solved iteratively:

%which is the $GW$ method. The DFT band structure can be used as the starting point for these calculations (see e.g. Refs.~\onlinecite{onida_electronic_2002,AULBUR20001}).
%
%In the latter, the  $\varepsilon_{n}$ obtained from the solution of the Kohn-Sham DFT equations are perturbatively corrected at the first order:
%\begin{equation}
%E_{n\boldsymbol{k}}=\varepsilon_{n\boldsymbol{k}}+\langle\psi_{n\boldsymbol{k}}|\Sigma^{GW}(E_{n\boldsymbol{k}})-V_{\rm XC}|\psi_{n\boldsymbol{k}}\rangle.
%\label{eq:gwperb}
%\end{equation}
%The closed set of equations that are to be solved a%In Eq.~\eqref{eq:gwperb}, $\Sigma^{GW}$ is the self-energy in the so-called $GW$ approximation.\cite{hedin,GW_aryasetiawan} The general expression for the self-energy, and related quantities, is given by:
\begin{align}
  \Sigma(1,2)&={\rm i}\int{\rm d}(34)~G(1,3^{+})W(4,1)\Gamma(3,2,4)\label{eq:selfE1}\\
  G(1,2)&=G_{0}(1,2) + \int{\rm d}(34)~G_{0}(1,3)\Sigma(3,4)G(4,2)\label{eq:GreenF}\\
 W(1,2)&=v(1,2)+\int{\rm d}(34)v(1,3)P(3,4)W(4,2)\label{eq:scrpot}\\
P(12)&=-{\rm i}\int{\rm d}(34)G(1,3)G(4,1^{+})\Gamma(3,4,2)\label{eq:irrpol}\\
  %-\frac{\delta G^{-1}(1,2)}{\delta V_{\rm tot}(3)}\label{eq:vert},
 \Gamma(1,2,3)&=\delta(1,2)\delta(1,3) + \nonumber \\&~~~\int {\rm d}(4567)\frac{\delta \Sigma(1,2)}{\delta G(4,5)} G(4,6) G(7,5)\Gamma(6,7,3)\label{eq:vert}\end{align}
where $G$ is the Green's function, $v(\boldsymbol{r},\boldsymbol{r}')=1/|\boldsymbol{r}-\boldsymbol{r}'|$ is the bare Coulomb interaction, $W$ is the screened Coulomb interaction, $\Gamma$ is the irreducible vertex function, $P$ is the irreducible polarizability (the functional derivative of the induced density with respect to the total potential) and $\Sigma$ is the self-energy operator. % This set of equations [\eqref{eq:selfE1}--\eqref{eq:vert}] known as Hedin's equations,~\cite{PhysRev.139.A796,schwinger,schwinger2} is completed by the equation for the irreducible polarizability (needed to determine $\epsilon^{-1}$):
%\begin{equation}
%P(1,2)=-{\rm i}\int {\rm d}(34)~G(1,3)\Gamma(3,4,2)G(4,1^{+}).
%\end{equation}
In Eqs.~\eqref{eq:selfE1} and~\eqref{eq:irrpol}, the indices subsume position and time and the $+$ superscript implies $t'=t+\eta$, with $\eta\rightarrow 0^{+}$.

In the standard GWA, also known as one-shot $GW$ or $G_0W_0$, Eq.~\eqref{eq:vert} is approximated as:
\begin{equation}
  \Gamma(1,2,3) \approx \delta(1,2)\delta(1,3) \label{eq:vertapp}
\end{equation}
in both the expressions for the self-energy [Eq.~\eqref{eq:selfE1}] and the irreducible polarization [Eq.~\eqref{eq:irrpol}]. In addition, $\Sigma$ and $P$ are both evaluated for $G=G_0$, the independent-particles Green's function,\footnote{Equivalent to performing one iteration, starting with $\Sigma = 0$ in Eq.~\eqref{eq:selfE1}}
that in the frequency-domain takes the form
\begin{equation}\label{eq:green}
G_{0}(\boldsymbol{r},\boldsymbol{r}',\omega)=\sum_{n}\frac{\psi_{n}(\boldsymbol{r})\psi_{n}^{*}(\boldsymbol{r}')}{\omega-\varepsilon_{n}\pm{\rm i}\eta}.
\end{equation}
In Eq.~\eqref{eq:green},  $\psi_{n}$ and $\varepsilon_{n}$ are the single-particle wavefunctions and energies; the index $n$ contains band, spin and wavevector indices and the $+(-)$ is for unoccupied(occupied) bands.

The approximation for the irreducible polarization obtained by neglecting the vertex is referred to variously as the
independent particle approximation, the time-dependent Hartree approximation, and the Random-Phase-Approximation
(RPA)~\cite{onida_electronic_2002}.  When using Eq.~\eqref{eq:green} in frequency space it takes the form:
\small \begin{equation}\label{eq:RPA_pol}
P^{\rm RPA}(\boldsymbol{r},\boldsymbol{r}';\omega)=\sum_{n_1n_2}(f_{n_2}-f_{n_1})\frac{\psi_{n_2}(\boldsymbol{r})\psi_{n_1}^{*}(\boldsymbol{r})\psi_{n_1}(\boldsymbol{r}')\psi_{n_2}^{*}(\boldsymbol{r}')}{\varepsilon_{n_2}-\varepsilon_{n_1}-\omega+{\rm i}(f_{n_2}-f_{n_1})\eta},
\end{equation}\normalsize
where $f_n$ are the single-particle occupations.
%% The $GW$ approximation to the self-energy corresponds to neglecting the vertex, $\Gamma(1,2,3) \approx \delta(1,2)\delta(1,3)$, and using the noninteracting Green's function, which -- in frequency space -- is

%% Note that . The random phase approximation (RPA) results from using the noninteracting Green's function and neglecting the vertex in the expression for the polarization, which leads to the following expression for the polarization

The Green's function in Eq.~\eqref{eq:green} can be constructed from the Kohn-Sham electronic structure, which is obtained from the self-consistent solution of Schr\"odinger-like equations with the single-particle Hamiltonian
\begin{equation}
H_0(\boldsymbol{r})=-\frac{1}{2}\boldsymbol{\nabla}^2+V_{\rm ext}(\boldsymbol{r})[\rho]+V_H(\boldsymbol{r})[\rho]+V_{\rm XC}(\boldsymbol{r})[\rho].
\label{eq:oneHam}
\end{equation}
In Eq.~\eqref{eq:oneHam}, $V_{\rm ext}(\boldsymbol{r})$ is the external potential due to nuclei and external fields, $V_H(\boldsymbol{r})$ is the Hartree potential describing the classical mean-field electron-electron interaction, and $V_{\rm XC}(\boldsymbol{r})$ is the exchange-correlation potential describing correlation effects missing in $V_{H}(\boldsymbol{r})$. $\rho$ is the electronic density calculated as $\sum_{\rm occ}|\psi_n(\boldsymbol{r})|^2$, that is constructed from the eigensolutions of $H_0$ (from which the self-consistent solution is obtained). Because of $V_{\rm XC}$, the corresponding single-particle $G_0$ effectively contains many-body effects. Then in Eq.~\ref{eq:GreenF} for the many-body Green's function, the self-energy is replaced by $\Delta\Sigma=\Sigma-V_{\rm XC}$ to avoid double counting.
Eq.~\ref{eq:GreenF} is rewritten as a nonlinear equation for the quasiparticle energies $E_{n\boldsymbol{k}}$
\begin{equation}
E_{n\boldsymbol{k}}=\varepsilon_{n\boldsymbol{k}}+\langle\psi_{n\boldsymbol{k}}|\Sigma(E_{n\boldsymbol{k}})-V_{\rm XC}|\psi_{n\boldsymbol{k}}\rangle,
\label{eq:gwEner}
\end{equation}
and solved after it is linearized:
\begin{equation}
E_{n\boldsymbol{k}}=\varepsilon_{n\boldsymbol{k}}+Z_{n\boldsymbol{k}}\langle\psi_{n\boldsymbol{k}}|\Sigma(\varepsilon_{n\boldsymbol{k}})-V_{\rm XC}|\psi_{n\boldsymbol{k}}\rangle,
\label{eq:gwlin}
\end{equation}
where the renormalization factor $Z_{n\kk}$ is:
\begin{equation}
Z_{n\boldsymbol{k}}=(1-\displaystyle \langle\psi_{n\boldsymbol{k}}|\partial\Sigma(\omega=\varepsilon_{n\boldsymbol{k}})/\partial\omega|\psi_{n\boldsymbol{k}}\rangle)^{-1}.
\label{eq:Zren}
\end{equation}
%In many $GW$ calculations, the $Z$ factor is usually employed in Eq.\eqref{eq:gwlin}. There are, however, several arguments for setting the $Z$-factor equal to $1$.  One argument relies on the $Z$-factor cancellation in the expression for the self-energy (the $Z$-factor and $\Gamma$ in Eq.~\ref{eq:selfE1} tend to cancel; for details, see Appendix A of Ref.~\cite{QSGW_paper}). Another argument relies on the formula for the derivative discontinuity of the DFT-RPA functional\cite{deriv_disc}, which is the same expression in Eq.~(\ref{eq:gwlin}), but for the $Z$ factor being equal to $1$. In this work we adopt the $Z=1$ choice and show that indeed this generally leads to a better agreement with experiment within the chosen framework.
\par
%A common technique is to obtain optical absorption spectra with excitonic effects using the Bethe-Salpeter equation\cite{BSE_paper,onida_electronic_2002} (derived from Eqs~\ref{eq:irrpol} and \ref{eq:vert}) and using the $G_0W_0$ energies from above when calculating the polarization.  This method has proved successful in producing optical absorption spectra in very good agreement with experiment.\cite{yambo}
%% Once the ground state energy has been minimized we have a set of Kohn-Sham energies and eigenfunctions, $\varepsilon_{n\boldsymbol{k}}$ and $\psi_{n\boldsymbol{k}}(\boldsymbol{r})$.  These can be used to generate the
%% where $\eta$ is a positive infinitesimal and the $+(-)$ sign is for unoccupied(occupied) states.  We have not included the spin index, it is contained in the index $n$.  We wish to determine the self-energy, which contains electronic exchange and correlation effects.

\subsection{Ladder diagrams in $W$}\label{ss:bse}

%%   A set of equations can be derived using the Green's function that allow an expression for the self-energy to be obtained.  This expression can be derived using the Schwinger functional derivative method, where an external perturbation is introduced and then set to zero once an expression for the self-energy is obtained.  The equations that allow one to calculate the self-energy are
%% %% \begin{align}
%% %% &\Sigma(1,2)={\rm i}\int{\rm d}(34)~G(1,3^{+})W(1,4)\Gamma(3,2,4)\label{eq:selfE}\\
%% %% &W(1,2)=\int{\rm d}3~\epsilon^{-1}(1,3)v(3-2)\\
%% %% &\Gamma(1,2,3)=-\frac{\delta G^{-1}(1,2)}{\delta V(3)}\label{eq:vert}
%% %% \end{align}
%% %% where $1=(\boldsymbol{r}_1,t_1,\sigma_1)$ and the $+$ implies $t'=t+\eta$.  $W(\boldsymbol{r},\boldsymbol{r}',\omega)=\int{\rm d}\boldsymbol{r}''\epsilon^{-1}(\boldsymbol{r},\boldsymbol{r}'',\omega)v(\boldsymbol{r}''-\boldsymbol{r}')$ is the screened Coulomb interaction and $\epsilon$ is the dielectric function introduced previously.  $\Gamma$ is the vertex function. For completeness, the polarization is generated from the Greens function and vertex

%% In many implementations of GW the vertex is ignored $\Gamma(1,2,3)=\delta(1-2)\delta(2-3)$.  In the single shot GW, once the diagonal elements of $\Sigma$ are determined these are used to correct the Kohn-Sham energies ; with the correction factor $Z_{n\boldsymbol{k}}=\displaystyle1/(1-\displaystyle \partial\Sigma(\omega)/\partial\omega|_{\omega=\varepsilon_{n\boldsymbol{k}}})$ arising from the Taylor series expansion of $\Sigma(\omega)$ about $\omega=\varepsilon_{n\boldsymbol{k}}$.
In previous works employing \qsgw\,\cite{QSGW_PRL,QSGW_paper,QSGW_prl1,opt_PRM} the RPA was used to make $W$. This leads
to errors noted in the introduction, e.g. a significant band gap overestimation.  Here we go beyond the RPA and include
ladder diagram corrections in $W$ through the BSE for the polarization~\cite{BSE_paper}.  To include the vertex in
Hedin's equations we need to determine the interaction kernel $\delta\Sigma/\delta G$ in Eq.~\ref{eq:vert}.  Suppose we
have adopted the $GW$ approximation for $\Sigma$\footnote{The vertex in Eq.~(\ref{eq:selfE1}) can been shown to
effectively cancel with the $Z$-factor, see for example Appendix~A in Ref.~\onlinecite{QSGW_paper}} and assume that
$\delta W/\delta G$ is negligible~\cite{onida_electronic_2002}, then $\delta \Sigma(12)/\delta G(45)={\rm
  i}W(12)\delta(1,4)\delta(2,5)$, where $W$ is determined in the $GW$ approximation. This expression is then inserted in
Eq.~\ref{eq:vert}.

Before we present the BSE for the polarization we will introduce the expansion of the two-point polarization to its four-point counterpart: $P(12)=P(1122)=P(1324)\delta(1,3)\delta(2,4)$. We are now able to present an expression for the polarization that goes beyond the RPA using Eqs~\ref{eq:irrpol} and \ref{eq:vert} and adopting the expression for the interaction kernel from above;
\begin{equation}
P(12)= P^{\rm RPA}(12)-\int{\rm d}(34)P^{\rm RPA}(1134)W(34)P(3422),\label{eq:BSE_PW}
\end{equation}
where $P^{\rm RPA}(1234)=-{\rm i}G(13)G(42)$. The $W$ that appears in the interaction kernel, $\delta \Sigma/\delta G$, is calculated at the level of the RPA and this is usually assumed to be static,\footnote{In few works this approximation has been relaxed, see e.g. Ref.~\onlinecite{PhysRevLett.91.176402}.} i.e., $\delta \Sigma/\delta G={\rm i}W^\mathrm{RPA}(\omega{=}0)$. To avoid confusion with $W$ in Eq.~\ref{eq:selfE1} we will refer to the $W$ in Eq.~\ref{eq:BSE_PW} from here on as $K$.
\par
The Dyson-like equation for the polarizability, Eq.~\ref{eq:BSE_PW}, can be transformed to an eigenproblem for an effective 2-particle Hamiltonian by introducing the basis of single particle eigenfunctions that diagonalize the RPA polarization.
Using the completeness of the eigenfunctions, any 4-point quantity can be expanded as
\begin{equation}
S(\boldsymbol{r}_1,\boldsymbol{r}_2,\boldsymbol{r}_3,\boldsymbol{r}_4)=\hspace{-0.15cm}\sum_{\tiny n_1n_2 n_3n_4\normalsize} S_{n_1n_2n_3n_4}\psi_{n_1}^{*}(\boldsymbol{r}_1)\psi_{n_2}(\boldsymbol{r}_2)\psi_{n_3}(\boldsymbol{r}_3)\psi_{n_4}^{*}(\boldsymbol{r}_4),
\label{Stoeig}\end{equation}
where we have again combined band, spin and wavevector indices, and $S_{n_1n_2n_3n_4}=\int{\rm d}(\boldsymbol{r}_1\boldsymbol{r}_2\boldsymbol{r}_3\boldsymbol{r}_4) S(\boldsymbol{r}_1,\boldsymbol{r}_2,\boldsymbol{r}_3,\boldsymbol{r}_4)\times$ $\psi_{n_1}(\boldsymbol{r}_1)\psi_{n_2}^{*}(\boldsymbol{r}_2)\psi_{n_3}^{*}(\boldsymbol{r}_3)\psi_{n_4}(\boldsymbol{r}_4)$.\vspace{0.1cm}\par
%The RPA polarization in this basis is ${\displaystyle{P_{n_1n_2n_3n_4}^{0}=(f_{n_2}-f_{n_1})\delta_{n_1n_3}\delta_{n_2n_4}/(\varepsilon_{n_2}-\varepsilon_{n_1}-\omega-{\rm i}\eta)}}$, see Eq.~\ref{eq:RPA_pol}.

Inserting the expression for the RPA polarization from Eq.~\eqref{eq:RPA_pol} in Eq.~\eqref{eq:BSE_PW}, one arrives at the following expression for the polarization %that contains local-field and excitonic effects
\begin{equation}
P_{\scriptsize\begin{array}{l}n_1n_2\boldsymbol{k}\\n_3n_4\boldsymbol{k}'\end{array}}(\boldsymbol{q},\omega)=\left[H(\boldsymbol{q})-\omega\right]_{\scriptsize\begin{array}{l}n_1n_2\boldsymbol{k}\\n_3n_4\boldsymbol{k}'\end{array}}^{-1}(f_{n_4\boldsymbol{k}'+\boldsymbol{q}}-f_{n_3\boldsymbol{k}'}),\label{eq:Pineig}
\end{equation}
whereby the conservation of momentum we have $\boldsymbol{k}_{2(4)}=\boldsymbol{k}_{1(3)}+\boldsymbol{q}$; and\footnote{Note that if we are determining the reducible polarizabililty $\chi$, such that $W=v+v\chi v$ then the Kernel becomes $K=W-V$ \cite{onida_electronic_2002}}
\begin{equation}\begin{array}{rl}
H_{\scriptsize\begin{array}{l}n_1n_2\boldsymbol{k}\\n_3n_4\boldsymbol{k}'\end{array}}(\boldsymbol{q})=&(\varepsilon_{n_2\boldsymbol{k}'+\boldsymbol{q}}-\varepsilon_{n_1\boldsymbol{k}'})\delta_{n_1n_3}\delta_{n_2n_4}\delta_{\boldsymbol{k}\boldsymbol{k}'}+(f_{n_2\boldsymbol{k}+\boldsymbol{q}}-f_{n_1\boldsymbol{k}})K_{\scriptsize\begin{array}{l}n_1n_2\boldsymbol{k}\\n_3n_4\boldsymbol{k}'\end{array}}(\boldsymbol{q}),\\&\\
K_{\scriptsize\begin{array}{l}n_1n_2\boldsymbol{k}\\n_3n_4\boldsymbol{k}'\end{array}}(\boldsymbol{q})=&\int{\rm d}\boldsymbol{r}_1{\rm d}\boldsymbol{r}_2\psi_{n_1}^{*}(\boldsymbol{r}_1)\psi_{n_3}(\boldsymbol{r}_1)\times W^{\rm RPA}(\boldsymbol{r}_1,\boldsymbol{r}_2;\omega=0)\psi_{n_2}(\boldsymbol{r}_2)\psi_{n_4}^{*}(\boldsymbol{r}_2).\end{array}
\end{equation}
The expression $(H-\omega)^{-1}$ can be expressed in the spectral representation as:
\begin{equation}
\left[H(\boldsymbol{q})-\omega\right]_{ss'}^{-1}=\sum_{\lambda\lambda'}\frac{A_{s}^{\lambda}(\boldsymbol{q})N_{\lambda,\lambda'}^{-1}(\boldsymbol{q})A_{s'}^{*\lambda'}(\boldsymbol{q})}{E_\lambda(\boldsymbol{q})-\omega\pm{\rm i}\eta},\label{eq:Hminone}
\end{equation}
where $A_{s}^{\lambda}(\boldsymbol{q})$ is element $s=n_1n_2\boldsymbol{k}$ of the eigenvector of $H(\boldsymbol{q})$
with corresponding eigenvalue $E_\lambda(\boldsymbol{q})$ and $N(\boldsymbol{q})$ is the overlap matrix. When the Tamm-Dancoff
approximation (TDA) is adopted,~\cite{TD_myrta} $H$ is Hermitian and Eq.~\eqref{eq:Hminone} reduces to
$\displaystyle\sum_{\lambda}\displaystyle\frac{A_{s}^{\lambda}(\boldsymbol{q})A_{s'}^{*\lambda}(\boldsymbol{q})}{E_\lambda(\boldsymbol{q})-\omega\pm{\rm
    i}\eta}$.\par The polarization in Eq.~\ref{eq:Pineig} can then be expressed in real-space according to
Eq.~\ref{Stoeig} and contracted to its two-point form. This two-point polarization is then used in Eq.~\ref{eq:scrpot}
to obtain $W$ with ladder diagrams included.  In what follows  we will denote
$W^\mathrm{RPA}$ with the symbol $W$, and refer to $W$ with ladders included as $\hat{W}$.  The updated $W$ or $\hat{W}$ is then used in the
expression for the self-energy with the vertex $\Gamma$ in the exact self-energy ($iGW\Gamma$) omitted.  The
justification for omission, and the consequences of it, is taken up in \S\ref{ss:preview}.

In works that report optical absorption $\alpha(\omega)$, we construct it using the relation
\begin{equation}
\alpha(\omega) = \frac{2\omega}{c}\mathrm{Im}\sqrt{\varepsilon(\omega)}
\label{eq:defalfa}
\end{equation}
where $\varepsilon$ is calculated from the macroscopic part of the dielectric matrix $\varepsilon=1-vP$.
\subsection{Self-consistency: \qsgw}\label{ss:qsgw}

In the DFT based $G_0W_0$ approximation, the $E_{n\boldsymbol{k}}$ are obtained as a first-order correction of the
Kohn-Sham single-particle energies. As mentioned in the introduction, the GWA works when the Kohn-Sham system gives a
qualitatively correct description of the physical system, i.e., when the Kohn-Sham single-particle energies are not `too
far' from the quasiparticle energies. When this is not the case, the GWA does not give accurate results.  It is often
improved in practice by choosing some other $G_0$ constructed, e.g. from a hybrid functional.  Another route is to
replace the corrected energy $E_n$ [Eq.~\eqref{eq:gwlin}] either in the Green's function~\cite{ferdi94}
[Eq.~\eqref{eq:green}], or in the RPA polarization [Eq.~\eqref{eq:RPA_pol}] entering the screened potential $W$, or in
both.

\begin{figure}[h!]
\begin{center}
\includegraphics[width=0.58\columnwidth]{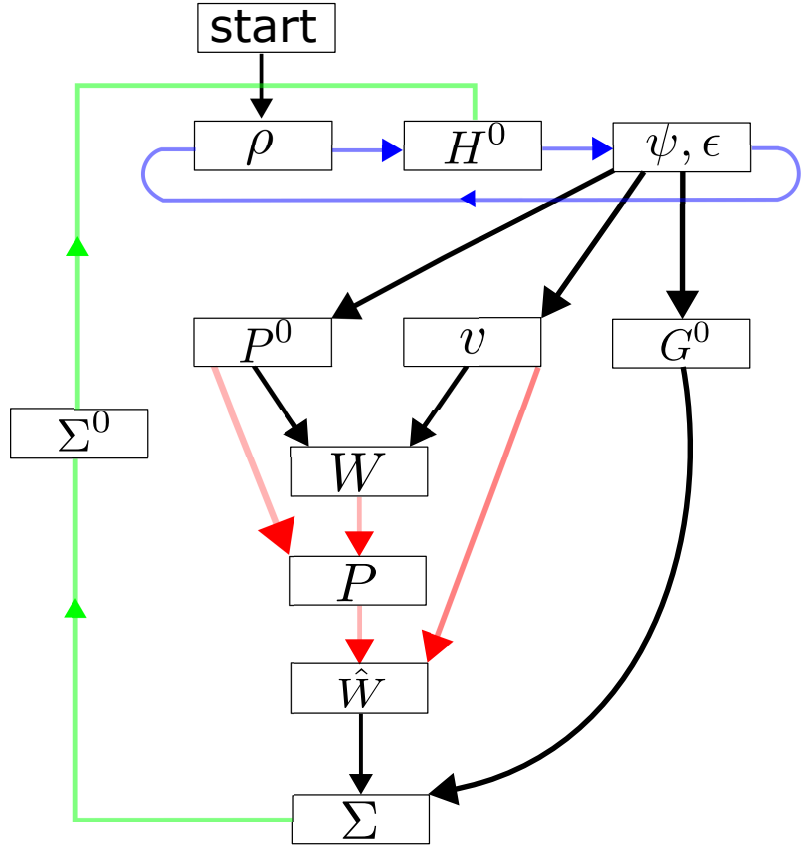}
\end{center}
\caption{\small\sffamily
  Flow chart of the QS\emph{GW} cycle. Non-interacting eigenfunctions and energies ($\psi,\epsilon$) are calculated self consistently (blue). These are used to construct the non-interacting Green's function $G^0$, Coulomb interaction $v$, RPA polarization $P^0{=}-iG^0G^0$, and $W{=}(1-vP^0)^{-1}v$. $W$ is used to make a vertex and better $P$ via Eq.~\ref{eq:BSE_PW}, which gives the improved $\hat{W}$. % (that can also be used self-consistently to further improve $\hat{W}$).
One cycle makes the static self-energy $\Sigma^0$, that is passed to $H^0$ (green) and the cycle repeated to self-consistency.
}
%\caption{\small\sffamily
%  Flow chart of the QS\emph{GW} cycle.
%  $P$ is made  in the RPA, $P{=}G^0G^0$, and $W$ from $W^{-1}{=}P+v^{-1}$, but optionally $W$ may be used to make
%  a vertex and a better $P$ via Eq.~\ref{eq:BSE_PW}, which improves $W$.
%  One cycle makes the static self-energy $\Sigma^0$, which can be passed again to $H^0$ and the cycle repeated to
%  self-consistency.
%}
\label{fig:timings}
\end{figure}

Here we use the \qsgw\ approach in which the starting point is chosen to effectively minimize $\Delta \Sigma$:
the difference between the dynamical self-energy and (static) quasiparticlized one.  In practice, once the self-energy
has been calculated within the $GW$ approximation, a new effective single-particle static potential is determined by
Eq.~\ref{eq:veff}.
Then, by substituting $V_{\rm XC}$ in Eq.~\eqref{eq:oneHam} with $\Sigma^0$, Eq.~\ref{eq:veff}, a new set of
single-particle energies and wavefunctions can be determined. In turn, those can be used to re-calculate the $GW$
self-energy, and the whole procedure can be repeated until self-consistency in the energies and eigenvalues is achieved.
In this procedure the resulting electronic structure does not depend on the quality of the Kohn-Sham DFT electronic
structure for the system, and equally important, it removes the arbitrariness in starting point~\cite{notea}.
Fig.~\ref{fig:timings} shows a flow chart of the process.

\subsubsection*{Why self-consistency is important}

Self-consistency is not typically performed in weakly correlated materials.  LDA-based \emph{GW} can do very well (see
description of {Bi\textsubscript{2}Te\textsubscript{3}}, \S\ref{ss:bi2te32}) but self-consistency improves the theory
and makes the discrepancies with experiment systematic.  Recent work shows this to be the case even for simple \emph{sp}
metals such as Li, Na, and Mg~\cite{Friedrich22} where errors in RPA for \emph{W} are likely to be less important than
in insulating systems.  Similarly excellent agreement is found for the Fermi liquid regime of Fe, in a detailed study
examining several properties~\cite{Sponza17}.

\begin{figure}[h!]
\begin{center}
\includegraphics[width=0.75\columnwidth]{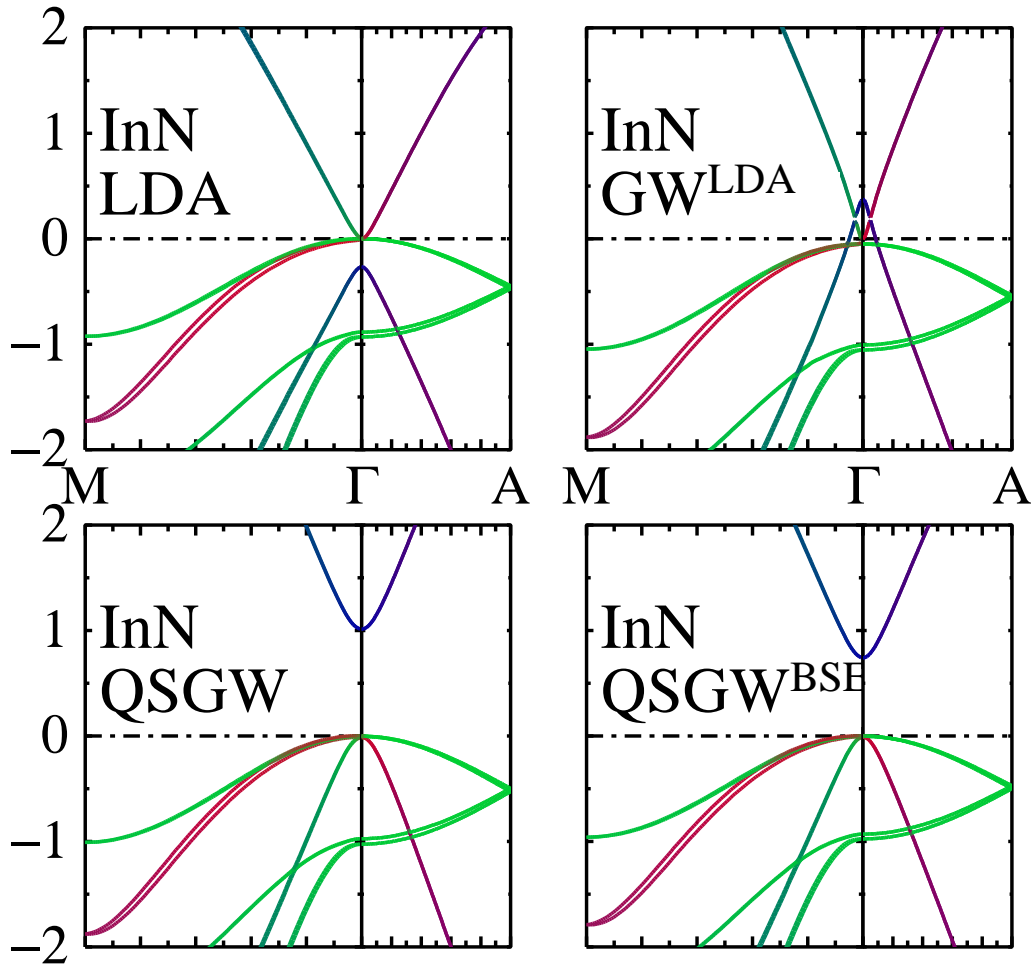}
\end{center}
\caption{\small\sffamily Low-energy band structure of wurzite InN at four different levels of approximation.  Colors
  depict orbital character of the bands: red for N $p_{xy}$ character, green for N $p_{z}$ character, blue for In $s$
  character.  LDA bands are shown in top left panel: the state of In $s$ character at $\Gamma$ lies below the three
  states of N $p$ character, reflecting an inverted gap.  Upper right panel shows the effect of \emph{GW} treated
  perturbatively from the LDA, i.e. Eq.~\ref{eq:gwlin} with $Z{=}1$ for reasons explained in the text.  \emph{GW}
  rectifies the inverted gap at $\Gamma$, but without off-diagonal parts of $\Sigma$ it cannot undo the wrong topology
  given by the starting point, and thus the bands cross near $\Gamma$.  Bottom left panel is the classic QS\emph{GW}
  result, Ref.~\onlinecite{QSGW_paper}.  It provides a good description of the InN energy bands; however the gap is
  overestimated (1.01\,eV) relative to experiment (0.68\,eV).  The QS\emph{GW} dielectric constant $\varepsilon_\infty$
  is calculated to be 6.1, about 3/4 of the measured value (8.4).  Bottom right panel is the QS\emph{GW} result with $W$
  augmented by ladder diagrams.  The gap (0.74\,eV) is slightly larger than experimental one, and differs by
  approximately the electron phonon interaction (estimated to be 0.07\,eV~\cite{Cardona05}).}
\label{fig:winn}
\end{figure}

Perhaps the first study applying \emph{GW} to a correlated material was the work of Aryasetiawan and
O. Gunnarsson~\cite{ferdi94}, in which case a starting point better than the LDA becomes essential.  This issue arises
for many kinds of narrow-band systems, and even in weakly or moderately correlated ones the starting point can be
important.  Narrow-gap semiconductors in which the LDA has a negative gap offer one notable illustration of this.  Using
\emph{GW} in the usual manner (correcting the reference eigenvalues via Eq.~\ref{eq:gwlin}) cannot correct the wrong
topology of the starting point~\cite{mark06adeq}.  InN is a classic example (Fig.~\ref{fig:winn}). Even while the
states at the $k$-point $\Gamma$ have the correct ordering, the improper initial ordering leads to unphysical
dispersions in the band structure in the vicinity of $\Gamma$. Other systems which fall into this class are Ge, PbTe,
InAs, and InSb.  (In PbTe, a gap appears at L within the LDA, but with L$_{6}^{+}$ and L$_{6}^{-}$ wrongly
ordered; see Fig.\,13 of Ref.~\cite{questaal_paper}).

Another effect of self-consistency is to modify the one-body part of $H$ or $G$.  This is because not only the
eigenvalues but the density is significantly renormalized relative to the LDA.  $GW$ induces a corresponding change in
the effective potential through the inverse of the susceptibility, $\chi^{-1}(x_1,x_2) = \delta V(x_1)/\delta n(x_2)$.
Starting from the perturbation $\delta V^0 = \Sigma^\mathrm{QSGW}[G^0_\mathrm{LDA}]-V^\mathrm{xc}[G^0_\mathrm{LDA}]$, and
if we assume that $\chi^{-1}(x_1,x_2)$ is adequately approximated by the LDA,  we can estimate the change in $n$
from $\delta{n} = \chi^0\delta{V^0}$, and from this obtain the attendant screening potential as
\begin{equation}
\delta V^\mathrm{scr} \approx V^\mathrm{Hxc}[n_\mathrm{LDA}+\delta{n}]-V^\mathrm{Hxc}[n_\mathrm{LDA}].
\label{eq:pertdeltan}
\end{equation}
Here $V^\mathrm{Hxc}$ is the combined Hartree + (LDA) $V^\mathrm{xc}$.  In practice the Questaal codes execute an
operation similar to this in the natural course of self-consistency: an internal loop is performed in the one-body code
adding the fixed $\Sigma^\mathrm{QSGW}$ as an external potential and making the density self-consistent.  This
accelerates convergence to self-consistency, but for the present we use that process to estimate the effect of $\delta
V^\mathrm{scr}$ on the bandgap.  In Table~\ref{tab:iteration0} we compare the bandgaps for \emph{GW} generated from the
LDA, in various forms.  It compares the usual Eq.~(\ref{eq:gwlin}) (with $Z{=}1$), \emph{GW} including the full
matrix structure of $\Sigma$ without updates to the one-body hamiltonian~\cite{Michiardi14}; an estimate for the change
in one-body potential as just described; and finally QS\emph{GW}.  A key take-away is that the off-diagonal parts of
$\Sigma$ are unimportant only in the simplest nearly homogeneous systems, such as Ge.  Even in SrTiO\textsubscript{3}, a
simple $d^0$ transition metal compound, they are significant, modifying the eigenvalues both directly, and indirectly
through changes in the density.  Another important finding is that both direct and indirect contributions vary widely in
both magnitude and sign, and indeed the change is often larger than the well-recognized need to account for the
electron-phonon contribution~\cite{Miglio20}.

One solution is to perform partial eigenvalue-only self-consistency: i.e. use Eq.~(\ref{eq:gwlin}) in a self-consistent
manner by updating the eigenvalues without changing eigenfunctions.  There is a simpler way to approximate
eigenvalue-only self-consistency by simply omitting the $Z$ factor in Eq.~(\ref{eq:gwlin}).  This was shown to be
rigorously true for a two-level system in the Appendix of Ref.~\onlinecite{mark06adeq}.  Eigenvalue-only self-consistency can
significantly reduce the discrepancies with experiments, but it cannot resolve the topology problems or the
modifications to the density noted above.  Further, the off-diagonal parts of $\Sigma$ can have nontrivial effects on
the quasiparticle spectrum, as noted for example in the discussion around Fig.~\ref{fig:winn}.  Another solution is to choose a better starting point, e.g. based on an extension of LDA --- as for example using a
hybrid DFT approach~\cite{PSSB:PSSB200945204} or LDA+U \cite{anisimov1997first} --- or the Coulomb-hole Screened
exchange approximation (COHSEX)~\cite{hedin}.  Since the starting-point dependence can be chosen freely, the theory
loses its \emph{ab initio} flavor.  This freedom is lost with QS\emph{GW}, and errors that appear better reflect the
nature of the approximations made.
% --- Table 0th iteration ----
\begin{center}{\begin{threeparttable}[h]
\caption{Dependence of the bandgap on various kinds of treatment in the off-diagonal parts of the self-energy.  In all cases
  the starting hamiltonian is the LDA.  $E_{G}(\Sigma^\mathrm{diag})$ is the outcome from a treatment similar to the
  usual way $GW$ is employed (Eq.~\ref{eq:gwlin} but with $Z{=1}$).  $E_{G}(n_0)$ adds the full
  $\Sigma-V^\mathrm{LDA}_{xc}$ to the LDA hamiltonian, including the off-diagonal elements, but without updating the
  density. $E_{G}(n_0{+}\delta{n})$ is similar to $E_{G}(n_0)$ but the density is updated in a ``small'' loop keeping
  $\Sigma$ fixed, as described in the text around Eq.~(\ref{eq:pertdeltan}). QS\emph{GW} is the quasiparticle
  self-consistent result (\qsgwl\ result in parenthesis).  Values reported for Ge, GaSb,
  and TiSe$_{2}$ are for the direct gap at $\Gamma$, with TiSe$_{2}$ in the high-temperature $P\bar{3}m1$ phase. In each
  of these three cases, the valence and conduction band edge states are inverted in the LDA, similar to
  Fig.~\ref{fig:winn}.}
\begin{tabular}{@{\hspace{0.6em}}l@{\hspace{0.6em}}|c|c|c|c}\hline
           & $E_{G}(\Sigma^\mathrm{diag})$
                  & $E_{G}(n_0)$
                           & $E_{G}(n_0{+}\delta{n})$
                                  &  QS\emph{GW} \cr
Ge         & 1.11 & 1.10   & 1.07 & 1.18 (1.06)  \cr %
GaSb       & 0.83 & 0.88   & 0.85 & 1.14 (1.01)  \cr %
CdO        & 0.58 & 0.53   & 0.63 & 1.52 (1.18)  \cr %
ZnO        & 3.13 & 3.04   & 3.15 & 4.12 (3.61)  \cr %
CaO        & 6.92 & 6.81   & 6.69 & 7.61 (7.06)  \cr %
MnO\tnote{1}
           & 1.54 & 1.55   & 1.98 & 3.77 (3.05)  \cr %
LiF\tnote{2}
           & 14.5 & 14.6   & 14.7 & 15.9 (14.6)  \cr %
MnTe       & 0.98 & 0.81   & 0.89 & 1.60 (1.36)  \cr %
SrTiO$_3$\tnote{3}
          & 2.54 & 2.19   & 1.89 & 4.56 (4.04)  \cr %
TiSe$_{2}$\tnote{4}
           & 0.23 & 0.30   & $-$.37 & $-$.25 ($-$.25)  \cr %
CeO$_{2}$\tnote{5}
          & 5.90 & 4.92   & 2.73  & 4.93 (4.24)  \cr %
{La\textsubscript{2}CuO\textsubscript{4}}\tnote{6}
          & 0.05 & 0.24   & 0.43  & 3.09 (1.67)  \cr %
\hline
\end{tabular}
\label{tab:iteration0}
\begin{tablenotes}[para,flushleft]\footnotesize
\item[1] \S\ref{ss:mno}
\item[2] \S\ref{ss:lif}
\item[3] \S\ref{ss:srtio3}
\item[4] See Ref~\cite{Acharya21a}
\item[5] \S\ref{ss:ceo2}
\item[6] \S\ref{ss:lsco}
\end{tablenotes}
\vbox{\vskip 10pt}
\end{threeparttable}}\end{center}
\subsection{Motivation for QS$G\hat{W}$}\label{ss:preview}
QS\emph{GW} is already a good approximation in many systems, but it is well known that discrepancies with experiment
appear.  They tend to be very systematic, and mostly related to the RPA approximation to \emph{W}.  Bandgaps being
systematically overestimated, the high-frequency dielectric constant $\varepsilon_\infty$ underestimated, and blue shifts
in peaks in Im\,$\varepsilon(\omega)$ - all fairly universal with QS\emph{GW} - are connected to the RPA approximation to
\emph{W}.  It has long been known, starting from independent work in the groups of Louie~\cite{Rohlfing98b} and of
Reining~\cite{Reining02}, that if the RPA is extended to include ladder diagrams, optical response is significantly
improved in simple semiconductors.

Our primary focus here is to determine to what extent ladder diagrams in \emph{W} ameliorate these discrepancies.  As we
will show here, when $W$ is extended to $\hat{W}$ and the cycle carried through to self-consistency, many of the systematic
errors in the QS\emph{GW}\textsuperscript{RPA} self-energy are ameliorated to a remarkable degree for a wide range of weakly
and strongly correlated insulators.  (We restrict outselves to insulators since that is where ladders are most
important~\cite{noteb}.) While this is encouraging, some discrepancies remain, and these form a major focus of this paper.  A very important
feature of QS\emph{GW}\textsuperscript{RPA} has been that when discrepancies with experiment appear, the origin can
often be clearly associated with a particular missing diagram, enabling the possibility for a systematic, hierarchical
extension of the theory.  We will show that this remains mostly true with $\Sigma=iG\hat{W}$: \qsgwl\ improves on
\qsgw\ but systematic errors remain.  The following omissions account for many of the shortcomings in results presented
in this paper.

\subsubsection{Shortcomings in \qsgwl}\label{ss:fidelity}
\begin{enumerate}[leftmargin=*]
\item \textbf{The electron-phonon interaction} is a well-identified contribution to the self-energy and, if lattice
  vibrations are in the harmonic approximation, consists of two contributions (Fan and Debye Waller terms)~\cite{hedin}.
  The diagram usually reduces insulating bandgaps; it also is needed to capture optical transitions between states of
  different wave numbers, e.g. in indirect gap semiconductors.  For its effect on the index of refraction, see
  \S\ref{ss:outliers}.

\item \textbf{Omission of $\Gamma$ in the exact self-energy $iGW\Gamma$}.  As noted in the last paragraph of the
  introduction, there is a partial renormalization $Z$-factor connecting $G$ to $G^0$ (see Appendix A of
  Ref.~\onlinecite{QSGW_paper}), which we rely on in the \qsgwl\ approximation.  Typically this vertex pushes down all
  the states in an approximately uniform manner with a minimal effect on the bandgap~\cite{Gruneis14}.  The effect is
  more pronounced for a nearly dispersionless \emph{d} or \emph{f} state, and when such a state comprises the valence
  band maximum, the gap is underestimated.  Semicore \emph{d} states in semiconductors such as CdTe and GaSb lie about
  0.7\,eV above photoemission experiments (Fig~\ref{fig:dcorelevels}).  Also, bandgaps in materials systems whose
  valence band consist of a 3\emph{d} state, or a strong admixture of it, tend to be too small
  (Table~\ref{tab:dvbmgaps}).  An extreme manifestation of this is EuO: the valence band maximum consists of a nearly
  dispersionless, atomic-like $f$ state, and as a result the QS\emph{GW} gap is underestimated~\cite{An11}.  From this
  calculation it was inferred that the Eu 4$f$ state should be pushed down by $\sim$0.7\,eV --- a somewhat larger shift
  than for a flat 3\emph{d} state (presumably it is even more atomic like).  A shift in core-like \emph{d} levels of
  order 0.5\,eV was first explicitly demonstrated by Gr\"uneis et al., who introduced a simple first-order vertex into
  $G^\mathrm{LDA}W^\mathrm{LDA}$~\cite{Gruneis14}.  Very recently Kutepov added a first-order vertex in a somewhat more
  rigorous manner~\cite{Kutepov22}.

  As regards the present work the most important error seems to occur with systems with shallow core-like levels,
  particularly when they occur near the valence band maximum.  See \S\ref{ss:otheroutliers} and also the discussion
  around Table~\ref{tab:dvbmgaps} for instances where this neglect is important.

\item \textbf{Higher order diagrams in the polarizability}.  The interaction kernel $W$ (Eq.~\ref{eq:BSE_PW}) is taken
  from the RPA and moreover it is assumed to be static.  Other diagrams, have been considered in a few works, e.g.  the
  second order screened exchange~\cite{Bechstedt18}.  This diagram when augmenting the RPA, was quite successfully used
  to predict total energies in chemical systems~\cite{Ren13}.  We consider only one additional diagram, namely to use
  $W^\mathrm{BSE}$ as the kernel in generating Eq.~\ref{eq:BSE_PW}, and note its effects on a few systems
  (\S\ref{ss:tda}).

\item \textbf{Inadequate treatment of spin fluctuations}.  In the theory presented here, the only spin contribution to
  the self-energy comes from the Fock exchange.  We present some spectral properties of correlated antiferromagnetic
  insulators (\S\ref{ss:antiferro}), and show that even in such correlated cases, the response in the charge channel
  seems to be reasonably described.  This is likely because, in contrast to spin fluctuations, charge fluctuations sense
  the long range coulomb interaction.  The situation may be different when the gap closes or becomes small on the scale
  of spin excitations ($\lesssim$0.1\,eV).  In such cases there may be cross coupling between spin and charge channels.
  Our solution to date has been to augment QS\emph{GW} with Dynamical Mean Field Theory (DMFT).  DMFT is a
  nonperturbative method and exact solutions are possible with e.g. Continuous Time Quantum Monte Carlo~\cite{gull} that
  include all diagrams.  However the vertex is assumed to be local, which is reasonable for spin fluctuations as the
  vertex is thought to predominantly reside on-site among the correlated orbitals where the fluctuations
  occur~\cite{Friedrich21}.  Indeed the {QS\emph{GW}\textsuperscript{\footnotesize{++}}} framework ($^{++}$ referring
  generally to extensions of QS\emph{GW}), augmented either by ladder diagrams or by DMFT, does seem to have
  unprecedented predictive power in a number of strongly correlated
  materials~\cite{Sponza17,Acharya19,Baldini20,Acharyalafeas,swagfese,AcharyaFeSenematicity,acharyasro2021}.  Yet there
  are places where a nonlocal vertex may be important, e.g. to explain the nematic phase of FeSe. Some approaches have
  been formulated to improve on DMFT, e.g. the ``D$\Gamma$A'' approximation --- a nonperturbative, semilocal approach
  ~\cite{DGAPhysRevB.75.045118}, but it is extremely demanding in practice.

  In addition to the T matrix~\cite{Friedrich21}, somewhat more sophisticated low-order diagrams that treat spin
  fluctuations on the same footing as charge fluctuations have been proposed~\cite{Stepanov19}, but this has not been
  attempted yet in an \emph{ab initio} context.  As spin fluctuations tend to be low energy, many channels are possible,
  so whether a low-order theory is sufficient or not remains an open question.  A low-order perturbation theory that
  could replace DMFT would be very advantageous, since DMFT has its own unique set of challenges.  We do not consider
  such cases in the present work, but it should be noted that the claim that charge fluctuations are well described
  already in low order is not universal~\cite{Gatti23} and whether a low-order perturbative theory can be sufficient
  remains an open question in low-density and strongly correlated metals.

  Perhaps surprisingly, this obvious deficiency does not seem to play a significant role in the systems we study here.
  The present work considers only systems with bandgaps, and the likely explanation is that spin wave frequencies are
  typically small energy compared to the optical gaps, which suppresses spin fluctuations.
%  ({Fe\textsubscript{3}O\textsubscript{4}}, \S\ref{ss:fe3o4}, may be a possible exception.)

\end{enumerate}

Other discrepancies with experiment in zincblende semiconductors will be presented that do not appear to have a simple
interpretation.  Perhaps the most notable unexplained error are the errors in the $k$-dependent dispersion of the conduction
band minimum in zincblende semiconductors (\S\ref{ss:semi}).  Such systems are weakly correlated and the origin of the
error is not readily explained.  One significant possibility is that Questaal's present implementation does not include
a scheme to make the basis set truly complete~\cite{Friedrich15}.  This would not be a limitation of the theory itself,
but in its implementation.

Finally, several outliers are noted often because the distinction between optical gaps and fundamental gaps is ignored,
e.g. in ScN (\S\ref{ss:scn}), SrTiO\textsubscript{3} (\S\ref{ss:srtio3}), {TiO\textsubscript{2}} (\S\ref{ss:tio2}), and
CuAlO\textsubscript{2} (\S\ref{ss:cualo2}), or are  likely artifacts of inaccurate experiments, e.g. \emph{h}BN (see
discussion around Table~\ref{tab:neph}), and in correlated systems where the experiments are less reliable.  FeO
seems to be an extreme example of this (\S\ref{ss:feo}).  The connection between discrepancies in one-particle
properties and those in two-particle properties is discussed in \S\ref{ss:otheroutliers}.

\subsubsection{Relation between L\qsgw\ and \qsgw}\label{ss:lqsgw}

We noted earlier that Kutepov constructed both a quasiparticlized scheme and a fully self-consistent one.  His
quasiparticlized scheme (L\qsgw) is somewhat different from \qsgw.  They are similar, but with the extensions to RPA
presented here and in his work, the fidelity becomes high enough that the difference can be significant.  To show this
we present here a brief analysis of the relation between L\qsgw\ and \qsgw.

Kutepov quasiparticlizes the self-energy with $\Sigma(\omega{=}0)$, but folds in an effective energy dependence through
$\Sigma'(\omega{=}0)$ (L\qsgw) while preserving the ability to construct a static hamiltonian.  The Appendix derives a
rough estimate for the expected difference in QP levels between L\qsgw\ and \qsgw, obtaining leading contribution from
the omitted quadratic term; (Eq.~\ref{eq:lqsgwE}).

\begin{figure}[h!]
\includegraphics[width=0.25\textwidth,clip=true,trim=0.0cm 0.0cm 0.38cm 0.0cm]{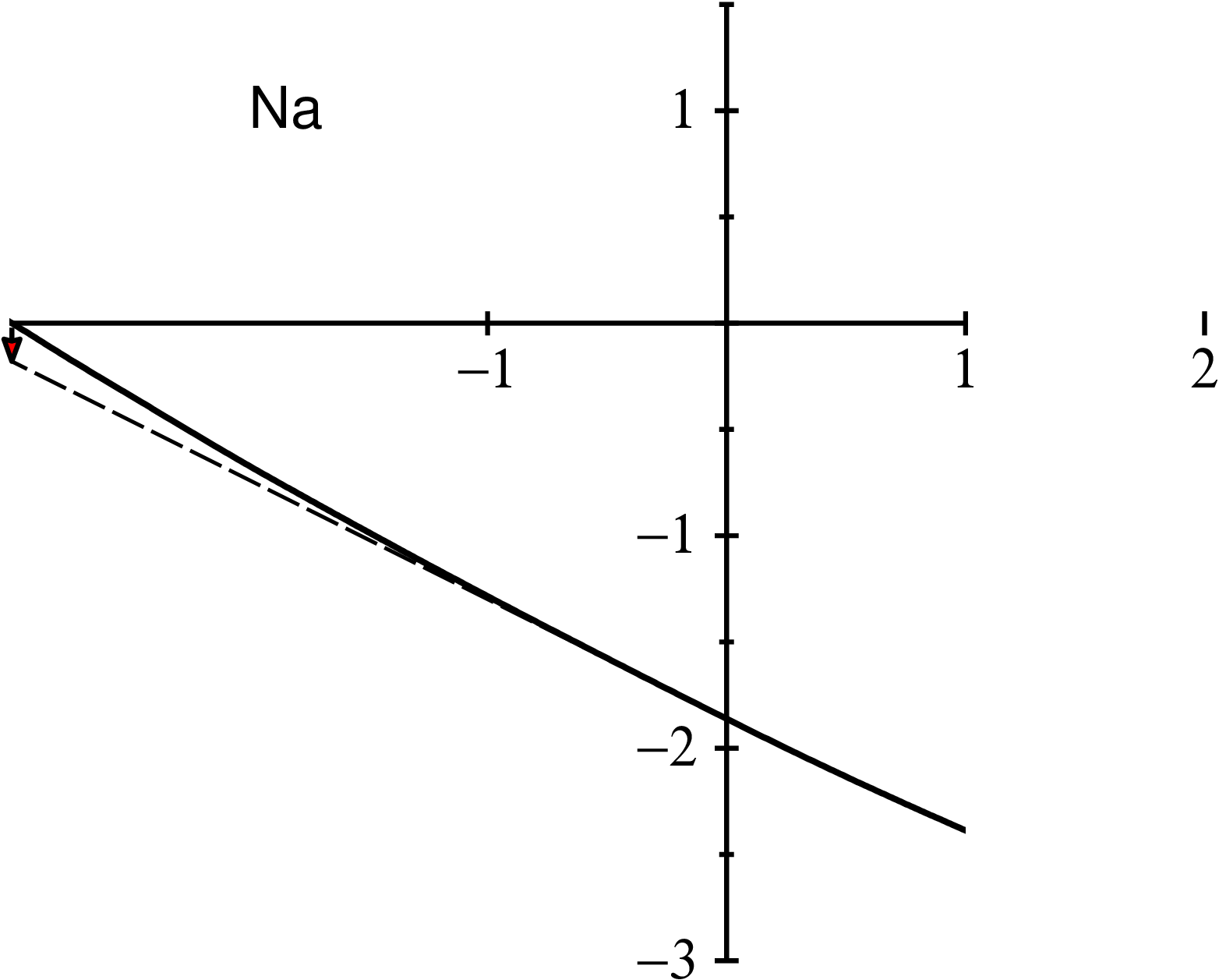}
\includegraphics[width=0.20\textwidth,clip=true,trim=0.0cm 0.0cm 0.0cm 0.0cm]{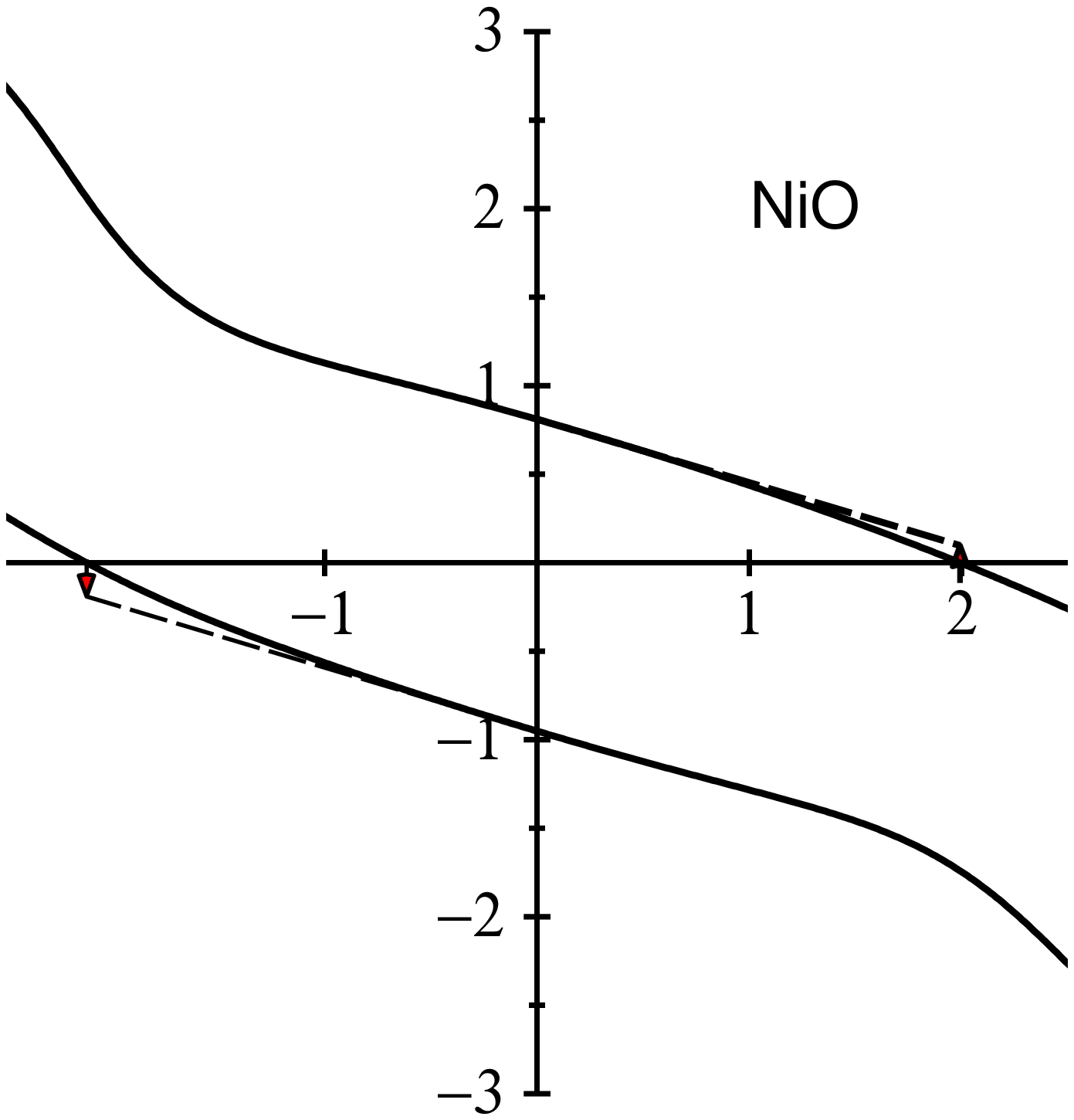}\\
\vbox{\vskip 12pt}
\begin{tabular}{|@{\hspace{0.6em}}c@{\hspace{0.6em}}|@{\hspace{0.6em}}c@{\hspace{0.6em}}|@{\hspace{0.6em}}c@{\hspace{0.6em}}|@{\hspace{0.6em}}c@{\hspace{0.6em}}|}\hline
3.16\footnote{L\qsgw, Ref.~\cite{Kutepov17}} &
2.90\footnote{\qsgw\ in LAPW basis Ref.~\cite{Friedrich22}} &
2.96\footnote{\qsgw, this work and Ref.~\cite{QSGW_PRL}} &
2.78\footnote{Experimental result of Ref.~\cite{Potorochin22}} \cr
\hline
\end{tabular}
\caption{(Left) Dynamical self-energy $\Sigma(\omega){-}\Sigma(\omega_\mathrm{QP})$ of the lowest valence band at the $\Gamma$
  point in Na, as function of $\omega$ (in eV).  (Right) The same for the highest valence band and lowest conduction band at the X point in NiO.  In all
  cases the QP level corresponds to the energy where $\Sigma(\omega){-}\Sigma(\omega_\mathrm{QP})$ crosses zero ($-$2.96\,eV in
  Na, $-$2.12\,eV and $+$1.99\,eV in NiO).  The arrows indicate the potential difference between L\qsgw\ and \qsgw for a
  particular state, at the \qsgw\ QP energy.  Table: some values for the Na bandwidth calculated from different variants
  of \qsgw, and the bandwidth from a recent experiment (in eV).  }
\label{fig:lqsgw}
\end{figure}

In the \qsgw\ scheme, the diagonal element of static (quasiparticlized) $\Sigma^0_{nn}$ is by construction equal to
dynamical self-energy at the quasiparticle level $\omega_\mathrm{qp}$, so
$\Sigma_{nn}(\omega)=\Sigma^\mathrm{QSGW}_{nn}$ at $\omega{=}\omega_\mathrm{qp}$.  For L\qsgw\ this is no longer true;
thus the quasiparticle levels of the static $G_{0}$ do not coincide with the poles of $G$.  We can estimate the
difference between the L\qsgw\ and \qsgw\ QP levels from the difference between $\Sigma_{nn}(\omega_\mathrm{qp})$ and the
linear approximation to it, $\Sigma_{nn}(\omega{=}0)+\omega_\mathrm{qp}\cdot\Sigma_{nn}'(\omega{=}0)$.  This is depicted in Fig.~\ref{fig:lqsgw} for
the first band in Na, and the highest valence band and lowest conduction band in NiO.  In Na, it corresponds to the
\qsgw-L\qsgw\ change in bandwidth; in NiO, the change is the \qsgw-L\qsgw\ difference in the direct gap at X.

For the Na case, according to the simple perturbative expression, Eq.~\ref{eq:lqsgwE}, L\qsgw\ and \qsgw\ should differ
by 0.11\,eV. A better estimate is the difference noted in the previous paragraph.  The graphs of Fig.~\ref{fig:lqsgw}
indicate that the L\qsgw\ bandwidth in Na should be slightly larger than \qsgw, and the NiO bandgap also slightly
larger.  Numerically the difference in self-energies in the Na case, at the \qsgw\ QP energy, is 0.17\,eV.  According to
first order perturbation theory, this is the expected difference between L\qsgw and \qsgw\ QP energies.  Indeed the
discrepancy between L\qsgw and \qsgw\ appears to be of this order: one L\qsgw\ and two \qsgw\ calculations have been
reported for Na\footnote{A precise comparison cannot be made because in the three implementations presented in the table
are all different, and yield slightly different results for obstensibly the same theory.  Perhaps the most rigorous
implementation of \qsgw\ is the SPEX implementation used in Ref.~\cite{Friedrich22}.} (see Table in
Fig.~\ref{fig:lqsgw}).  0.17\,eV is similar to the spread between \qsgw\ and a recent photoemission
measurement~\cite{Potorochin22}; see the Table in Fig.~\ref{fig:lqsgw}.  As \emph{GW} is known to break down at
sufficiently low densities, an accurate determination of the bandwidth in Na is important since it is one of the best
realizations of a nearly homogeneous low-density metal.

%\qsgw\ should also be better than L\qsgw\ in another sense: it provides an optimal path for computing the total energy
%through the adiabatic connection formalism~\cite{QSGW_paper}.

\subsubsection{Hybrid \qsgw\ self-energies}\label{ss:hybrid}

As Questaal has no implementation of the electron-phonon vertex as yet, or the vertex modifying $GW$, we cannot
evaluate its effect \emph{ab initio}.  However, by perturbing slightly the \qsgwl\ self-energy with an admixture of the
\qsgw\ $\Sigma$ when the gap is underestimated or LDA $V^\mathrm{xc}$ when it is overestimated, we can modify to $\Sigma^h$ to
reach a target bandgap without affecting the eigenfunctions too severely.  That permits us to assess the effect of the
error in $E_{G}$ on $\varepsilon_\infty$.  Table~\ref{tab:neph} presents cases where both $E_{G}$ and
$\varepsilon_\infty$ are well known, and it establishes that discrepancies in the two are intimately connected for
several systems.  That provides an independent confirmation that the one-body hamiltonian would be of high fidelity if
this perturbation were properly included.

To make a reasonable proxy to the \qsgwl\ self-energy, e.g. for the electron-phonon self-energy and missing vertex noted
in \S\ref{ss:fidelity} (points 1 and 2), we will construct a hybrid one-body self-energy $\Sigma^h$ defined as
\begin{align}
  \label{eq:hybrid}
  \Sigma^h &= \alpha{\cdot}\Sigma[\mathrm{QS}G\hat{W}] + \beta{\cdot}\Sigma[\mathrm{QS}GW] + \gamma{\cdot}V^\mathrm{LDA}_\mathrm{xc}\\
  1        &= \alpha + \beta  + \gamma \nonumber
\end{align}
This equation has often been used with $\alpha$=0, $\beta$=0.8, $\gamma$=0.2, because $\varepsilon_\infty$ computed from
\qsgw\ has been found empirically to be very nearly 80\% of the true value for a wide range of semiconductors (see
Fig.~\ref{fig:gapsandepsinfty}\emph{b}).  This formula has been empirically found to yield very good bandgaps in
many kinds of materials sytems~\cite{Chantis06a,Deguchi16}.  In \S\ref{ss:otheroutliers}
we use it to show how the errors in $\Sigma$ (whatever their origin) are closely connected to discrepancies in
the dielectric function.  A caveat should be noted here: even while the bandgap can be rendered accurate with
such a hybrid self-energy, \epsi\ computed from the RPA is not similarly improved, so the theory cannot capture both
quantities in a consistent manner.  \S\ref{ss:consistency} discusses this at greater length.

We will also assume ths connection to hold in cases where the fundamental gap is uncertain while $\varepsilon_\infty$ is
better known.  By aligning $\varepsilon_\infty$, or the frequency-dependent $\varepsilon(\omega)$ with measured data, we
can make a reasonable estimate for the fundamental gap.  This is done for several systems, e.g.  {TiO\textsubscript{2}}
(\S\ref{ss:tio2}) and FeO (\S\ref{ss:feo}).

\subsection{Numerical evaluation of the kernel matrix elements}\label{ss:lmto}

Our numerical implementation of the BSE relies on a generalization of the linear muffin-tin orbital
basis~\cite{methfessel_lmto,QSGW_paper,fusion}. The eigenfunctions are expanded in Bloch-summed muffin-tin orbitals in
spheres around atom centers.  The radial part of the eigenfunctions in these spheres is expanded by numerical solutions
of the radial Schr\"{o}dinger equation.  In the region between the spheres, the eigenfunctions are then expanded in
either smoothed Hankel functions~\cite{fusion} and/or plane waves.  Expanding the interstitial in plane waves, the
eigenfunctions are
\begin{equation}\label{eq:eig_exp}
\Psi_{n\boldsymbol{k}}(\boldsymbol{r})=\sum_{\boldsymbol{R}u}\alpha_{\boldsymbol{R}u}^{\boldsymbol{k}n}\varphi_{\boldsymbol{R}u}^{\boldsymbol{k}}(\boldsymbol{r})+\sum_{\boldsymbol{G}}\beta_{\boldsymbol{G}}^{\boldsymbol{k}n}P_{\boldsymbol{G}}^{\boldsymbol{k}}(\boldsymbol{r}),
\end{equation}
where $\boldsymbol{R}$ denotes the atomic site and $u$ is a composite index that contains the angular momentum of the
site along with an index that denotes either: a numerical solution of the radial Schr\"{o}dinger equation at some
representative energy; its energy derivative (since the energy dependence has been linearized by expanding in a Taylor
series about the representative energy\cite{OKA}); or a local orbital which is a solution at an energy well above or
below the representative energy.  In $GW$ and the BSE a basis is required that expands the product of eigenfunctions.
Expanding the interstitial in plane waves, the product eigenfunctions will also be expanded in plane waves, and within
the spheres the basis is expanded by $\varphi_{Ru}(\boldsymbol{r})\times\varphi_{Ru'}(\boldsymbol{r})$.  This mixed
product basis (MPB) is denoted $M_I^{\boldsymbol{k}}(\boldsymbol{r})$.

Using the notation in Ref.~\onlinecite{QSGW_paper}, the kernel $K$ in the MPB is read as
\begin{equation}\label{kernel_eq}
%\displaystyle \bar{v}_{{\scriptsize\begin{array}{l}n_1n_2\boldsymbol{k}\\n_3n_4\boldsymbol{k}'\end{array}\normalsize}}(\boldsymbol{q})&=\displaystyle\sum_{I,J}&\langle\psi_{n_2,\boldsymbol{k}+\boldsymbol{q}}|\psi_{n_1,\boldsymbol{k}}\widetilde{M}_I^{\boldsymbol{q}}\rangle \bar{v}_{IJ}(\boldsymbol{q})\\&&\times\langle\widetilde{M}_J^{\boldsymbol{q}}\psi_{n_3,\boldsymbol{k}'}|\psi_{n_4,\boldsymbol{k}'+\boldsymbol{q}}\rangle\\&&\\
\displaystyle K_{{\scriptsize\begin{array}{l}n_1n_2\boldsymbol{k}\\n_3n_4\boldsymbol{k}'\end{array}\normalsize}}\hspace{-0.1cm}(\boldsymbol{q})\hspace{-0.1cm}=\hspace{-0.1cm}\displaystyle\sum_{I,J}\hspace{-0.12cm}\langle\psi_{n_3,\boldsymbol{k}'}|\psi_{n_1,\boldsymbol{k}}\widetilde{M}_I^{\boldsymbol{k}'-\boldsymbol{k}}\rangle W_{IJ}^\mathrm{RPA}(\boldsymbol{k}'-\boldsymbol{k};\omega=0)\langle\widetilde{M}_J^{\boldsymbol{k}'-\boldsymbol{k}}\psi_{n_2,\boldsymbol{k}+\boldsymbol{q}}|\psi_{n_4,\boldsymbol{k}'+\boldsymbol{q}}\rangle,
\end{equation}
where the matrix elements and $W_{IJ}^\mathrm{RPA}$ are calculated as in Ref.~\onlinecite{QSGW_paper}.

Owing to the huge computational demands of the BSE only a subset of transitions that occur between bands within a selected
energy range about the Fermi level are considered. Contributions from transitions not included in the BSE are, however,
included at the level of the RPA. To include such contributions, we effectively have a matrix $H$ that is diagonal
except for a block corresponding to the coupled transitions discussed above. To calculate the polarization in this case,
the RPA contribution from the subset of states that are treated at the level of the BSE are not included in the full
$P^\mathrm{RPA}$ [calculated according to Eq.~(32) in Ref.~\onlinecite{QSGW_paper}] and the contribution from $P^\mathrm{BSE}$ is
added to $P^\mathrm{RPA}$. The corrected polarization is then transformed into the MPB and the dielectric matrix
$\epsilon=1-vP$, and hence $W=\epsilon^{-1}v$, are thus obtained. The so-obtained $W_{IJ}(\boldsymbol{q},\omega)$ is
used as in Eq.~(34) of Ref.~\onlinecite{QSGW_paper} to calculate the correlation part of the self-energy.

\subsection{Divergence of the macroscopic dielectric function at $\boldsymbol{q}=0$}

The macroscopic ($\boldsymbol{G}=\boldsymbol{G}'=0$) dielectric function (head of the dielectric matrix) is constructed
from the divergent bare Coulomb interaction ($4\pi/|\boldsymbol{k}|^2$) and polarization function. Since the dielectric
matrix contains a three dimensional integral over $\kk$, the dielectric matrix for $\boldsymbol{k}=0$ itself remains
finite but angular dependent; resulting in the dielectric tensor. In this work we employ the offset $\Gamma$
method~\cite{QSGW_paper,kotani_MPB,friedrich_paper}, to treat the divergent part of $W$, where an auxillary mesh is
introduced that is shifted from the original $\Gamma$ centered mesh. The averaged macroscopic dielectric function
calculated in a small cell around $\Gamma$ is then used to calculate the macroscopic part of the screened Coulomb
interaction for $\boldsymbol{k}\rightarrow 0$, as in Ref.~\onlinecite{kotani_MPB}.

The \textbf{G}=\textbf{G'}=0 component of the irreducible polarizability should vanish at \emph{q}=0.  Owing to
numerical errors this is not exactly the case, so to correct for this its value is subtracted from the irreducible
polarizability for all $q$.  This adjustment stabilizes the calculations and also improves on the \textbf{k} convergence
of the polarizability and self-energy.  We performed careful checks for the \textbf{k}-convergence in
$\varepsilon_\infty$ in the RPA, and found for example in zincblende semiconductors an 8$\times$8$\times$8 mesh was
reasonably good, and a 12$\times$12$\times$12 mesh converged $\varepsilon_\infty$ to $\sim$1\% in all cases
but the smallest gap semiconductors.

\subsection{Including the Fr\"olich contribution to the band gap}\label{ss:frolich}

To correct the value for the band gap in this method due to the neglection of electron-phonon interactions we can
include an approximation for the contribution from the Fr\"{o}lich contribution to the Fan term, which -- in polar
insulators such as LiF---should be the dominant part. We include lattice polarization corrections (LPC) using the method
outlined in Ref.~\onlinecite{phonons_walt}. The energy shift is determined from
\begin{align}\label{eq:phon}
\Delta E_{n\boldsymbol{k}} &=\frac{e^2}{4a_p}\left(\epsilon_{\infty}^{-1}-\epsilon_0^{-1}\right)\\
a_P &= \sqrt{\frac{\hbar}{2\omega_\mathrm{LO} m^*}} = a_0\left(\frac{m}{m^*}\frac{e^2}{2a_0\hbar\omega_\mathrm{LO}}\right)^{1/2}
\end{align}
$a_P$ is the polaron length scale, which, in the effective mass approximation is computed from the optical mode phonon
frequency $\omega_\mathrm{LO}$ and the effective mass $m^*$.  $a_P$ is different for electrons and holes, and we take an
average of the electron and hole contributions, following Ref.~\onlinecite{phonons_walt}. \epsi\ is the ion-clamped
static (optical) dielectric constant and $\epsilon_0$ contains effects accounting for nuclear relaxations. The values
for $\epsilon_0$, \epsi\ and $\omega_{\rm LO}$ used in this work are taken from
Refs~\onlinecite{phonons_walt,PhysRevMaterials.2.013807} for materials discussed there.  For many of the systems studied
here a more rigorous calculations of the gap shift has been published (Ref.~\cite{Miglio20}).  Where available we use
those results.

For a given shift, we use a proxy $\Sigma^h$ (Eq.~\ref{eq:hybrid}) to estimate the effect on the band structure and
\epsi.

\subsection{Effective Oscillator model for Index of Refraction}\label{ss:oscillators}

In Ref.~\cite{Wemple71} it was established that the frequency-dependent index of refraction of many compounds can be fit
reasonably well by a single oscillators model.  The model has the form
\begin{align}\label{eq:oscillators}
n^2-1 = \frac{E_d E_0}{E_0^2 - (\hbar\omega)^2}
\end{align}
where $E_{0}$ is the oscillator energy, and $E_{d}$ is a measure of the strength of interband optical transitions.
Empirically, $(n^2{-}1)^{-1}$ has been found to be a mostly linear function of $(\hbar\omega)^2$, for a wide range of
ionic materials, which lends credence to the model.  In some experiments where $n(\omega)$ is tabulated, we use
Eq.~\ref{eq:oscillators} to extrapolate to $\omega$=0.

\section{Results and Discussion}\label{sec:res}

%We apply the \qsgwl\ to calculate the electronic structure, the dielectric properties of a broad range of materials that
%were listed in the introduction. We first discuss in detail a few prototypical materials: \textcolor{green}{FINISH}
%
%Si (\S\ref{ss:si}),
%SrTiO$_3$ (\S\ref{ss:srtio3}), LiF (\S\ref{ss:lif}), NiO and CoO (\S\ref{ss:coonio}). We then analyze the
%performance of \qsgwl\ in predicting the band gap and the static dielectric constant (\S\ref{ss:prfmnc}).

\subsection{Computational Details}

All results have been obtained using Questaal~\cite{questaal_web}. Table~\ref{tb:comp_details} contains the relevant
parameters used in the calculations.  The $\ell$ cut-off for partial waves inside muffin-tin spheres was set to $4$ and
an \emph{spdfg-spdf} basis was used in all calculations, except in some lighter systems where the \emph{g} orbitals were
omitted.  Local orbitals were also used in some systems as indicated in table~\ref{tb:comp_details}. Empty sites were
used as placements for additional site-centered Hankel functions (to $\ell_\mathrm{max}=2$ or 3) without augmentation,
to improve the basis in systems with large interstital voids.  When calculating the polarization within the RPA, the
tetrahedron method is employed for integration over the Brillouin zone~\cite{QSGW_paper}.  In the BSE implementation, a
broadening was applied according to Eq.~\ref{eq:Hminone} and set to 0.01~Ha for vertex calculations.

The TDA was also adopted due the huge increase in compution required to store, calculate and diagonalize the
non-Hermitian matrix that has twice as many rows and columns as the Hermitian TDA one. We would, however, not expect going beyond
the TDA too have to much of an impact on the systems investigated in this work~\cite{TD_myrta}.  We did remove the TDA in
a few cases, e.g. InSb, ScN, and MgO, and found the effect to be minor, as anticipated (\S\ref{ss:tda}).

\subsubsection*{Treatment of the Screened Coulomb interaction}\label{ss:q0limit}

For numerical reasons the Questaal codes compute Fock matrix elements not of the bare coulomb interaction $1/q^2$ but a
slightly screened one, $1/(q^2+\kappa^2)$.  A small value for $\kappa$ is taken, between $10^{-5}$ and $10^{-4}$.  The
QS\emph{GW} self-energy is not usually sensitive to the value of $\kappa$; however, the dielectric constant
$\varepsilon_\infty$ can vary by a few percent for $\kappa$ ranging between $10^{-5}$ and $10^{-4}$.  For that reason,
we compute $\varepsilon_\infty$ for three values of $\kappa$ between $10^{-5}$ and $3{\times}10^{-5}$ and extrapolate to
$\kappa$=0.

Also, to avoid evaluating matrix elements at $q=0$, we use the offset-$\Gamma$ method~\cite{QSGW_paper}, which requires
generating the polarizability for small values of $q$ near zero.  For $\varepsilon_\infty$ we evaluate
$\varepsilon(\omega,\mathbf{q})$ for three small finite values of $q$ and extrapolate to $q{=}0$.  The direction of
approach to $q{=}0$ gives us the orientation dependence of $\varepsilon(q=0)$.

Both kinds of extrapolations are done in one process.  The difference between extrapolated and finite-($q,\kappa$)
values can differ by a few percent.

\begin{figure}[h!]
\includegraphics[width=0.36\textwidth,clip=true,trim=0.0cm 0.0cm 0.0cm 0.0cm]{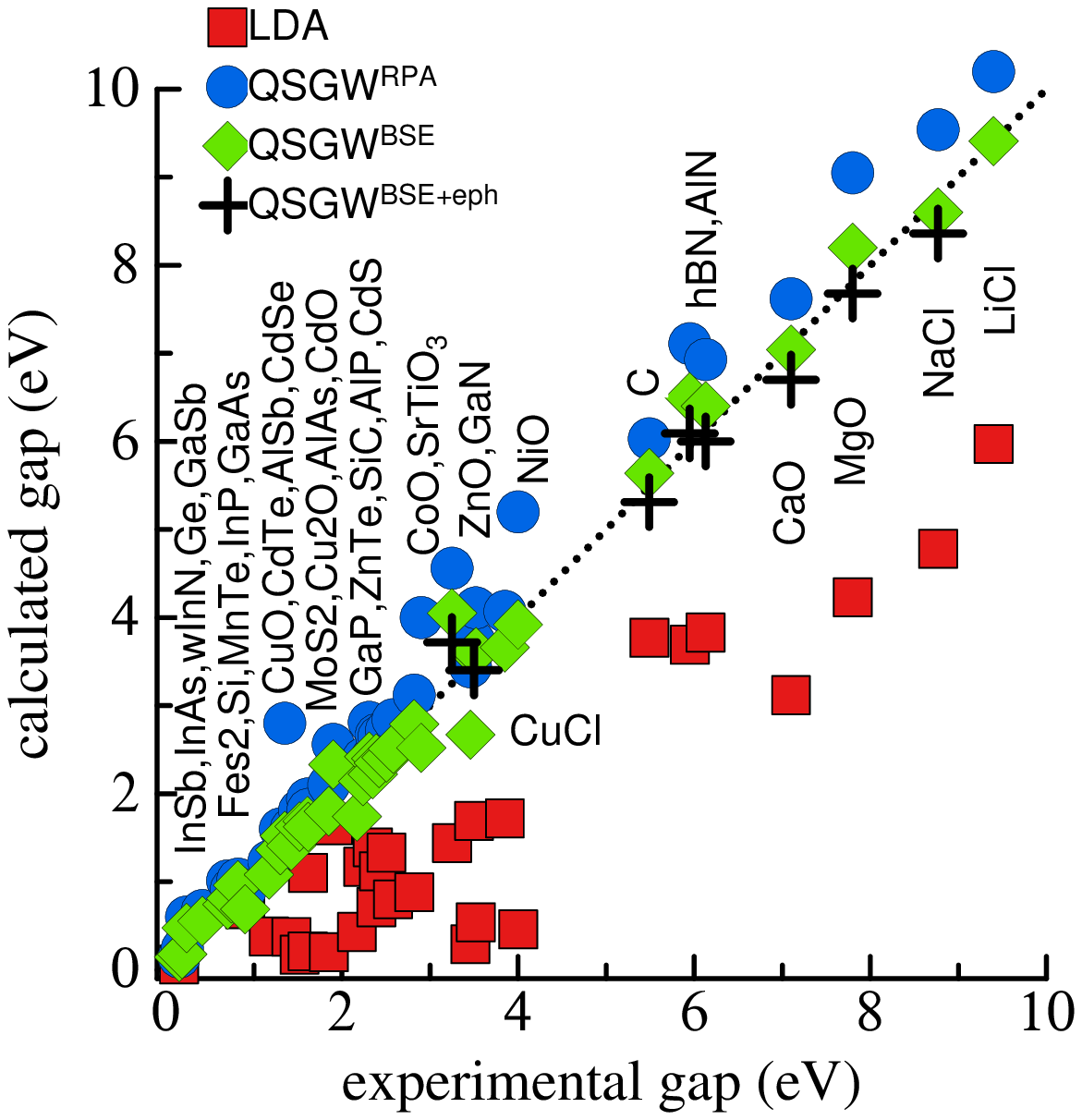}
\includegraphics[width=0.36\textwidth,clip=true,trim=0.0cm 0.0cm 0.0cm 0.0cm]{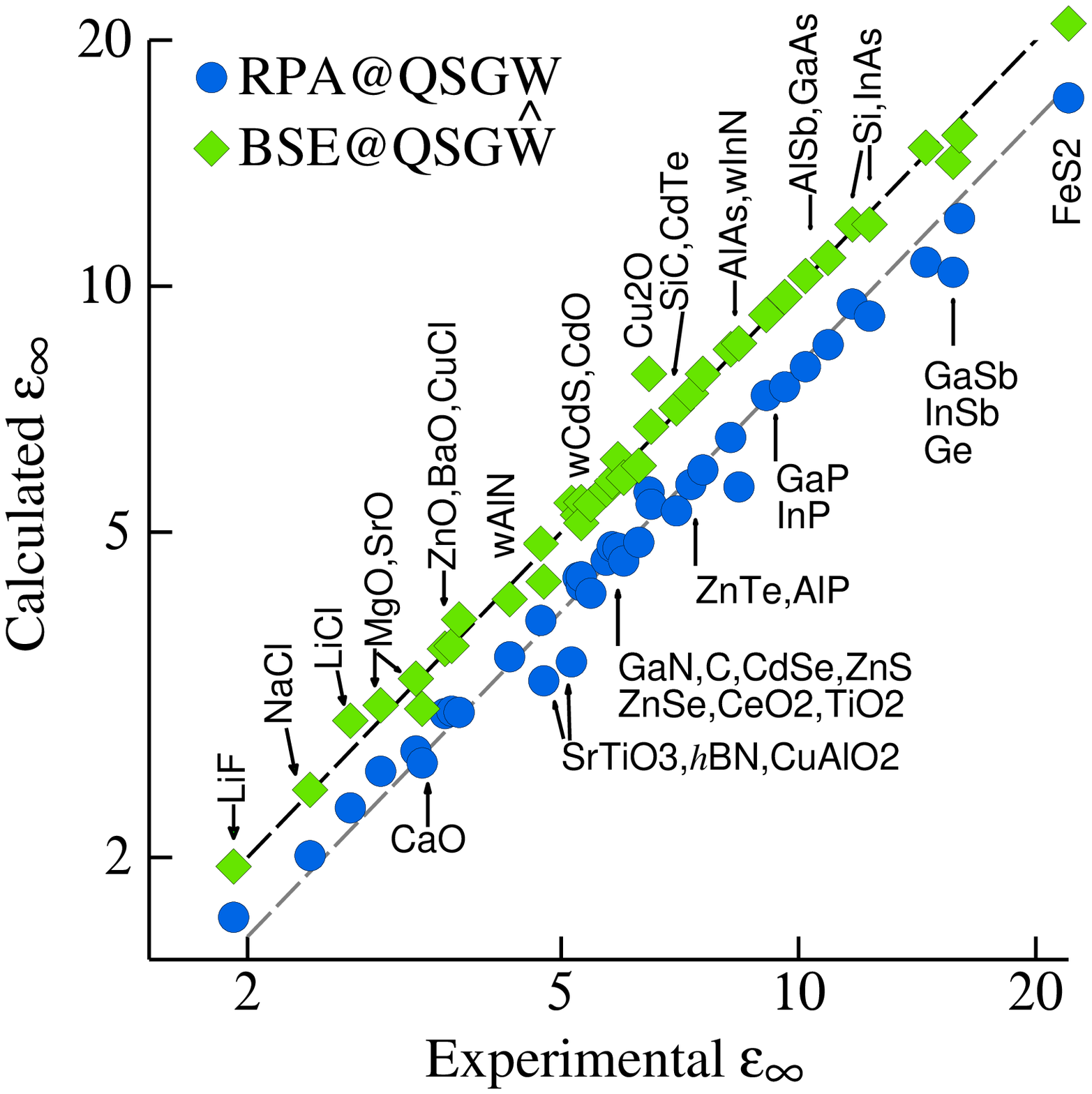}
\caption{
  (top) Fundamental bandgap for selected materials calculated within the LDA (red squares), \qsgw\ (blue circles), and
  \qsgwl\ (green diamonds).  The black crosses add to \qsgwl\ an estimate for the gap correction from the electron-phonon interaction
  when it exceeds 0.19\,eV.
  Where available, this was taken from Ref.~\cite{Miglio20}; otherwise it was estimated from the Fr\"olich expression,
  Eq.~(\ref{eq:phon}).\\
  (bottom) $\varepsilon^\mathrm{RPA}_\infty$ calculated from $G_{0}$ generated QS\emph{GW} (blue circles) and
  $\varepsilon^\mathrm{BSE}_\infty$ calculated from $G_{0}$ generated \qsgwl\ (green diamonds).  The dark dashed line corresponds
  to perfect agreement with experimental data; The light dashed line corresponds to
  $\varepsilon_\infty^\mathrm{th}/\varepsilon_\infty^\mathrm{exp}=0.8$.  For \emph{h}BN, we used $\varepsilon$ in the
  basal plane; see Table~\ref{tab:birefringence}.  Bencharking $\varepsilon_\infty$ in the antiferromagnetic oxides CuO,
  MnO, FeO, CoO, NiO and Cu is more complex.  They're omitted here but discussed in \S\ref{ss:antiferro}.
  }
\label{fig:gapsandepsinfty}
\end{figure}

\subsection{Survey of Results}\label{ss:survey}

We begin with a birds-eye view of some key results.  Fig.~\ref{fig:gapsandepsinfty} shows the fundamental bandgaps
($E_{G}$) and high-frequency dielectric constant ($\varepsilon_\infty$) for a wide variety of materials, comparing
classical \qsgw\ results to \qsgwl.  This figure elucidates the general trends: \qsgw\ tends to overestimate bandgaps,
and underestimate $\varepsilon_\infty$ by an almost universal constant factor 0.8.  As anticipated, addition of ladder
diagrams ameliorates both of these discrepancies.  Apart from a few exceptions (see discussion in
\S\ref{ss:otheroutliers}).  \qsgwl\ greatly improves on \qsgw.  On the wide scale of the figure the ability of
\qsgwl\ to predict optical properties (Fig.~\ref{fig:gapsandepsinfty}) looks stellar, but discrepancies appear on closer
inspection.  A main theme of this paper is to seek out these deviations, and associate them, where possible, with the
missing diagrams noted in Sec~\ref{ss:fidelity}.  Fortunately, most of the discrepancies with measured data are fairly
systematic, which opens the possibility that the shortcomings can be rectified with relatively simple low-order
diagrams.

\subsubsection{Index of refraction}

%e.g. in NiO reported values range between 5.43~\cite{SpicerNiOCoO} and 6.0~\cite{Chern92},

Typically, $\varepsilon_\infty$ is obtained by extrapolating the frequency-dependent index of refraction
$\varepsilon(\omega)$ to zero using, e.g., Eq.~\ref{eq:oscillators}.  Its value is known only to a resolution of a few
percent even in the best cases, and the uncertainty is often larger.  An extreme case is AlN, where several values have
been reported ranging between 3.8~\cite{Brunner97} and 4.8~\cite{Dean67}, and hBN is another instance
(\S\ref{ss:outliers}). When several experimental values are available for the compounds in Fig.~\ref{fig:gapsandepsinfty},
we use an average value.  Reported values for $\varepsilon_\infty$ for antiferromagnetic transition metal oxides (not shown
in Fig.~\ref{fig:gapsandepsinfty}) also show variations, and calculations show larger deviations from the average value.
They are discussed in \S\ref{ss:antiferro}.

There is a small indeterminacy on the theory side also.  Besides the extrapolation noted in \S\ref{ss:q0limit}, care
must also be taken to converge the uniform $k$ mesh entering into numerical Brillouin zone integration: for narrow-gap
semiconductors pushing this mesh beyond $12\times12\times12$ divisions was not feasible, leading to a slight tendency to
underestimate $\varepsilon_\infty$.  These approximations lead to an uncertainty of a few percent.

The bulk of the remaining paper focuses on discrepancies where either $E_{G}$ or $\varepsilon_\infty$ are outside the
experimental uncertainty, which appear in some systems.  One primary aim of this work is to draw a connection between
$E_{G}$ and $\varepsilon_\infty$. Generally speaking, in well-characterized systems the discrepancies in $E_{G}$ and
$\varepsilon_\infty$ occur largely at the same time: when the gap is accurately described, $\varepsilon_\infty$ is also.
We believe this to be a significant finding, and it is taken up in \S\ref{ss:consistency}.  \S\ref{ss:outliers} and
\S\ref{ss:otheroutliers} present cases where $E_{G}$ deviates the most strongly from experiment. We use hybrid
self-energies (Eq.~\ref{eq:hybrid}) to correct the gap, to see how the change in $\varepsilon_\infty$ tracks it.

\subsubsection{NiO as archetype system}\label{ss:archetype}

\begin{figure}[h!]
\includegraphics[width=0.20\textwidth,clip=true,trim=0.0cm 0.0cm 0.0cm 0.0cm]{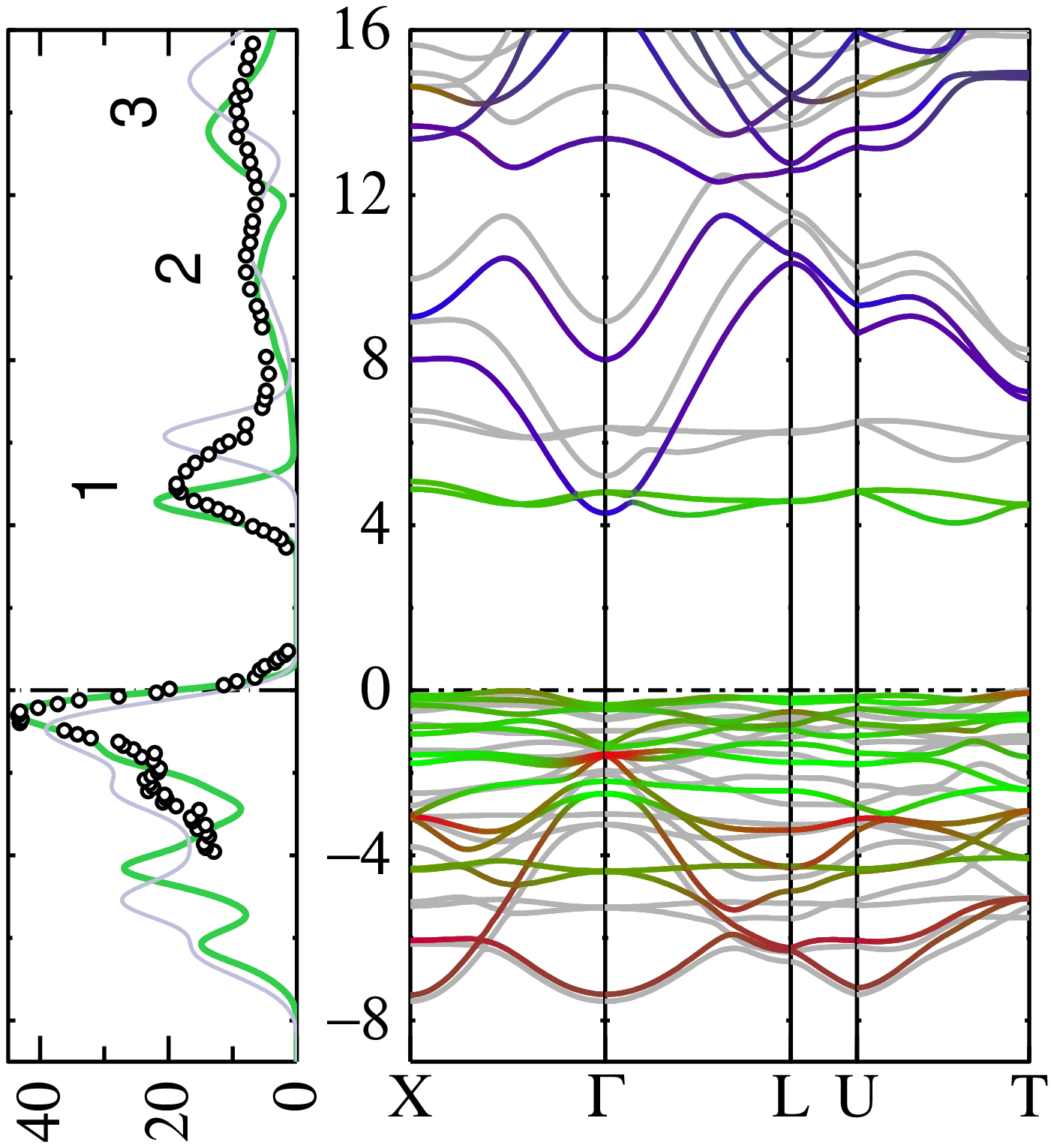}
\raisebox{-2mm}[0pt][0pt]{
\includegraphics[width=0.26\textwidth,clip=true,trim=0.0cm 0.0cm 0.0cm 0.0cm]{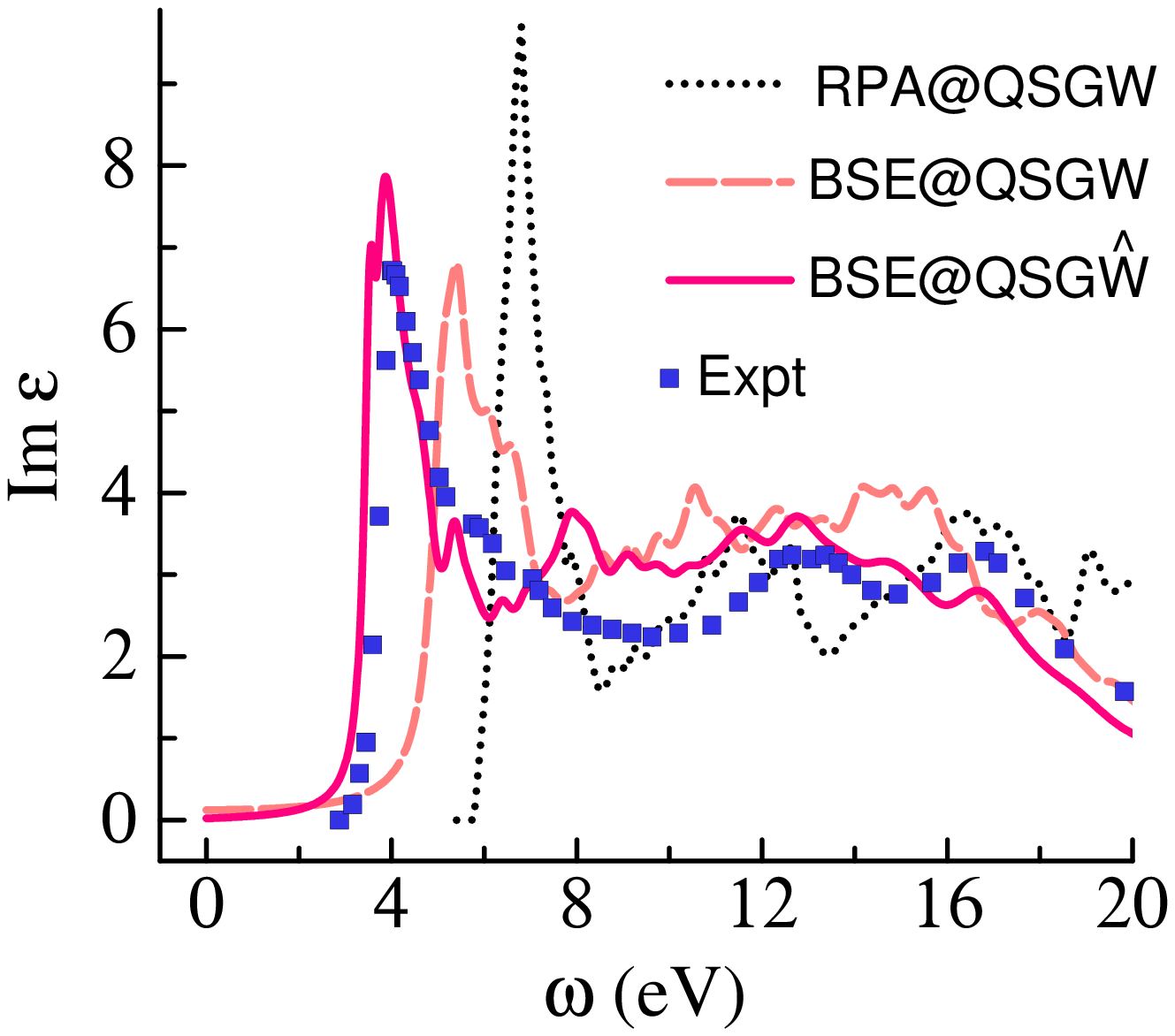}
}
\caption{
  Middle panel: Energy bands in NiO, within the \qsgwl\ approximation (colored bands), with valence band maximum at 0.  Green
  projects onto Ni \emph{d} character, blue onto Ni \emph{sp} character and red onto O \emph{p} character.  Shown for
  comparison are the \qsgw\ results (light gray bands).  Left panel: \qsgwl\ (green) and \qsgw\ (gray) DOS, compared
  against PES data ($E{<}E_{F}$) and BIS data ($E{>}E_{F}$) from Ref.~\onlinecite{Sawatzky84} (circles).
  Right: experimental dielectric function $\mathrm{Im}\,\epsilon(\omega)$ from Ref.~\cite{SpicerNiOCoO}, compared to
  results calculated at three levels of approximation: RPA@\qsgw, BSE@\qsgw, and BSE@\qsgwl.  As is typical, the
  shoulder of RPA@\qsgw\ is blue-shifted, by roughly 2\,eV in this case.  Adding ladders (BSE@\qsgw) shifts the shoulder
  towards the experiment, but it is still $\sim$1\,eV too high, as a consequence of overestimate of the
  \qsgw\ fundamental gap (Table~\ref{tab:afminsulators}).  BSE@\qsgwl\ describes $\mathrm{Im}\,\epsilon(\omega)$ rather
  well, including peaks around 6\,eV, 13.5\,eV and 17\,eV.  However the shoulder around 3.5\,eV is slightly red-shifted
  compared to experiment, indicating that the fundamental gap is underestimated.  $\varepsilon_\infty$ is also
  overestimated.  See \S\ref{ss:nio} for more details.
}
\label{fig:niobands}
\end{figure}

NiO is an archetype system that exhibits many of the phenomena that are the subject of this work.
Fig.~\ref{fig:niobands} shows in greater detail how ladder diagrams renormalize the \qsgw\ self-energy in NiO.
This manifests as shifts in \qsgwl\ energy bands and peaks in the density-of-states (DOS).  DOS are compared to bremsstrahlung-isochromat-spectroscopy (BIS) and x-ray
photo-emission (XPS) measurements in the left panel~\cite{exptcaution}.

\begin{itemize}[leftmargin=*]

\item BIS data exhibits three peaks between 0 and 9\,eV, which the \qsgwl\ DOS captures quite well, except for a small
  underestimate of the fundamental gap seen in both BIS and optics (see \S\ref{ss:nio}).  This shows that ladders
  do an excellent job of capturing the frequency dependence of the local (\emph{k}-integrated) spectral function.

\item The corresponding \qsgw\ peaks are blue shifted relative to experiment, but in varying amounts.  Peak 1, which is
  composed almost entirely of flat Ni \emph{d} states, is shifted about 1.5\,eV while Peak 2, derived essentially of
  dispersive Ni \emph{sp} states, is shifted by about half of that.  This reflects a universal tendency: flat bands are
  affected by ladders more than dispersive ones. {Fe\textsubscript{3}O\textsubscript{4}} offers a particularly striking
  example (\S\ref{ss:fe3o4}).

\item \qsgwl\ significantly narrows the occupied Ni \emph{d} bands relative to \qsgw.  Red bands (depicting O character)
  are almost unaffected, while there is a significant narrowing of the green bands relative to \qsgw.  This is a
  potentially important finding.  It is well known that the LDA severely overestimates \emph{d} band widths in
  narrow-band transition metal compounds.  Further, it has been shown in several works, e.g. Ref.~\cite{Tomczak12}, that
  \qsgw\ narrows \emph{d} bands relative to the LDA, but nevertheless continue to overestimate these bandwidths, especially in
  systems with strong spin fluctuations such as BaFe$_{2}$As$_{2}$ and FeSe.  In cases we have studied where
  experimental information is also available, e.g. in Sr\textsubscript{2}RuO4\textsubscript{4}~\cite{acharyasro2021},
  this overestimate is remedied very well by augmenting \qsgw\ with DMFT, which includes vertices in both charge and
  spin channels.  Whether the bandwidth can be captured entirely by a combination of low-order diagrams in both spin and
  charge channels remains an intriguing possibility.  To the extent it is true, this greatly simplifies the complexity of the
  electronic structure problem in correlated systems.  This will be explored in a future work.

\end{itemize}

Some more details for NiO are presented in \S\ref{ss:nio}.  Also, there are some strong parallels with
{La\textsubscript{2}CuO\textsubscript{4}}; see \S\ref{ss:lsco}.

\subsubsection{Consistency between one and two particle properties}\label{ss:consistency}

The consistency between benchmarks for one- and two-particle quantities ($E_{G}$ and $\varepsilon_\infty$ in
Fig.~\ref{fig:gapsandepsinfty}) is striking.  Apart from some outliers to be discussed in \S\ref{ss:outliers}, the
calculated values $\varepsilon_\infty$ agree with measured ones to within the available resolution.  When this is not
the case, usually there is a corresponding discrepancy in the fundamental gap: discrepancies in $E_{G}$ and
$\varepsilon_\infty$ occur largely at the same time: overestimate of $E_{G}$ yields underestimate of
$\varepsilon_\infty$, and underestimate of $E_{G}$ yields overestimate of $\varepsilon_\infty$.

The internal consistency between one- and two-particle properties is a signature of consistency of the theory, since the
same quantities (\emph{G} and \emph{W}) construct both $\varepsilon(\omega)$ and the potential $\Sigma(\omega)$ that
makes \emph{G}.

If we assume the fidelity of the theory is sufficient for this principle to be universally applicable, the extra
information provides an ansatz to predict optical properties in materials with stronger correlations, where benchmarking
is less simple.  In such cases there is often a large uncertainty in the benchmark itself, not only owing to a wide
variation in reported experimental data, but also the extraction from one-particle properties (e.g. fundamental gap)
from two-particle response functions. This is reasonable for tetrahedral semiconductors where excitonic effects are
small (see Fig.~\ref{fig:e01esemi}), but has less validity in general.  For these more correlated cases our approach
will be to compare optical experiments directly with calculated response functions.  Combining such a comparison with
the observed relation between calculated one-particle and two-particle properties, we can benchmark the theory, and
sometimes provide values of quasiparticle levels where not well known, or new interpretation of accepted values.

\begin{figure}[h!]
\includegraphics[width=0.40\textwidth,clip=true,trim=0.0cm 0.0cm 0.0cm 0.0cm]{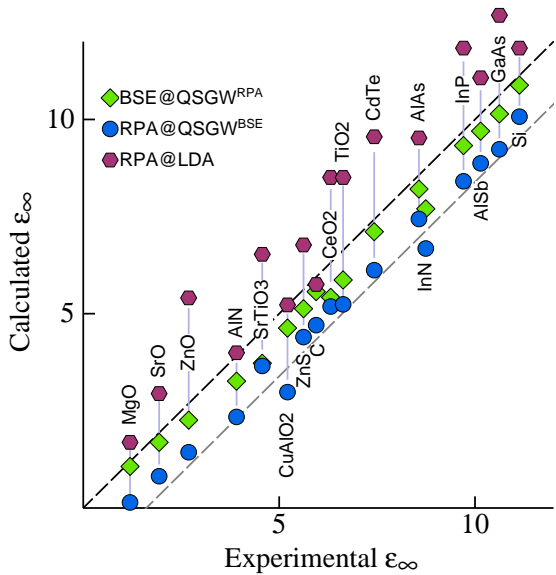}
\caption{$\varepsilon_\infty$ generated by two inconsistent approximations: $\varepsilon^\mathrm{RPA}_\infty$@\qsgwl
  (blue circles) and $\varepsilon^\mathrm{BSE}_\infty$@\qsgw\ (green diamonds), to be compared against
  Fig.~\ref{fig:gapsandepsinfty}.  Also shown are LDA results: $\varepsilon^\mathrm{RPA}_\infty$@LDA (red squares).
  Agreement is best for indirect gap semiconductors, where gaps for vertical transitions are relatively large.
}
\label{fig:epsinfty2}
\end{figure}

Fig.~\ref{fig:gapsandepsinfty} was generated from two consistent approximations:
$\varepsilon^\mathrm{RPA}_\infty$@\qsgw, and $\varepsilon^\mathrm{BSE}_\infty$@\qsgwl.  Consider by contrast two
inconsistent approximations, $\varepsilon^\mathrm{BSE}_\infty$@\qsgw\ and
$\varepsilon^\mathrm{RPA}_\infty$@\qsgwl\ (Fig.~\ref{fig:epsinfty2}).  Both of these approximations show more randomness
than either $\varepsilon^\mathrm{RPA}_\infty$@\qsgw, or $\varepsilon^\mathrm{BSE}_\infty$@\qsgwl of
Fig.~\ref{fig:gapsandepsinfty}.  Yet there is a strong similarity between the green diamonds and the blue circles in the
two figures.  The green diamonds in Fig.~\ref{fig:epsinfty2} fall slightly below the ideal line, showing a modest but
non-negligible effect of improving the reference $G_{0}$.  The blue circles rise slightly above the 80\% showing that
the RPA continues to underestimate $\varepsilon_\infty$, even with a nearly ideal reference $G_{0}$.  This affirms that
most---but not all---of the underestimate of $\varepsilon_\infty$ originates from the RPA itself.

%This is a consequence of the effect of ladders weakening as the gap closes, a point we shall return to later.

This sheds light on the commonly observed fact that $\varepsilon^\mathrm{RPA}_\infty$, when computed from the
$G_\mathrm{LDA}$, often provides a rather good estimate for $\varepsilon^\mathrm{expt}_\infty$, e.g. in \emph{sp}
semiconductors.  The obvious, naive reason for this is a fortuitous error cancellation: LDA underestimates bandgaps,
which tends to overestimate $\varepsilon_\infty$, while the RPA's neglect of electron-hole attraction tends to underestimate
the screening, and thus tends to underestimate $\varepsilon_\infty$.  However there has been some speculation that the good
agreement is not accidental, but a consequence of characteristics inherent in the RPA and the LDA.  In particular a
recent work \cite{Loon21} asserts that $\varepsilon^\mathrm{RPA}$@LDA should be a good approximation for insulators,
based on two arguments.  First, ladders involve tunneling processes, and are effective at short range but not long
range; thus the long-range screening that predominately controls $\varepsilon(q{=}0)$ is well described by the
RPA~\cite{Yan00,Irmler19}.  The first argument is rather appealing, and consistent with prior work establishing that the
largest corrections to the RPA occur at short distances~\cite{Yan00,Ruzsinszky11,Irmler19}. This argument can be
  rigorously checked by comparing blue circles in Fig.~\ref{fig:epsinfty2} to green diamonds in
  Fig.~\ref{fig:gapsandepsinfty}: both share the same eigenfunctions generating $\varepsilon$; the only difference being
  the presence or absence of ladder diagrams.  Agreement is fairly good, differing by 15-20\%, which explains in part
  why \qsgw\ is a good and consistent theory.  The argument of van Loon et al.~\cite{Loon21} is only partially true:
  even if the vertex part of $P^{0}$ is short range, $v$ is long range so $\varepsilon{=}1-vP^{0}$ can have a long range
  contribution from the short ranged part of $P^{0}$.  Notably, the difference does not get smaller as the gap becomes
  large, as Ref.~\cite{Loon21} asserted based on the tunneling argument.  This is apparently because as the gap closes
  the screening becomes large, so the long range contribution from a short range vertex becomes relatively less
  important.  It is nevertheless striking that the BSE correction is so insensitive to the bandgap.

The second argument of Ref.~\cite{Loon21} is that the local vertex in insulators is approximately accounted for by using
LDA eigenfunctions.  The argument is based on a connection between the LDA derivative discontinuity and the missing
vertex, which emerges in a model.  To examine this proposition, $\varepsilon^\mathrm{RPA}_\infty$@LDA is also presented
in Fig.~\ref{fig:epsinfty2} (red hexagons).  The second argument is more difficult to assess quantitatively because
screening modifies both $G_{0}$ and $\varepsilon_\infty$, but roughly speaking the difference between blue circles and
green diamonds
%(not discriminating whether to use figure Fig.~\ref{fig:gapsandepsinfty} or Fig.~\ref{fig:epsinfty2})
should be similar to the difference between red hexagons and blue circles.  One might attribute the poor agreement to
inadequacy of the LDA functional (distinct from the derivative discontinuity in the exact functional), but at least in a
few systems where it has been tested, the primary gap error has been shown to originate largely from the derivative
discontinuity, and not inadequacy of the functional~\cite{doi:10.1063/1.2189226}.

%Perhaps because it is based on a model, the argument does not apply well to real materials, as can be seen from the red
%hexagons in Fig.~\ref{fig:epsinfty2}.  $\varepsilon^\mathrm{RPA}_\infty$@LDA does best for weakly correlated systems
%with large direct gaps that are not too severely underestimated (e.g. MgO, AlN, and diamond).  When the gap closes as in
%InN, $\varepsilon_\infty$ has an unphysical divergence.  Even in relatively weakly correlated wide-gap \emph{sp} systems
%(ZnO, CdTe), $d^{0}$ systems (SrTiO\textsubscript{3}, TiO\textsubscript{2}), and an $f^{0}$ system
%(CeO\textsubscript{2}), $\varepsilon^\mathrm{RPA}_\infty$@LDA is poor.

\begin{figure}[h!]
\includegraphics[width=0.30\textwidth,clip=true,trim=0.0cm 0.0cm 0.0cm 0.0cm]{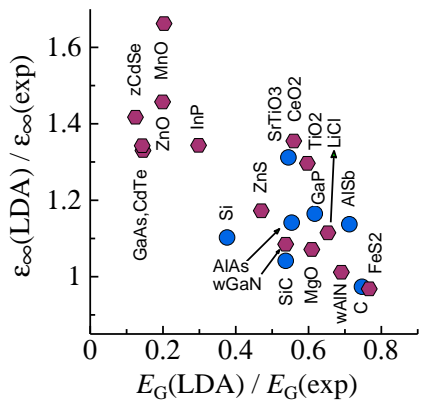}
\caption{Ratio $\varepsilon^\mathrm{RPA}_\infty\mathrm{@LDA}/\varepsilon^\mathrm{exp}_\infty$, plotted
  as a function of the LDA relative gap error, $E^\mathrm{LDA}_G/E^\mathrm{exp}_G$.
  Blue circles are indirect gap semiconductors; red hexagons are direct-gap.}
\label{fig:epsinfty3}
\end{figure}

There is a distinct tendency for LDA to better predict $\varepsilon_\infty$ for indirect gap tetrahedral semiconductors
than for direct-gap ones: compare diamond, AlAs, AlSb, GaSb in Fig.~\ref{fig:epsinfty2} to ZnO, ZnS, CdTe, InP and GaAs.
Since the derivative discontinuity does not vary wildly between direct- and indirect-gap materials, this is a hint that
some other parameter controls the errors in $\varepsilon^\mathrm{RPA}_\infty\mathrm{@LDA}$.  Note also that in the
one-oscillator model~\cite{Wemple71}, \S\ref{ss:oscillators}, the effective oscillator energy $E_{0}$ tends to better
align with the smallest direct gap than the fundamental one.  To disentangle the various effects,
Fig.~\ref{fig:epsinfty3} plots the relative error in $\varepsilon^\mathrm{RPA}_\infty$@LDA against the relative error in
the gap, which is a proxy for the derivative discontinuity.  Excepting the $d^{0}$ and $f^{0}$ systems, the relationship
between $E^\mathrm{LDA}_G/E^\mathrm{exp}_G$ and
$\varepsilon^\mathrm{RPA}_\infty\mathrm{@LDA}/\varepsilon^\mathrm{exp}_\infty$ is roughly linear.  The sensitivity of
$\varepsilon^\mathrm{RPA}_\infty\mathrm{@LDA}$ to the derivative discontinuity, together with the tendency of RPA to
underestimate bandgaps established earlier, provides strong indication that $\varepsilon^\mathrm{RPA}_\infty$@LDA yields
reasonable $\varepsilon^\mathrm{expt}_\infty$ only sometimes, and since it generally produces values larger than
experiment while the RPA underestimates the screening as we have shown, there is an additional hidden benefit from
fortuitous error cancellation.

\subsection{Benchmarks in weakly correlated semiconductors}\label{ss:semi}

\begin{figure}[h!]
\includegraphics[width=0.30\textwidth,clip=true,trim=0.0cm 0.0cm 0.0cm 0.0cm]{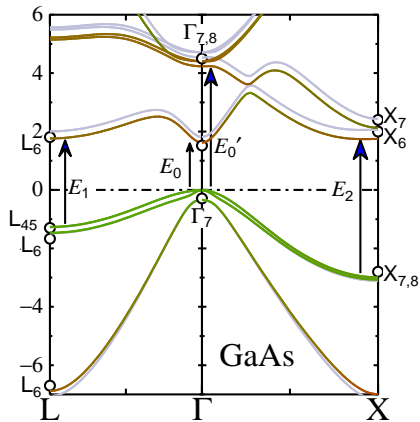}
\caption{
  Energy bands in GaAs, depicting vertical transitions $E_{0}$, $E'_{0}$, $E_{1}$ and $E_{2}$ that can be measured by ellipsometry.
  Circles depict measurements of states at high-symmetry points.  They have been determined by ARPES~\cite{PhysRevB.21.3513}
  to a resolution of about 0.1\,eV.  $E_{0}$, $E'_{0}$, $E_{1}$ and $E_{2}$ are reported by Lautenschlager et al,
  in Ref.~\cite{PhysRevB.35.9174}.  Combining this data provides one way to determine levels at X and L
  in the conduction band.  Colored bands are taken from \qsgwl\ calculations, with red and green showing projections onto Ga and As, respectively.
  Gray lines show results of \qsgw\ calculations.  In the valence, \qsgwl\ and \qsgw\ are nearly indistinguishable.
  \qsgwl\ and \qsgw\ dispersions in the conduction band are very similar, with \qsgw\ slightly higher in energy.
}
\label{fig:gasbands}
\end{figure}

The tetrahedrally coordinated $sp^3$ compounds form a good benchmark for weakly correlated systems in part because they
are the best characterized of any family of materials, but also because weak correlations make it possible to well
identify transitions between single-particle levels, especially associating peaks in ellipsometry measurements with
them.  The valence band maximum falls at or very near $\Gamma$ for all tetrahedrally coordinated semiconductors, which
simplifies the analysis.  Besides the lowest $\Gamma{-}\Gamma$ transition $E_{0}$, the next $\Gamma{-}\Gamma$ transition
$E'_{0}$ has been measured for some materials.  Ellipsometry also measures $E_{1}$ and $E_{2}$ shown in
Fig.~\ref{fig:gasbands}.  $E_{0}$ and $E_{1}$ are easier to measure accurately, because there is larger volume in
\emph{k} space where the valence and conduction bands are parallel.  $E_{2}$ has been measured for most semiconductors,
but its determination is less certain (excepting compounds such as Si, C, and SiC where the global conduction band
minimum lies near X). Some data for $E'_{0}$ are available, but their values are also less well known.

The wider conclusions we draw from the detailed analysis to be described below are as follows.

\begin{enumerate}[leftmargin=*]

\item Bandgaps in light (and especially polar) materials are overestimated (MgO, LiF, LiCl, NaCl, TiO\textsubscript{2},
  SrTiO\textsubscript{3}, C).  The primary cause is the electron-phonon interaction (\S\ref{ss:outliers}).  A
  diagrammatic electron-phonon contribution to $\Sigma$ has long been known~\cite{hedin} though historically speaking,
  reliably determining its magnitude has posed a challenge.  A fairly high fidelity calculation of it has recently
  appeared (Ref.~\cite{Miglio20}), and we use their results to estimate this term where available.  In other cases we
  make a simple estimate using the Fr\"olich approach of Ref~\onlinecite{PhysRevMaterials.2.013807}
  (\S\ref{ss:frolich}).  The fundamental gaps with these adjustments are shown as black crosses in
  Fig.~\ref{fig:gapsandepsinfty}.  See also \S\ref{ss:otheroutliers}.

\item The gap in compounds with shallow, nearly dispersionless $d$ levels are too small (Table~\ref{tab:dvbmgaps}), and
  semicore \emph{d} levels are too shallow (Fig~\ref{fig:dcorelevels}).  This is a consequence of the imperfect $Z$
  factor cancellation noted in \S\ref{ss:fidelity}, Point 2.  To correct it would require the missing vertex $\Gamma$ in
  the exact self-energy, $GW\Gamma$.  Several instances of this are presented in \S\ref{ss:otheroutliers}.

\item $k$-dispersions in the conduction bands of zincblende semiconductors show systematic errors of the order
  $\pm$0.1\,eV (See discussion around Tables~\ref{tab:Xpointgaps}, \ref{tab:E0gaps}).  There is no obvious diagram that
  explains this discrepancy.

\end{enumerate}

\begin{figure}[h!]
\includegraphics[width=0.20\textwidth,clip=true,trim=0.0cm 0.0cm 0.0cm 0.0cm]{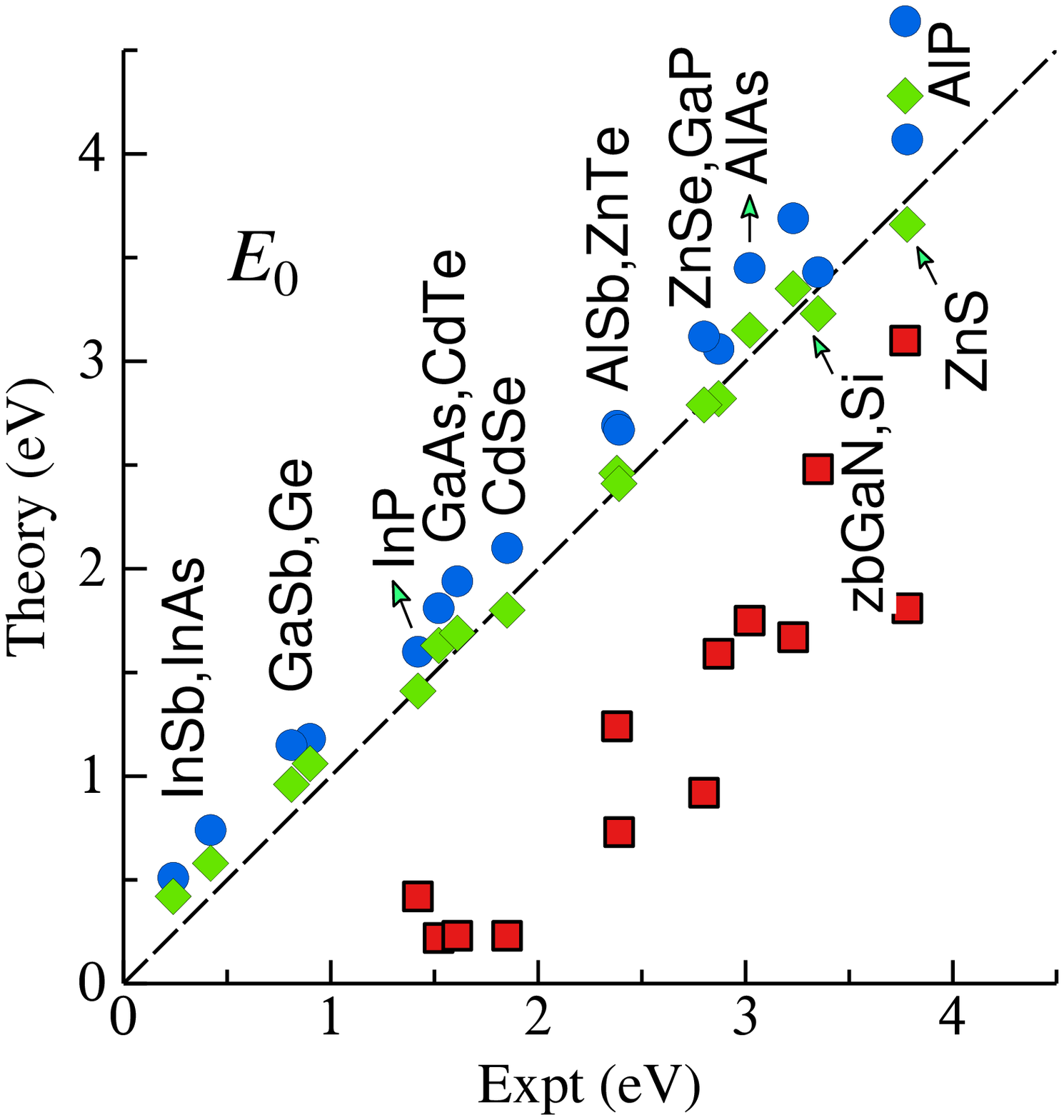}\ \
\includegraphics[width=0.20\textwidth,clip=true,trim=0.0cm 0.0cm 0.0cm 0.0cm]{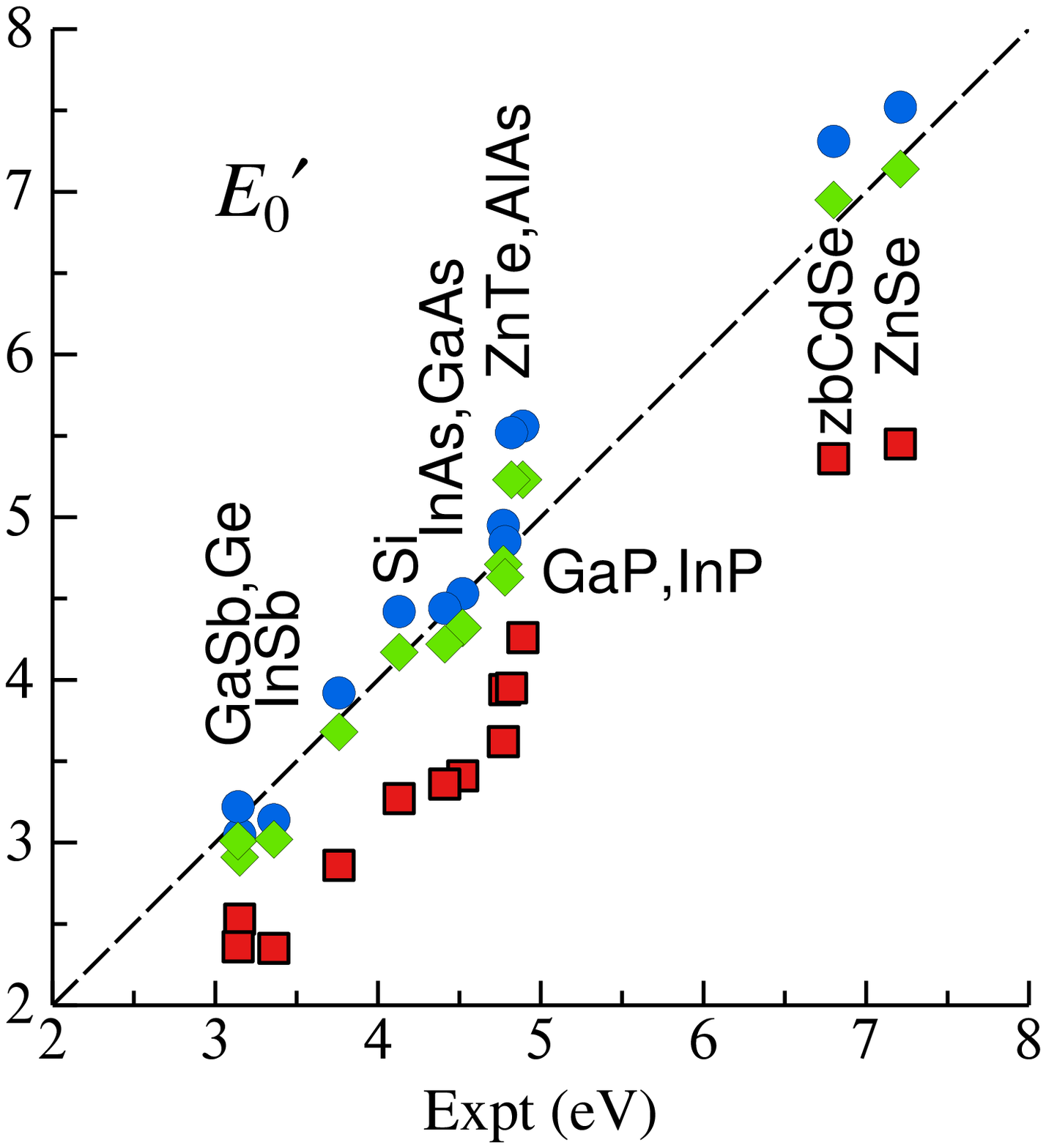}
\includegraphics[width=0.20\textwidth,clip=true,trim=0.0cm 0.0cm 0.0cm 0.0cm]{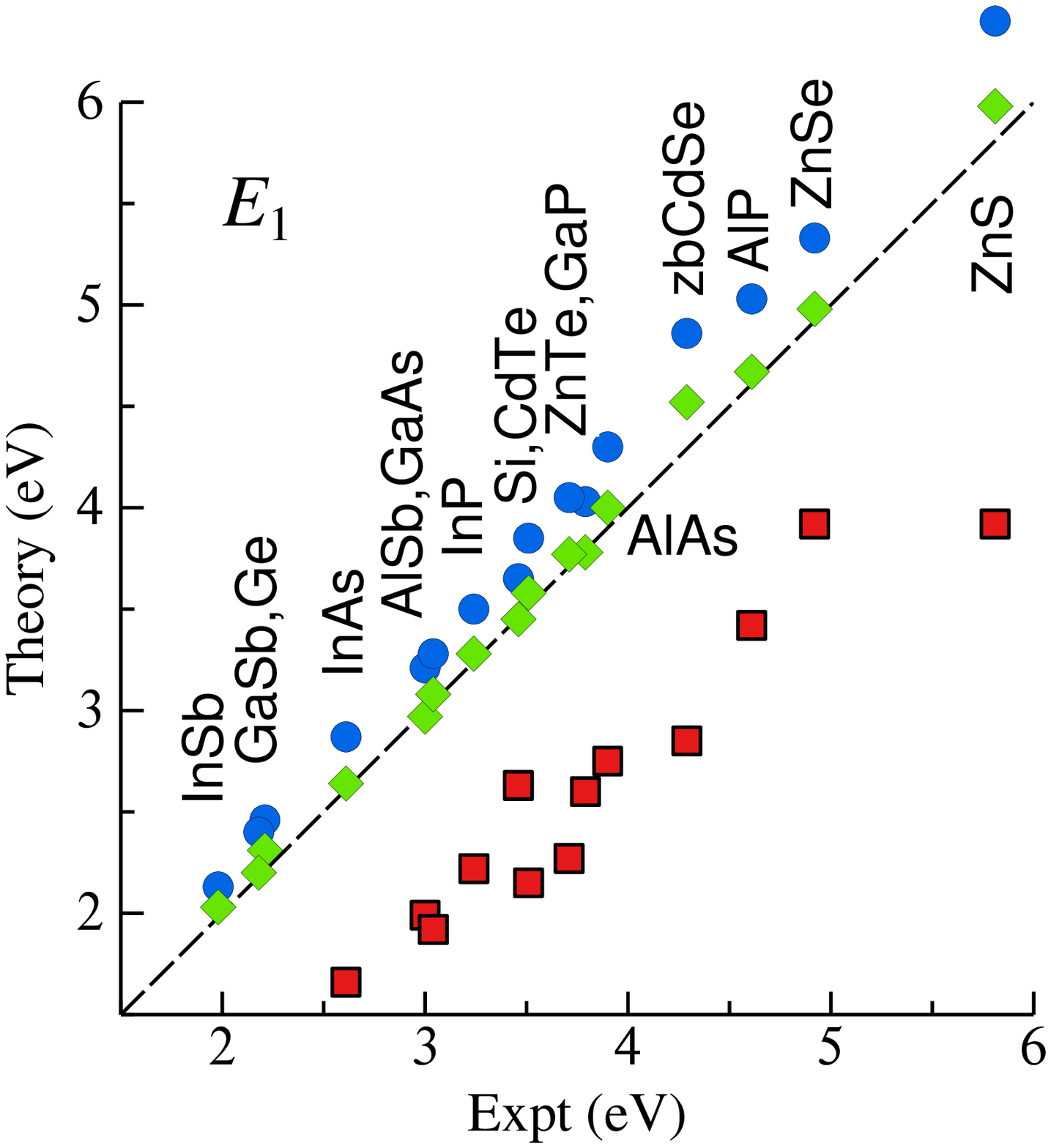}\ \
\includegraphics[width=0.20\textwidth,clip=true,trim=0.0cm 0.0cm 0.0cm 0.0cm]{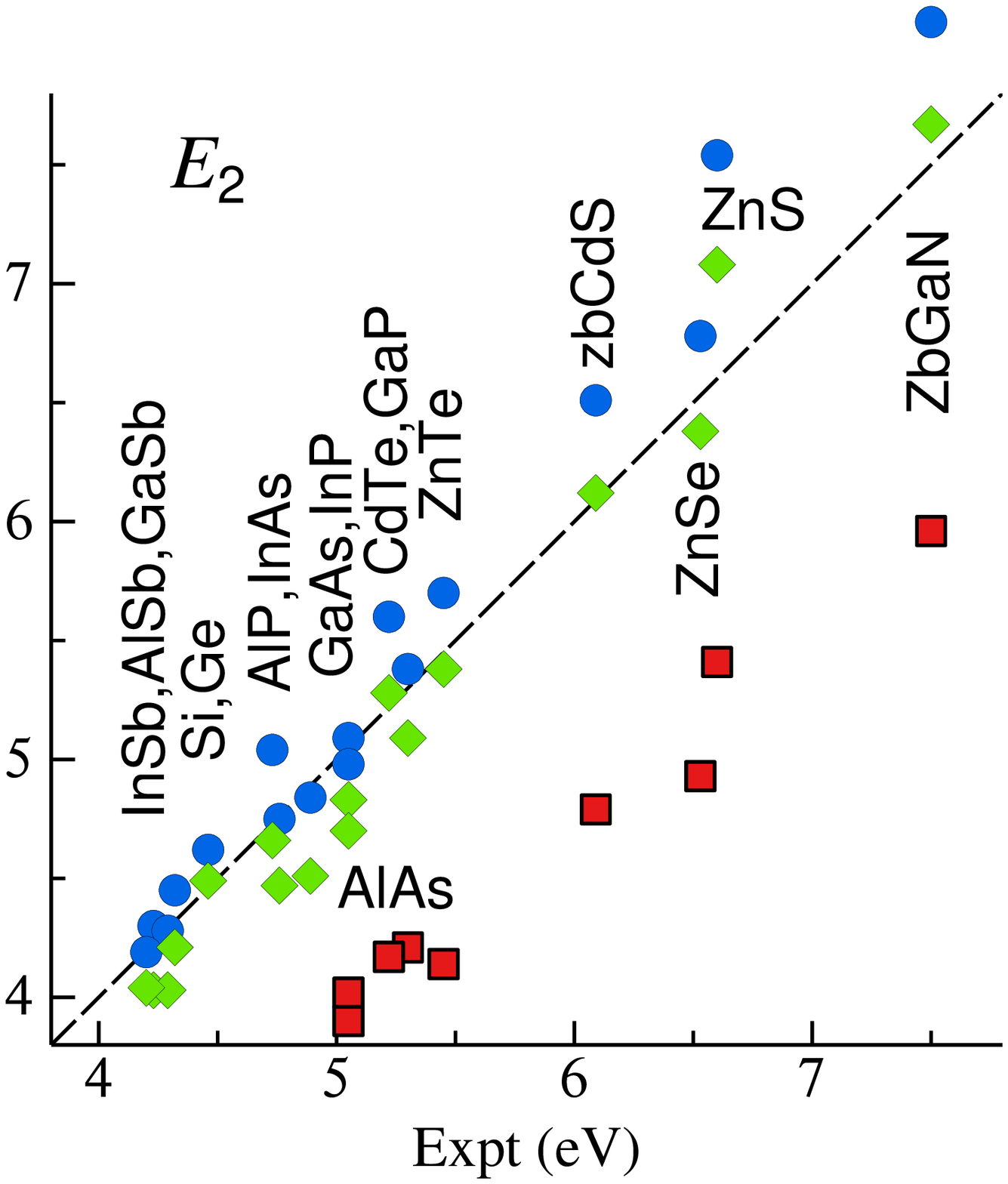}
\caption{
  $E_{0}$, $E'_{0}$, $E_{1}$ and $E_{2}$ transitions in zincblende semiconductors, where experimental data is available.
  Red squares, blue circles, and green diamonds correspond to LDA, \qsgw and \qsgwl\ matching Fig.~\ref{fig:gapsandepsinfty}.
}
\label{fig:e01esemi}
\end{figure}

Fig.~\ref{fig:e01esemi} benchmarks $E_{0}$, $E'_{0}$, $E_{1}$ and $E_{2}$ transitions in zincblende semiconductors where
ellipsometry data is available.  $E_{1}$ shows close agreement, but $E_{0}$ and $E_{2}$ exhibit discrepancies with
distinct patterns:

\begin{figure}[h!]
\includegraphics[width=0.40\textwidth,clip=true,trim=0.0cm 0.0cm 0.0cm 0.0cm]{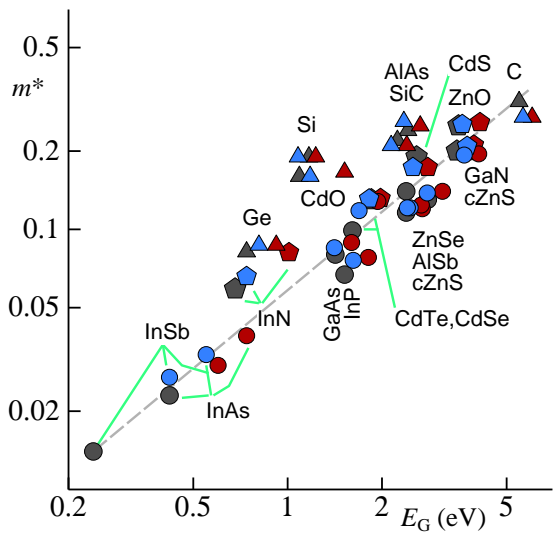}
\caption{
  Effective masses in weakly (and mostly tetrahedralloy) coordinated semiconductors.  Mass $m^*$ is plotted as a
  function of bandgap $E_G$.  Black denotes experimental data, blue \qsgwl\ results, red \qsgw\ results.  Circles denote
  direct gaps in zincblende structures, pentagons denote direct gaps in wurtzite compounds, and triangles denote
  indirect gaps.  The figure provides two independent kinds of information: the bandgap discrepancy can be seen
  comparing blue (or red) against black on the abscissa; the mass discrepancy compare the ordinate.  The light dashed
  gray line shows a linear function $m^*(E_g)$.  This is the dependence of mass on the gap in $k{\cdot}p$ theory,
  assuming the matrix element is fixed.\\ Data for III-V semiconductors are taken from Ref.~\onlinecite{Vurgaftman01};
  other data taken from Adachi's compilation~\cite{Adachi07}.  CdTe, ZnTe, and GaAs are taken from a two-photon
  magnetoabsorption experiment~\cite{Neumann88} which is thought to be reliable.  AlSb data is for
  the conduction band at $\Gamma$.
}
\label{fig:mass}
\end{figure}

% --- Table Xpointgaps ----
\begin{table}[h]
\caption{Bandgaps at X or L for some zincblende semiconductors.
  All the semiconductors listed above have a global conduction band minimum near X, except for Ge and GaSb.
  When the electron-phonon interaction is taken into account, gaps at X are systematically underestimated
  by $\sim$0.15\,eV, while those at L are not.}\begin{tabular}{c|c|c|c|c|c}\hline
          & $E_{G}$ (expt) &               & \qsgwl & ZP (est) & \qsgwl-ZP\\
C         & 5.40           & {\quad}X{\quad} & 5.64   &  -0.40   &  5.24 \cr
Si        & 1.17           & {\quad}X{\quad} & 1.08   &  -0.06   &  1.02 \cr
SiC       & 2.42           & {\quad}X{\quad} & 2.35   &  -0.15   &  2.20 \cr
GaP       & 2.35           & {\quad}X{\quad} & 2.22   &  -0.09   &  2.13 \cr
AlAs      & 2.24           & {\quad}X{\quad} & 2.14   &  -0.04   &  2.10 \cr
Ge        & 0.74           & {\quad}L{\quad} & 0.81   &  -0.05   &  0.76 \cr
GaSb      & 0.88           & {\quad}L{\quad} & 0.91   &  -0.03   &  0.88 \cr
\hline
\end{tabular}
\label{tab:Xpointgaps}
\end{table}
Tables~\ref{tab:Xpointgaps} and \ref{tab:E0gaps} establish that there is a systematic, $k$-dependent error in the
conduction band in zincblende semiconductors on the order of 100\,meV.  The consequences can be significant: note for
example, that \qsgwl\ predicts GaSb to have a global minimum at L, with $E_{\Gamma}{-}E_\mathrm{L}{=}0.05$\,eV, while
experimentally at 0K it is a direct gap, with $E_{\Gamma}{-}E_\mathrm{L}{=}-0.09$\,eV~\cite{Wu92}.  Also, where gaps are
overestimated, effective masses are too large (Fig.~\ref{fig:mass}).

\begin{table}[h]
\caption{$E_{0}$ in narrow gap zincblende semiconductors is overestimated by 0.1\,eV.  The tendency
  does not hold for narrow-gap semiconductors that form in other structures, shown as the entries in the second half of
  the Table. An estimate for low-temperature bandgap for Ti$_{2}$Se$_{2}$ (24-atom $P\bar{3}c1$ CDW structure) is taken from
  Ref.~\cite{Rasch08}.
}
\begin{tabular}{c|c|c|c|c}\hline
                 & $E_{G}$ (expt) & \qsgwl & ZP (est) & \qsgwl-ZP\\
Ge               & 0.90           & 1.06   &  -0.05   &  1.01 \cr
GaSb             & 0.81           & 0.96   &  -0.03   &  0.93 \cr
InAs             & 0.42           & 0.53   &  -0.02   &  0.51 \cr
InSb             & 0.24           & 0.42   &  -0.02   &  0.40 \cr
\hline
InN              & 0.70           & 0.74   &  -0.07   &  0.67 \cr
Bi$_{2}$Te$_{3}$ & 0.15           & 0.15   &          &       \cr
PbTe             & 0.19           & 0.18   &          &       \cr
TiSe$_{2}$       & 0.15           & 0.15   &          &       \cr
\hline
\end{tabular}
\label{tab:E0gaps}
\end{table}

The \textbf{k}-dependent gap error is further discussed in \S\ref{ss:tda}, but we can find no obvious explanation for
it.  One possibility is that Questaal's implementation of QS\emph{GW} contains an error not inherent in QS\emph{GW}
itelf.  In particular the incomplete basis noted by Betzinger et al. in the $G^\mathrm{LDA}W^\mathrm{LDA}$
context~\cite{Friedrich-dynamical-sternheimer} may be a factor.  It cannot be ruled out that the nearly perfect
agreement for so many systems is a fortuitous artifact of the implementation, or fortuitous cancellation of higher order
diagrams.  At all events there is no simple explanation that reconciles these inconsistencies.

Fig.~\ref{fig:mass} benchmarks for effective masses and bandgaps in tetrahedral semiconductors.  For the direct gap
systems (circles and hexagons), the discrepancy in $m^*$ compared to the experimental value scales approximately in
proportion to the discrepancy in $E_G$ (compare to the light dashed gray line).  $k{\cdot}p$ theory predicts a $m^*$ to
be proportional to $E_G$, assuming a fixed matrix element coupling valence and conduction band, showing that errors in
$m^{*}$ have the same origin as whatever causes the gap to be too large.

%\begin{threeparttable}[h]
%  \caption{3\emph{d} and 4\emph{d} core levels in semiconductors.  $GW^\mathrm{LDA}$ refers to a perturbation to the LDA
%    with $Z{=}1$ and keeping retain the off-diagonal parts of the self-energy.
%}
%\begin{tabular}{|@{\hspace{0.2em}}c@{\hspace{0.4em}}|@{\hspace{0.2em}}c@{\hspace{0.2em}}|@{\hspace{0.2em}}c@{\hspace{0.2em}}|@{\hspace{0.2em}}c@{\hspace{0.2em}}|@{\hspace{0.2em}}c@{\hspace{0.2em}}|@{\hspace{0.2em}}c@{\hspace{0.2em}}|}\hline
%                       & $E_{d}$ (expt)      & LDA     & $GW^\mathrm{LDA}$ & \qsgw   & \qsgwl  \cr\hline
%Ga in GaAs\tnote{1}    &  18.6               &  14.5   &  17.2                       &  17.6   &  17.4 \cr
%As in GaAs\tnote{1}    &  40.0               &  34.6   &  38.6                       &  39.3   &  39.0 \cr
%Ga in GaSb\tnote{1}    &  18.7               &  14.8   &  17.5                       &  17.9   &  17.7 \cr
%Sb in GaSb\tnote{1}    &  31.7               &  28.0   &  30.2                       &  31.4   &  31.1 \cr
%Zn in ZnSe\tnote{2}    &   9.0               &   6.5   &   8.0                       &   8.4   &   8.2 \cr
%Cd in CdTe\tnote{2}    &  10.4               &   8.2   &   9.5                       &   9.9   &   9.7 \cr
%mean error             &                     &   3.6   &   1.2                       &   0.65  &   0.9 \cr
%\hline
%\end{tabular}
%\label{tab:coredlevels}
%\begin{tablenotes}[para,flushleft]\footnotesize
%\item[1] 5/2 state, Ref.~\cite{Eastman80}
%\item[2] Band center, Ref.~\cite{Vesely72}
%\end{tablenotes}
%\vbox{\vskip 12pt}
%\end{threeparttable}

\begin{figure}[h!]
\includegraphics[width=0.45\textwidth,clip=true,trim=0.0cm 0.0cm 0.0cm 0.0cm]{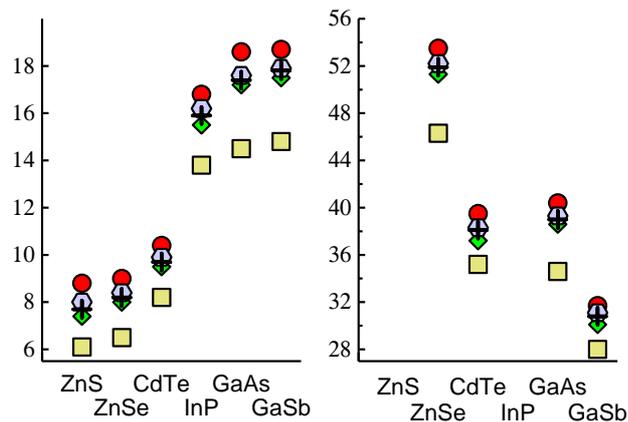}
\caption{(Left) cation \emph{d} core levels relative to the valence band maximum, in eV.
         (Right) anion \emph{d} core levels for systems where they are present.
         The center of gravity of the $d_{5/2}$ level was taken, except for ZnSe where it was not reported.
         Red circles are photoemission data taken from Refs~\cite{Vesely72,Ley74,Eastman80}.  Yellow squares, green
         diamonds, blue hexagons, and crosses are results from LDA, {$(GW)^\mathrm{LDA}$}, \qsgw, and \qsgwl,
         respectively.  Calculated anion $d_{5/2}{-}d_{3/2}$ splittings (1.5\,eV, CdTe; 0.8\,eV, GaAs; 1.3\,eV, GaSb) are in
         close agreement with photoemission data.  The table below shows the mean difference with the photoemission data,
         and the RMS fluctations about the mean, for the different levels of approximation.
}
\vskip 12pt
\begin{tabular}{|c|cc|cc|cc|cc|}
         \hline
\vbox{\vskip 12pt}
         &\multicolumn{2}{c|}{LDA}&\multicolumn{2}{c|}{$(GW)^\mathrm{LDA}$}&\multicolumn{2}{c|}{\qsgw}&\multicolumn{2}{c|}{\qsgwl}\cr
         &  cation &   anion & cation  & anion  & cation & anion & cation &   anion \cr \hline
mean     &  3.1    &   5.2   &  1.2    &  2.0   &  0.7   &  1.1  &  0.9   &   1.3  \cr
RMS      &  1.5    &   2.5   &  0.4    &  0.6   &  0.4   &  0.5  &  0.4   &   0.4   \cr
\hline
\end{tabular}
\
%\label{tab:coredlevels}
\label{fig:dcorelevels}
\end{figure}

\emph{Core levels}

Fig.~\ref{fig:dcorelevels} presents \emph{d} core levels at different levels of theory and compares to photoemission
results.  As is well known, their position is too shallow in the LDA.  $(GW)^\mathrm{LDA}$ improves agreement with
experiment, but levels remain too shallow.  Self-consistency (\qsgw) shows further improvements. \qsgwl\ fares slightly
worse than \qsgw\ on average, but the consistency improves with the level of theory.

Table~\ref{tab:dvbmgaps} shows some materials system where the valence band maximum is nearly flat and dispersionless.
The bandgap is consistently underestimated in \qsgwl.  (\qsgw\ fares better, but this is an artifact of fortuitous error
cancellation).

Shortcomings shown in Fig.~\ref{fig:dcorelevels} and Table~\ref{tab:dvbmgaps} have a common origin, the missing vertex
(\S\ref{ss:fidelity}, point 2).  Note that the discrepancy with experiment increases with distance from the Fermi level:
from $\sim$0.5\,eV for the valence states near $E_{F}$, $\sim$0.9\,eV for cation levels around $-10$\,eV, and
$\sim$1.3\,eV for the deeper anion levels.

\begin{table}[h]
\caption{Bandgaps in systems with valence band maximum formed from a core-like $d$ state.
The bandgap of FeS\textsubscript{2} is not well known, but the \qsgwl\ gap lies below the most likely value of
0.9\,eV.
}
\begin{tabular}{|@{\hspace{0.6em}}c@{\hspace{0.6em}}|@{\hspace{0.6em}}c@{\hspace{0.6em}}|@{\hspace{0.6em}}c@{\hspace{0.6em}}|@{\hspace{0.6em}}c@{\hspace{0.6em}}|@{\hspace{0.6em}}c@{\hspace{0.6em}}|}\hline
                 & $E_{G}$ (expt)                           & \qsgwl & \qsgw    \cr\hline
CuCl             & \ \ 3.46\footnote{Two-photon absorption Ref.~\cite{Reimann94}}%
                                                            & 2.67   &   3.44    \cr
Cu$_{2}$O         & \ \ 2.17\footnote{Inferred from interpretation of experiment with effective mass model and exciton observed at 2.0\,eV, Ref.~\cite{Uihlein81}}.
                                                            & 1.74   &   2.27    \cr
FeS$_{2}$         & $\sim$0.9\footnote{An estimate from many measurements, from Ref.~\cite{Ferrer90}} & 0.69   &   0.81    \cr
VO$_{2}$          & $\sim$0.7\footnote{See \S\ref{ss:vo2}}       & 0.43   &  0.76     \cr
FeO              & $\sim$1.1\footnote{See \S\ref{ss:feo}} & 0.64   &   1.9     \cr
%EuO \cr
\hline
\end{tabular}
\label{tab:dvbmgaps}
\end{table}

%Crystal field waln : .230 compare to Rinke-band-parameters-III-N-PhysRevB.77.075202.pdf
%Crystal field wgan : .017 compare to Rinke-band-parameters-III-N-PhysRevB.77.075202.pdf
%Crystal field winn : .015 compare to Rinke-band-parameters-III-N-PhysRevB.77.075202.pdf

\emph{Valence band parameters}

The structure of the valence band around $\Gamma$ provide less reliable benchmarks because of experimental uncertainty in
the parameters.  Key parameters are the effective masses, and in the wurtzite structure, the crystal field splitting
arising from inequivalence of the $z$ and $xy$ directions.  As regards the masses, the matter is considerably
complicated by intermixing three states at the valence band maximum near the $\Gamma$ point and the nontrivial role of
spin-orbit coupling that splits the threefold degeneracy at $\Gamma$ and pushes the band maximum slightly off it.

To encapsulate the many different masses, a Luttinger model is typically used, which has only three independent
parameters.  The Luttinger parameters can be generated from the following effective masses:

\begin{align}
\gamma_{1}&={{1}\over{2m_\mathrm{lh}^{001}}}+{{1}\over{2m_\mathrm{hh}^{001}}} \nonumber \\
\gamma_{2}&={{1}\over{4m_\mathrm{lh}^{001}}}-{{1}\over{4m_\mathrm{hh}^{001}}} \\
\gamma_{3}&={{1}\over{4m_\mathrm{lh}^{001}}}+{{1}\over{4m_\mathrm{hh}^{001}}}-{{1}\over{2m_\mathrm{hh}^{111}}} \nonumber
\end{align}

% Other relations: p56 γ1 = −1/3(A + 2B), γ2 = −1/6 (A − B), and γ3 = −1/6(C)

$m_\mathrm{lh}$ and $m_\mathrm{hh}$  denote light-hole and heavy-hole masses.
Table~\ref{tab:vbparameters} shows Luttinger parameters for a few systems where they are best known.  The range of
values shown in the experimental columns correspond to the range collated from different measurements.  In the two cases
where the bandgap is close to experiment (Si and InP) the calculated Luttinger parameters fall within the range of experimental
data.  In the other two cases (Ge and GaAs) the parameters are underestimated for the same reason the conduction band
effective masses are overestimated (see Fig.~\ref{fig:mass}): the direct gap is somewhat overestimated; (see
e.g. Table~\ref{tab:E0gaps}) In Ge, for example, the conduction band mass at $\Gamma$ was measured to be
0.037~\cite{Roth59}, while our \qsgwl\ mass is 0.047.

Table~\ref{tab:vbparameters} also show shows crystal-field splitting $\Delta_\mathrm{cr}$ in the III-N semiconductors
(splitting between states of $p_z$ and $p_{xy}$ character at $\Gamma$) in the absence of spin-orbit coupling.  The
\qsgwl\ result is within $\sim$0.01\,eV of the measured values, which is quite satisfactory.  This quantity is rather
sensitive to find details of the potential.  To obtain $\Delta_\mathrm{cr}$ reliably, a fine $k$ mesh of
$9{\times}9{\times}6$ divisions was needed: its value increased by 0.005\,eV compared to the standard
$6{\times}6{\times}4$ mesh.  Note that OEP-based \emph{GW} reported in Ref.~\cite{Rinke08} yields quite different values
for $\Delta_\mathrm{cr}$.

\begin{table}[h]
  \caption{Left: valence band Luttinger parameters in selected zincblende semiconductors.
Right: crystal field splitting parameter in III-N compounds.
}
\noindent
\begin{tabular}{|r@{\hspace{0.2em}}c|ccc|ccc|}
\hline
\vbox{\vskip 2pt}
         & & \multicolumn{3}{c|}{\qsgwl}
                                       &\multicolumn{3}{c|}{Expt}\\
         & $E_{G}$ &$\gamma_1$ &$\gamma_2$ &$\gamma_3$ &$\gamma_1$ &$\gamma_2$  &$\gamma_3$\\
\hline
\vbox{\vskip 12pt}
Si\footnote{compilation in Ref.~\cite{LandoltBornsteinVol41A1b}} $\vert$
                               & 1.22    & 4.25      &  0.25     &  1.31     & 4.26-4.29 &  0.34-0.38 &  1.45-1.56 \\
Ge$^\mathrm{a}$ $\vert$
                               & 0.81    & 10.2      &  2.81     &  4.12     & 13.2-13.4 &  4.20-4.24 &  5.56-5.69 \\
GaAs\footnote{compilation in Ref.~\cite{Vurgaftman01}} $\vert$
                               & 1.63    & 6.73      &  1.82     &  2.77     & 6.79–7.20 &  1.90–2.88 &  2.68–3.05 \\
InP$^\mathrm{b}$ $\vert$
                               & 1.41    & 5.41      &  1.52     &  2.37     & 4.61–6.28 &  0.94–2.08 &  1.62–2.76 \\
\hline\\
\vspace{-0.8cm}\end{tabular}
%\begin{tablenotes}[para,flushleft]\footnotesize
%\item[1] compilation in Ref.~\cite{LandoltBornsteinVol41A1b}
%\item[2] compilation in Ref.~\cite{Vurgaftman01}
%\end{tablenotes}
\noindent
\begin{tabular}{|r@{\hspace{0.3em}}|@{\hspace{0.3em}}cc@{\hspace{0.5em}}|@{\hspace{0.3em}}cc|@{\hspace{0.3em}}cc|}
\hline
& \multicolumn{2}{c|}{AlN} & \multicolumn{2}{c|}{GaN} & \multicolumn{2}{c|}{InN}\\
\vbox{\vskip 2pt}
& $E_{G}$ & $\Delta_\mathrm{cr}$ & $E_{G}$ & $\Delta_\mathrm{cr}$ & $E_{G}$ & $\Delta_\mathrm{cr}$ \\
\hline
\vbox{\vskip 12pt}
\qsgw{\hspace{0.2em}}         & 6.93   & -0.224 & 3.93 & 0.023       & 1.02 & 0.010 \\
\qsgwl{\hspace{0.2em}}        & 6.40   & -0.228 & 3.60 & 0.021       & 0.74 & 0.010 \\
$G_{0}W_{0}$\rlap{\footnote{Experimental data and $G_{0}W_{0}$ results taken from Ref.~\cite{Rinke08}.}}{\hspace{0.2em}}
              & 6.47   & -0.295 & 3.24 & 0.034       & 0.69 & 0.066 \\
Expt{\hspace{0.2em}}          & 6.13   & -0.230 & 3.5  & 0.009-0.038 & 0.67 & 0.019-0.024 \\
\hline
\end{tabular}
\begin{tablenotes}[para,flushleft]\footnotesize
\end{tablenotes}
\noindent
\label{tab:vbparameters}
\end{table}

\subsubsection{Two extensions to the theory and their effect on zincblende semiconductors}\label{ss:tda}

As two possible sources of error on the QP levels of zincblende semiconductors, we first considered eliminating the Tamm-Dancoff
approximation (TDA).  Here we focus on InSb as it has the largest relative gap error.  Removing the TDA reduced the
\qsgwl\ $E_{0}$ gap by 0.03\,eV --- considerably less than the discrepancy with experiment.  We also considered whether
eliminating the TDA improves the \emph{k}-dispersion, in particular the wrong prediction of the global minimum in GaSb
noted above.  Removing the TDA reduces the gap in GaSb by 0.03\,eV (similar to InSb), but the shift was essentially
independent of \emph{k} and did not rectify this shortcoming.

We also considered the effect of using a better kernel in the BSE.  In all the calculations presented here, we used
$W^\mathrm{RPA}$ for the kernel (Eq.~\ref{kernel_eq}).  It is possible that the dispersion errors in these compounds is
a consequence of $W^\mathrm{RPA}$ being too removed from the exact vertex.  We can assess this effect by using a better kernel, namely
BSE $W$ as the kernel for the BSE.  If we assume naively that the main effect of BSE is to reduce $W(q=0,\omega=0)$
(i.e. the change in $\epsilon_\infty$) then the substitution $W^\mathrm{RPA}{\rightarrow}W^\mathrm{BSE}$ in
Eq.~\ref{kernel_eq} would reduce the strength of the electron-hole attraction and shift the electronic structure to
(e.g. the bandgap) something intermediate between \qsgw\ and \qsgwl.  This is roughly what happens in some cases,
e.g. CrX$_{3}$~\cite{Acharya21b}.  In \emph{sp} semiconductors, however, using a better \emph{W} in the vertex causes
the gap to decrease still further by a small amount, e.g. by 0.03\,eV in InSb.  This is another manifestation of vertex
corrections being short ranged, as noted earlier.

To conclude, the combined effect of eliminating the TDA and better \emph{W} in the vertex, are not sufficient to explain
the tendency to overestimate the direct gap in small-gap zincblende semiconductors, or errors in the band dispersion.

\subsection{Response functions in semiconductors}\label{ss:outliers}

\subsubsection{Birefringence}\label{ss:birefringence}

Birefringence occurs when the refractive index depends on the polarization and propagation direction of light.  It is
normally measured as a difference in the principal axes of the ellipsoid's index of refraction, sometimes called the
``ordinary'' and ``extraordinary'' indices when there are two inequivalent ones.  We consider a few materials where
\emph{n} in the basal plane differs from \emph{n} normal to it:
\begin{align}\label{eq:birefringence}
\overline{n} &= (n_{\parallel}{+}n_{\perp})/2; \;\; &\Delta{n} = n_{\parallel}{-}n_{\perp} \\
\overline{\varepsilon} &= (\varepsilon_{\parallel}{+}\varepsilon_{\perp})/2; \;\; &\Delta{\varepsilon} = \varepsilon_{\parallel}{-}\varepsilon_{\perp}
\end{align}
Birefringence is measured as the difference $\Delta n$.  Table~\ref{tab:birefringence} compares \qsgw\ and
  \qsgwl\ predictions against experiment.  With the possible exception of \emph{h}BN, \qsgwl\ predicts
  $n$ and $\Delta{n}$ approximately within the available resolution of the experiment.

% --- Table birefringence ----
\begin{table}[h]
\caption{Birefringence in selected insulators.}
\begin{tabular}{l|c c|c c|c c}\hline
          &\multicolumn{2}{c|}{\qsgwl}&\multicolumn{2}{|c|}{\qsgw}&\multicolumn{2}{|c}{Expt} \cr \hline \hline
          & $\overline{n}$ & $\Delta{n}$  & $\overline{n}$ & $\Delta{n}$   & $\overline{n}$ & $\Delta{n}$ \cr \hline
ZnO       & 1.91      & 0.011                        & 1.77      &  0.010                        &  1.92     & 0.012%
\footnote{Extrapolated from Ref.~\cite{Bond65}, using Eq.~\ref{eq:oscillators}} \cr
          &           &                              &           &                               &  1.92     & $-$%
\footnote{Ref.~\cite{Wemple71}}\cr
CdS       & 2.29      & 0.015                        & 2.10      &  0.010                        &  2.30     & 0.016%
\footnote{Ref.~\cite{Lisitsa69}}\cr
TiO$_2$   & 2.45      & 0.22                         & 2.37      &  0.21                         &  2.50     & 0.26%
\footnote{Ref.~\cite{Cardona67}} \cr
          &           &                              &           &                               &  2.55     & 0.24%
\footnote{Ref.~\cite{Toyoda85}} \cr
\emph{h}BN &1.83      & 0.49                         & 1.79      &  0.44                         &  1.89     & 0.48%
\footnote{Single crystal, Ref.~\cite{Ishii83}} \cr
          &           &                              &           &                               &  2.12     & 0.20%
\footnote{Polycrystalline, Ref.~\cite{Geick66}} \cr
AlN       & 2.04     & 0.038                         & 1.82      &  0.048                        &  2.06     & 0.046%
\footnote{Infrared frequencies, Ref.~\cite{Moore05}} \cr
          &           &                              &           &                               &  2.16     & 0.040%
\footnote{Ref.~\cite{Roskovcova67}} \cr
\end{tabular}
\label{tab:birefringence}
\end{table}

\subsubsection{Relation between gap and dielectric function}\label{ss:otheroutliers}

%0.9$\Sigma$(\qsgwl)+0.1$\Sigma$(\qsgw), the gap increases by 0.25\,eV \textcolor{green}{CHECK}, and

Fig.~\ref{fig:gapsandepsinfty}(b) appears to predict $\varepsilon_\infty$ very well, but there are discrepancies.  Here
we focus on systems for which $\varepsilon^\mathrm{BSE}_\infty$ falls outside the uncertainty of experimental values
(estimated by the variation in reported values), and show that these errors directly correlate with errors in the
fundamental gap.

Several known potential sources of error in $\Sigma$ were enumerated in \S\ref{ss:preview}.  Among them, the electron
phonon interaction is significant for wide-gap, light-element compounds, especially polar ones where the narrow valence
band enhances the Fr\"olich interaction, Eq.~\ref{eq:phon}.  The electron-phonon interaction usually reduces gaps, by as
much as 0.5\,eV in an extreme case such as MgO.  Table~\ref{tab:neph} selects some materials where this reduction
exceeds 0.3\,eV.  In such cases $\varepsilon_\infty$ is slightly underestimated.  As we noted previously, at present
Questaal does not have the capability to incorporate the electron-phonon self-energy into the \qsgw\ cycle; however, we
can make a proxy by making a hybrid of the LDA and \qsgwl\ potentials to reduce the gap (Eq.~\ref{eq:hybrid}).  We
choose $\beta$=0 and pick the mixing parameter $\alpha$ to approximate the gap change from electron-phonon interaction
calculated in Ref.~\cite{Miglio20}.  This should be a reasonable proxy for $\Sigma^\mathrm{e-ph}$ since for these
systems the LDA and \qsgwl\ bands differ mostly in a simple rigid shift of the conduction band.  Materials in
Table~\ref{tab:neph} above the dividing line show systems for which the electron-phonon interaction exceeds 0.3\,eV, and
where both $E_{G}$ and $\varepsilon_\infty$ are thought to be reliably known.  Renormalization causes a modest increase
in $\varepsilon_\infty$, and the systematic tendency to underestimate it is reduced to approximately the experimental
uncertainty.\footnote{MgO is a mild anomaly: \qsgwl\ already overestimated $\varepsilon_\infty$ and the gap reduction worsens
the discrepancy to about 7\%.  Several independent experiments place $\varepsilon_\infty$ at 2.95, so the experimental
number is likely reliable.}

For compounds in the bottom half of the table, benchmarking becomes murkier, because the electronic structure is not
known or is poorly understood.  \emph{h}BN might have been put in the top half of the table, if so it would present a
severe anomaly.  Data for \emph{h}BN in Tables~\ref{tab:birefringence} and \ref{tab:neph} were computed from an average
of Refs.~\cite{Ishii83} and \cite{Geick66}.  To suggest the possible source of the anomaly, Table~\ref{tab:neph} also
shows an entry where experimental data is taken only from Ref.~\cite{Ishii83}, and by such a comparison the agreement is
in line with other materials.  Further experiments are needed to determine the true values (both ordinary and extraordinary)
for $\varepsilon_\infty$ in \emph{h}BN.

For less well characterized systems, if we make the ansatz that the calculated $\varepsilon_\infty$ should coincide with
the experimental one when $E_{G}$ also coincides, we can assess the effect of the error in the fundamental gap if
$\varepsilon_\infty$ is better known (this is a common situation).  We can estimate what $E_{G}$ should be by matching
$\varepsilon_\infty$ (more generally $\varepsilon(\omega)$) to experiment. In later sections we apply this technique to
several materials systems, e.g. CuAlO$_{2}$, \S\ref{ss:cualo2} and FeO, \S\ref{ss:feo}.

\begin{table}[h]
\caption{Estimated change to $\varepsilon_\infty$ induced by adjusting the \qsgwl\ self-energy according to
  Eq.~(\ref{eq:hybrid}), using $\alpha$ given in the table. ${\Delta}E_{G}{<}0$ shows the change in fundamental bandgap,
  in eV: ${\Delta}E_{G}{<}0$ indicates the gap is reduced by taking $\beta{=}0$ and $\gamma{=}1{-}\alpha$.
  ${\Delta}E_{G}{>}0$ indicates the gap is increased by taking $\gamma{=}0$ and $\beta{=}1{-}\alpha$.
  $\overline{\varepsilon}$ and $\Delta{\varepsilon}$ are defined in Eq.~\ref{eq:birefringence}.  Column ``org''
  indicates the probable predominant physical origin of $\Delta E_G$: one of eph (electron-phonon); $\Gamma$ (missing
  vertex in $\Sigma$), or * (unknown).  Top box displays systems where both $E_{G}$ and \epsi\ are fairly reliably
  known, or reliably known.  Bottom box cotains entries where $E_{G}$, and to some extent \epsi, are not well known.
  For CoO and MnO no adjustment was made owing to uncertainty in \epsi, and lack of information about the effect of the
  electron-phonon interaction.}
\begin{tabular}{|l|c c|c c c c c|c c|}\hline
  &\multicolumn{2}{c|}{\qsgwl}&\multicolumn{5}{|c|}{$[\alpha\cdot\mathrm{QS}G\hat{W}+\Delta{\Sigma}]$%
  \footnote{$\alpha$ determines $\Delta{\Sigma}$ from Eq.~\ref{eq:hybrid}, as described in the figure caption}}
                                                                               & \multicolumn{2}{|c|}{Expt}\cr \hline \hline
          & $\overline{\varepsilon}$ & $\Delta{\varepsilon}$
          & $\overline{\varepsilon}$ & $\Delta{\varepsilon}$ & $\alpha$ & $\Delta E_G$ & org
          & $\overline{\varepsilon}$ & $\Delta{\varepsilon}$ \cr \hline
AlN        & 4.14     &   .155    & 4.33      &  .172     & 0.80\footnote{to make ${\Delta}E_{G}$ correspond approximately to shift given in Ref.~\cite{Miglio20}}
                                                                               & $-0.44$                    & eph      &  4.47%
\footnote{Average of Refs.~\cite{Moore05,Roskovcova67}}                                                                            & .185 \cr
TiO$_2$    & 6.03     &   1.09    & 6.40      &   1.20    & 0.90$^\mathrm{b}$   & $-0.37$                    & eph      &  6.39%
\footnote{Average of Refs.~\cite{AIPHandbook,Cardona67}}                                                                           & 1.25 \cr
SrTiO$_3$  & 4.84     &           & 5.11      &           & 0.90$^\mathrm{b}$   & $-0.35$                    & eph      &  5.17%
\footnote{Average of Refs.~\cite{Toyoda85,Cardona67}}                                                                              & \cr
C          & 5.64     &           & 5.82      &           & 0.80$^\mathrm{b}$   & $-0.31$                    & eph      &  5.70%
\footnote{Ref.~\cite{Ruf00}}                                                                                                       & \cr
CaO        & 3.04     &           & 3.28      &           & 0.90$^\mathrm{b}$   & $-0.37$                    & eph      &  3.28%
\footnote{Ref.~\cite{Wemple71}}                                                                                                    & \cr
MgO        & 3.07     &           & 3.16      &           & 0.85$^\mathrm{b}$   & $-0.53$                    & eph      &  2.95%
\footnote{Ref.~\cite{Lines90}}                                                                                                     & \cr
InSb       & 14.2     &           & 15.5      &           & 0.80\footnote{so that $E_{G}$ approximately match known gap, 0.24\,eV at 0K}%
                                                                   & $-0.23$                                & *        & 15.7%
\footnote{Ref.~\cite{Burstein67}. See discussion around Tab.~\ref{tab:E0gaps}}                                                     & \cr
CuCl       & 3.91     &           & 3.69      &          &  0.0\footnote{adjusted gap approximately matches 3.46\,eV gap from two-photon absorption, Ref.~\cite{Reimann94}} %
                                                                   & $+0.77$                                & $\Gamma$  & 3.71\footnote{Average of Refs.~\cite{Feldman69} and ~\cite{Sueta70}}
                                                                                                                                   & \cr
NiO        & 6.14     &          & 5.97       &          &  0.9\rlap{\footnote{adjust gap to average of Refs.~\cite{Sawatzky84,Hufner84}}}%
                                                                   & $+0.23$                                & $\Gamma$ &  5.73\footnote{Average of Refs.~\cite{SpicerNiOCoO,Chern92,Pecharroman94}, variation $\pm$0.3.  See \S\ref{ss:nio}}
                                                                                                                                   & \cr\hline
                                                                                                                       &           & \cr\hline
\emph{h}BN & 3.42      &  1.78    & 3.56      &  1.97     & 0.80$^\mathrm{b}$
                                                                               & $-0.48$                    & eph      &  4.08%
\footnote{Average of Refs.~\cite{Ishii83,Geick66}}                                                                                 &1.33 \cr
\emph{h}BN &           &          &           &           &        &                                        &          &  3.63%
\footnote{Data from Ref.~\cite{Ishii83} alone}                                                                                     &1.82 \cr
Cu$_{2}$O  & 7.81     &           & 6.80       &          &  0.0\rlap{\footnote{\qsgw\  gap approximately matches
    2.17\,eV reported in Ref.~\cite{Uihlein81}, and is slightly smaller than the 2.3\,eV gap reported by Zimmermann~\cite{Zimmermann99}}}%
                                                                   & $+0.53$                                & $\Gamma$ & 6.46%
\footnote{Ref.~\cite{Okeeffe63}}                                                                                                   & \cr
CuAlO$_{2}$ & 5.43    &          & 5.22       &          &  0.8\rlap{\footnote{adjust gap to approximately match measured $\varepsilon_\infty$}}%
                                                                   & $+0.18$                                & *        & 5.13%
\footnote{Average of Refs.~\cite{Pellicer09,Vu21}, with variation $\pm$0.15. See \S\ref{ss:cualo2}}                             & \cr
CuO        & 7.86     &          & 7.12       &          &  0.8    & $+0.25$    & $\Gamma$ &  6.5\ %
\footnote{Average of Refs~\cite{Ito98,Tahir12}.  See \S\ref{ss:cuo}.}                                                           & \cr
FeO        & 17.6     &          & 12.8       &          &  0.7    & $+0.26$    & $\Gamma$ &  10.2%
\footnote{Average of Refs~\cite{Prevot77,Kugel77}, variation $\pm$0.9.  See \S\ref{ss:feo}.}                                     & \cr
CoO        & 5.15     &          &            &          &         &            &          &  5.05%
\footnote{Average of Refs.~\cite{SpicerNiOCoO,Rao65,Gielisse65}, variation $\pm$0.3.  See \S\ref{ss:coo}}                        & \cr
MnO        & 4.76     &          &            &          &         &            &          &  4.95%
\footnote{Ref.~\cite{Plendl69}.  See \S\ref{ss:mno}}                                                                             & \cr\hline
\end{tabular}
\label{tab:neph}
\end{table}

\subsection{Band Structure and dielectric function in selected nonmagnetic materials}\label{ss:eps}

In this section we present a variety of selected materials. Where sufficient experimental information is available
(e.g. LiF) those results are used to benchmark the theory.  For most of the systems presented here, the available
experimental information is partial, confused, or contradictory.  For these systems we use a mix of theory and what
experimental information seems sufficiently reliable, to arrive at a consistent picture where it seems reasonable to
do so.  In a few cases it is not fully possible (see CuAlO\textsubscript{2}, \S\ref{ss:cualo2}).

The analysis relies on the ansatz stated in \S\ref{ss:otheroutliers}, namely that if $G_{0}$ is good enough to well
characterize one-particle properties, it also well characterizes two-particle properties provided an adequate theory
for the vertex is used; and moreover, that ladder diagrams are sufficient for the vertex.  This hypothesis was affirmed
in nearly every case in the present study where reliable information is available.

\subsubsection{LiF}\label{ss:lif}

The macroscopic dielectric function of the polar insulator LiF was recently calculated in Ref.~\onlinecite{opt_PRM}
within the BSE using \qsgw\ as the starting $G_{0}$.  Since the vertex corrections are omitted in \qsgw\, the screening
of the exchange was underscreened and the gap too large.  Combined with the neglect of the electron-phonon self-energy,
this results in a greatly overestimated band gap of $\sim$16.2~eV, i.e. about 2~eV larger than the experimental
value. The underscreening also caused an overestimation of about 0.5--1~eV of the exciton binding energy. As a result of
the partial cancellation of these errors, producing the optical absorption spectrum using the BSE with the
\qsgw\ electronic structure results in a blue shift of $\sim$~0.9~eV with respect to experiment.

\begin{figure}[h!]
\hspace{0.0cm}
\includegraphics[width=0.4\textwidth,clip=true,trim=0.0cm 0cm 0.0cm 0.0cm]{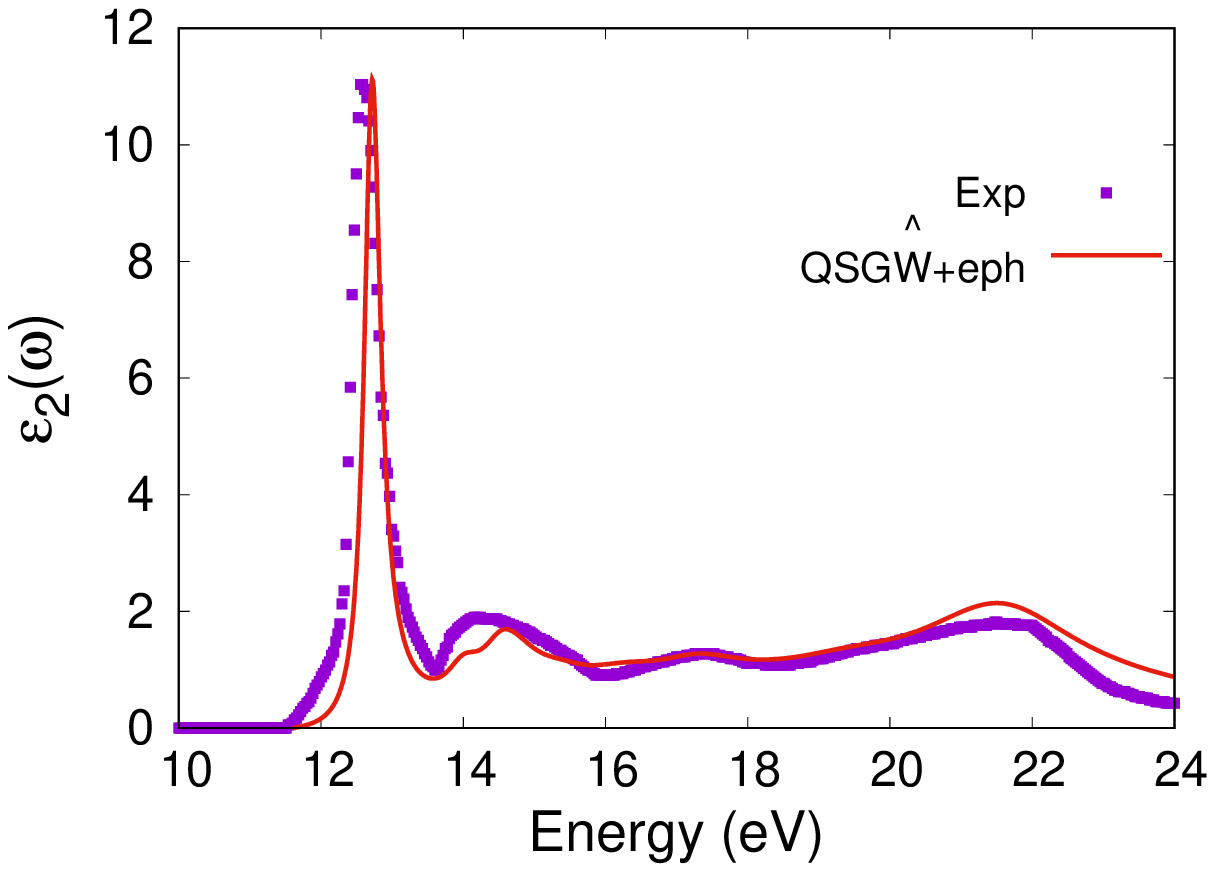}
\caption{Imaginary part of the macroscopic dielectric function for LiF. The experimental data (blue squares) \cite{Roessler:67} is compared with the results from the \qsgwl\ method. The spectrum is red-shifted by 0.483~eV to account for lattice polarization effects (see text). The spectral broadening was increased linearly to match that of the experimental spectrum: the broadening at the first excitonic peak is 0.15~eV and at the X exciton peak ($\sim$22~eV) is 1.25~eV.}
\label{image:lif}
\end{figure}

Here, we repeat the BSE calculation of the optical spectrum but on top of the \qsgwl\ electronic structure and also
consider the lattice polarization effects. Including ladder-diagram vertex-corrections produces a fundamental band gap
of 14.7~eV; a reduction of over 1.4~eV, in agreement with the vertex correction calculated in
Ref.~\onlinecite{PhysRevLett.99.246403}. Including the 0.48~eV polaron shift correction from
Ref.~\onlinecite{phonons_walt} gives then a band gap in excellent agreement with the experimental value of
14.2$\pm$0.2~eV~\cite{PhysRevB.13.5530}. The exciton binding energy is around 2~eV, also in agreement with the
experimental value~\cite{PhysRevB.13.5530} and as a consequence Fig.~\ref{image:lif} shows an excellent overall agreement
between the theoretical and experimental spectra.  The BSE \epsi\ (1.95) is close to the experimental one
(1.96, Ref.~\cite{Lee77} and 1.92, Ref.~\cite{Adair89}).

%Table~\ref{tb:seps} reports the values for the ion-clamped static dielectric constant, \epsi, at the
%various levels of theory as in Ref.~\onlinecite{opt_PRM}. Calculating the macroscopic dielectric function at the BSE
%level improved noticeably the agreement with the experimental value when comparing to the RPA value. When the electronic
%structure is calculated at the \qsgw\, rather than $G_0W_0$ level, the agreement with experiment is improved. When the
%electronic structure is then calculated with the \qsgwl\ electronic structure, the value is slightly overestimated
%($1.96$ versus the experimental value of $1.90-1.95$).
%\footnote{This value could perhaps be improved upon with better convergence in the $\kk$-grid. Since band gaps tend to converge from below (as discussed for Si) \epsi\ will then tend to converge from above.}
%\textcolor{green}{\bf As calculations are supposed to be at convergence as much as possible, the referee can ask us to, either calculate with a better grid, either extrapolate the value or estimate the error. The trend is a bit strange, one would expect qsgwl to lie between g0w0 and qsgw, right... - to note as well missing LPC}

\subsubsection{Bi\textsubscript{2}Te\textsubscript{3}}\label{ss:bi2te32}

\begin{figure}[h!]
\includegraphics[width=0.40\textwidth,clip=true,trim=0.0cm 0.0cm 0.0cm 0.0cm]{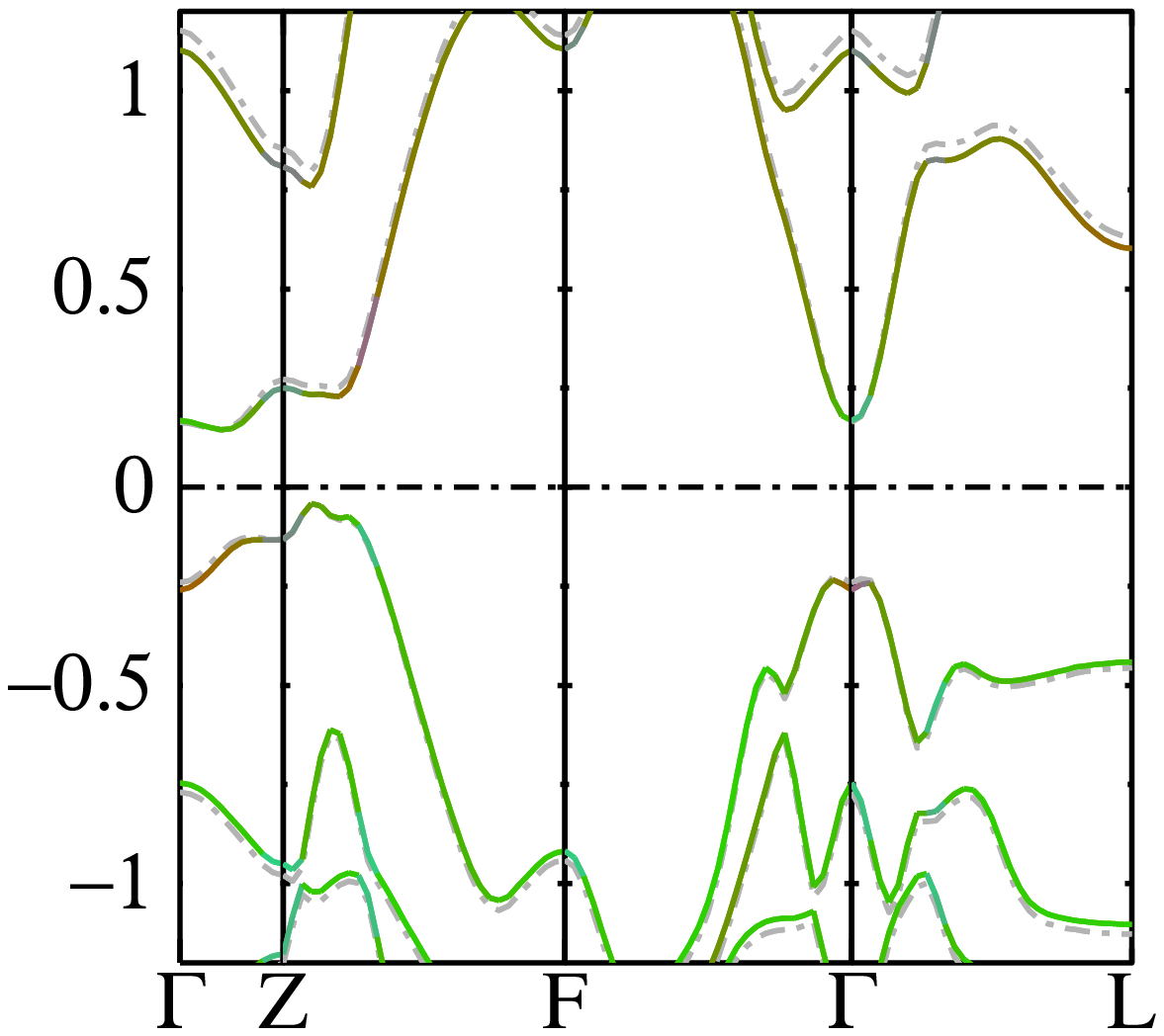}
\caption{
  Energy band structure for Bi\textsubscript{2}Te3\textsubscript{3} computed by \qsgwl\ (colored lines) and by
  \qsgw\ (grey dashed lines).
}
\label{fig:bi2te3}
\end{figure}

{Bi\textsubscript{2}Te\textsubscript{3}} is widely studied because it has topological surface states protected by
time-reversal symmetry~\cite{Fu07}.  It is a narrow-gap system with reported energy gaps between 130 and
170\,meV~\cite{Austin58,Li61,Sehr62,Thomas92}.  The \qsgw\ and \qsgwl\ bands are shown in Fig.~\ref{fig:bi2te3}, and are
seen to be nearly identical.  This system was studied previously~\cite{Michiardi14} within the LDA and
$G^\mathrm{LDA}W^\mathrm{LDA}$ approximations (albeit including the off-diagonal parts of $\Sigma$).
Fig.~\ref{fig:bi2te3} is in close agreement with Fig. 1 of Ref~\onlinecite{Michiardi14}.  That the three many-body
calculations ($G^\mathrm{LDA}W^\mathrm{LDA}$ of Ref~\onlinecite{Michiardi14}, \qsgw\ and \qsgwl) are so similar suggests
that $W$ is already well described by the LDA.  Evidently ladder diagrams have almost no effect.  This is perhaps not
suprising, since the LDA and \emph{GW} bands are also similar, with the LDA gap slightly smaller at
50\,meV~\cite{Michiardi14}.  \qsgw\ and \qsgwl\ energy gaps are both 145\,meV, slightly larger than 120\,meV reported in
Ref.~\cite{Michiardi14} (presumably because of self-consistency), and within the range of reported
experiments~\cite{Austin58,Li61,Sehr62,Thomas92}.

\subsubsection{ScN}\label{ss:scn}

ScN is a material of considerable interest in opto-electronics applications, especially as a buffer layer.  It is an
indirect gap material with the conduction band minimum at X.  Its bandgap has been controversial with many reported
values ranging from 2.03-3.2 eV for the direct gap and 0.9-1.5 eV for the indirect one.  Theoretical predictions
similarly vary, with predictions ranging between 1.82–2.59 eV (direct) and 0.79-1.70 (indirect) (see Ref.~\cite{Deng15}
for a summary and detailed discussion).

\begin{figure}[h!]
\includegraphics[width=0.22\textwidth,clip=true,trim=0.0cm 0.0cm 0.0cm 0.0cm]{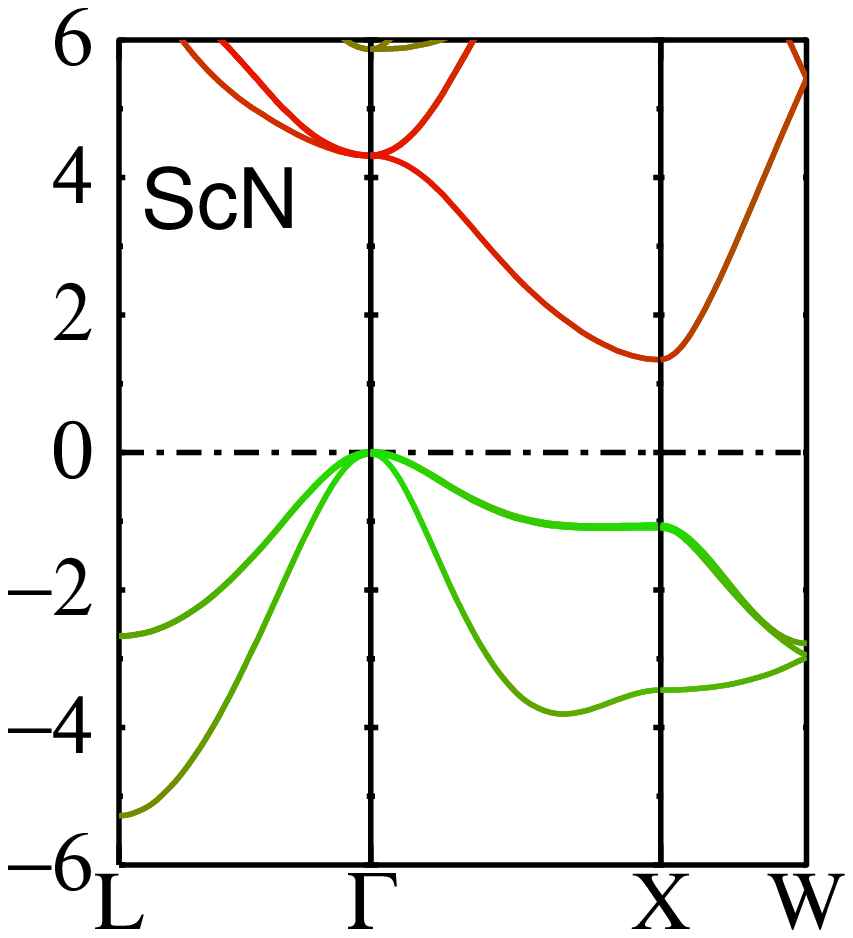}
\includegraphics[width=0.24\textwidth,clip=true,trim=0.0cm 0.0cm 0.0cm 0.0cm]{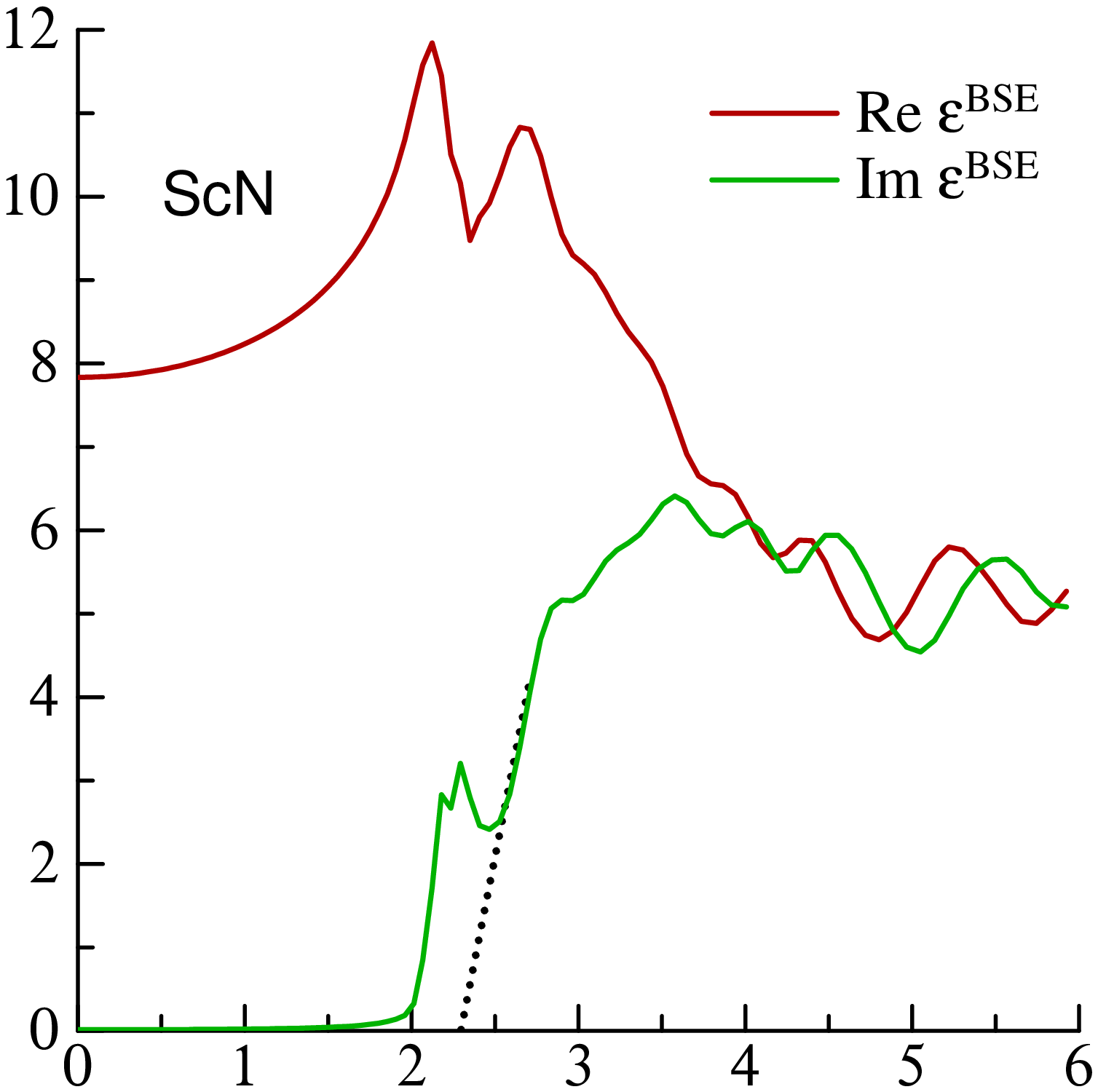}
\label{fig:scn}
\caption{(Left) energy bands for ScN.  Red and green depict projection onto Sc and N orbitals, respectively.  (Right)
  real and imaginary parts of the BSE dielectric function.  The grey dashed line depicts the RPA dielectric function
  computed from the \qsgwl\ self-energy.  The dotted line is a guide to the eye, extrapolating the second shoulder
  in $\mathrm{Im}\,\epsilon$ to zero.  Table below: Experimental and \qsgwl\ gaps and values for
  $\epsilon_{\infty}$.
}
\vskip 12pt
\begin{tabular}{|@{\hspace{0.6em}}l|c|c|c|c|}\hline\hline
                 & $E_{G}$      & $E_{G}$ (dir) & $\Delta{E}$ (eph) & $\varepsilon_\infty$ \cr\hline
Expt\footnote{Ref.~\cite{Deng15}.}
                 & 0.92           & 2.07   &          &  7.7 \cr
\qsgwl           & 1.27           & 2.33   &  -0.10\footnote{Based on analogy with InN, as described in the text}
                                                      &  7.8  \cr
\qsgwl+eph       & 1.17           & 2.23   &          &       \cr
\hline
\end{tabular}
\label{tab:scn}
\end{figure}

The most recent and detailed experimental study taking into account prior work (Ref.~\cite{Deng15}) yields an optical
indirect gap of 0.92$\pm$0.05\,eV (Table in Fig.~\ref{tab:scn}).  \qsgwl\ predicts a larger fundamental gap, 1.27\,eV.
The latter should be reduced by the electron-phonon interaction; Unfortunately no information is available in the
literature, but it is likely to be similar to InN.  The longitudinal and transverse mode phonon frequencies are similar,
while the electron effective masses in ScN are heavier (0.34\,eV for ScN, 0.07 for InN).  Thus according to the
Fr\"olich formula, Eq.~\ref{eq:phon}, the electron-phonon renormalization should be larger, by a factor between 1 and 2.
A reasonable estimate is 0.1\,eV, which is used in the Table in Fig.~\ref{tab:scn}.

A study of $\epsilon^\mathrm{BSE}(\omega)$ yields an exciton at 2.19\,eV; thus there is a spread of 0.14\,eV between
fundamental and optical gaps.  This is apparent in the dielectric function (right panel of Fig.~\ref{tab:scn})).  The
shoulder between 2 and 2.3\,eV are the subgap excitonic transitions.  The dotted line is a guide to the eye,
extrapolating the onset of the second shoulder to zero.

Combining this shift with an (admittedly crude) estimate of 0.1\,eV for the electron-phonon interaction, brings the
\qsgwl\ and ellipsometry direct optical gaps to within 0.1\,eV (approximately the uncertainty in both theory and
experiment).  $\varepsilon_\infty$ also agrees well with Ref.~\cite{Deng15} (Table in Fig~\ref{tab:scn}).  As
$\varepsilon^{\mathrm{BSE}}(\omega)$ does not include indirect couplings via the electron-phonon interaction, and we did
not consider contribution from $q{\ne}0$ transitions, and cannot determine the exciton binding for the indirect
transition, but it is reasonable to assume it is similar to the direct one.

To summarize, a consistent picture emerges in close agreement with the recent work of Ref.~\cite{Deng15}, with the
proviso that the one-particle and two-particle gaps must be distinguished.

\subsubsection{CeO\textsubscript{2}}\label{ss:ceo2}

\begin{figure}[h!]
\includegraphics[width=0.23\textwidth,clip=true,trim=0.0cm 0.0cm 0.0cm 0.0cm]{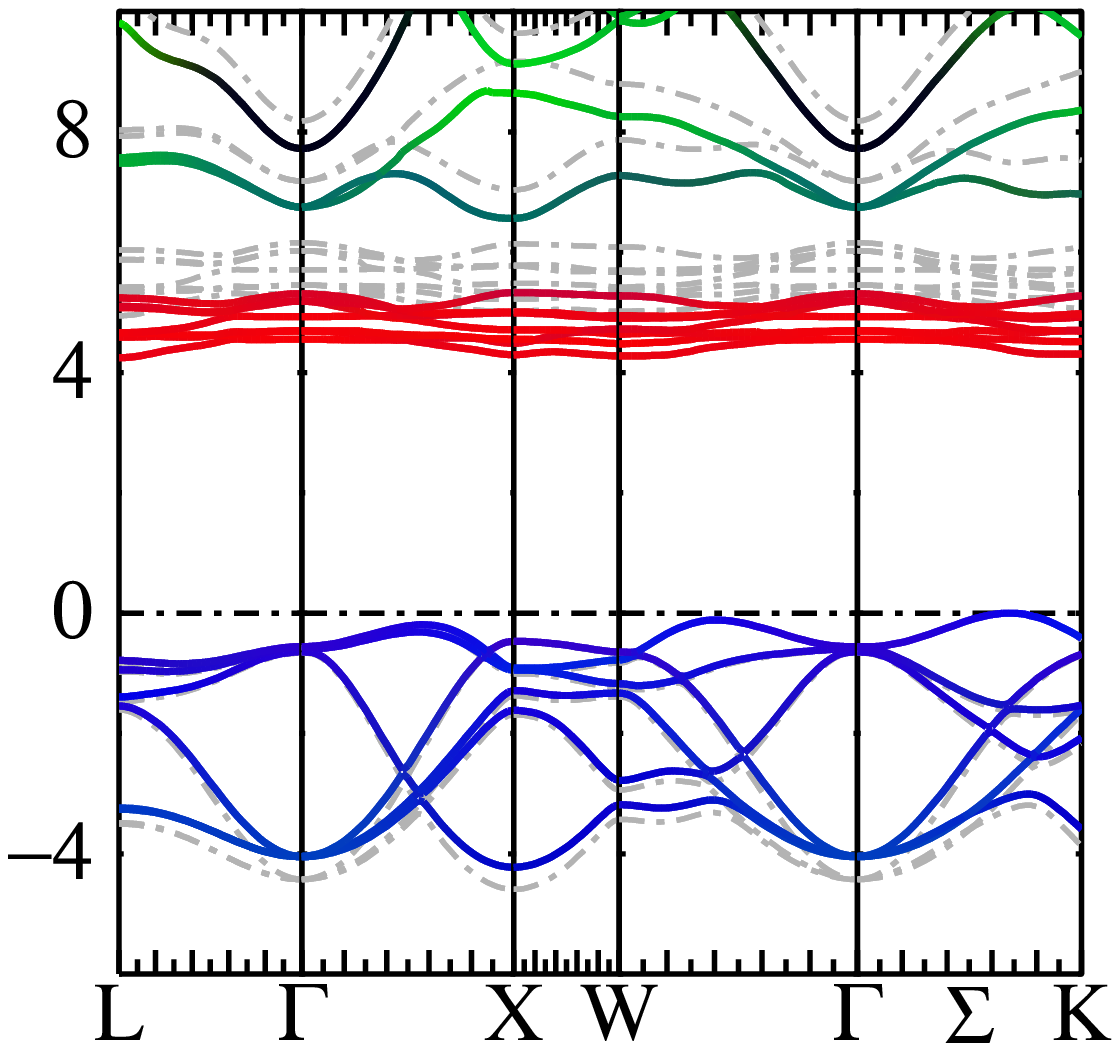}
\includegraphics[width=0.23\textwidth,clip=true,trim=0.0cm 0.0cm 0.0cm 0.0cm]{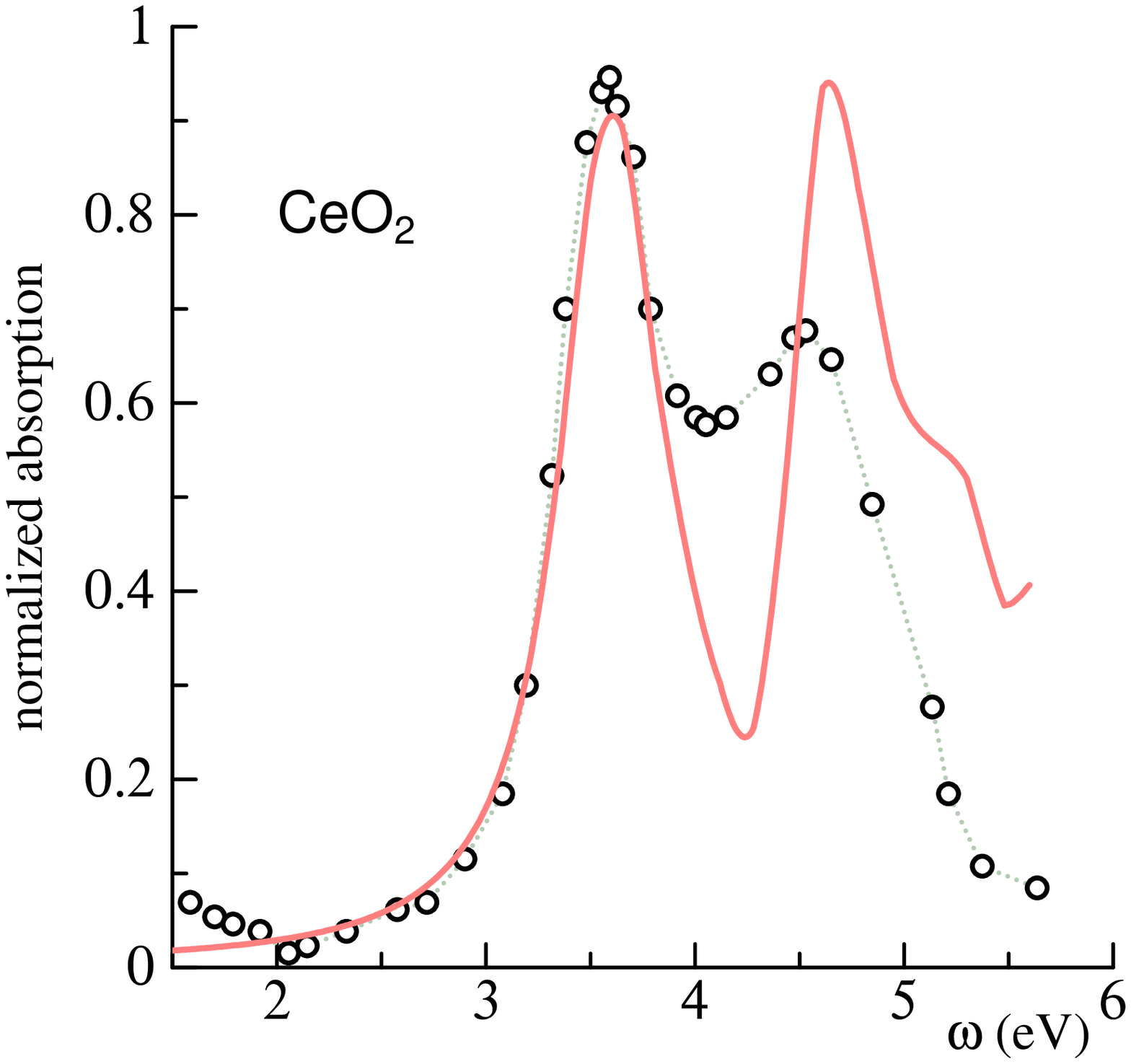}
\caption{
  Left: energy band structure in CeO\textsubscript{2} computed by \qsgwl.  Red, green, and blue correspond to
  Ce-\emph{f}, Ce-\emph{d}, and O-\emph{p} character, respectively.  Light gray dashed lines show corresponding
  \qsgw\ bands.  Right: Corresponding dielectric function.  Circles are absorpance data digitized from
  Ref.~\cite{BRITO20101821}.  Red curve is the BSE absorption $\alpha$ generated from the \qsgwl\ hamiltonian
  ($\alpha{=}4\pi{n}/{\lambda}$, $n^2$=dielectric function). The energy axis for the calculated function is redshifted
  by 0.5\,eV as described in the text.
}
\label{fig:ceo2}
\end{figure}

Electric conduction in CeO\textsubscript{2} takes place both by ionic and electronic conduction, and can be controlled
by changing the O\textsubscript{2} pressure.  Its unusual electrical properties make it a promising candidate anode in
solid oxide fuel cells, or for intermediate temperature electrolytes~\cite{Thangadurai10}.  Its energy bandgap has been
measured optically by a number of groups, by absorption~\cite{BRITO20101821,SANDHYAKUMARI2008509} or
reflectance~\cite{ZHANG2009671,Thangadurai10,CHEVIRE20063184}.  They vary in details, but all report optical bandgaps
ranging between 3.1 and 3.3\,eV.  We compare against absorption data in Ref.~\cite{BRITO20101821} because the
reported energy range was wide enough to show two peaks (Fig.~\ref{fig:ceo2}).

The \qsgwl\ bandgap is computed to be 4.24\,eV (4.93\,eV in \qsgw).  The valence band is almost pure O-\emph{p}
character, the lowest conduction bands are nearly dispersionless Ce-4\emph{f} bands (see left panel,
Fig.~\ref{fig:ceo2}).  Above the narrow Ce-4\emph{f} bands are states of mixed Ce-5\emph{d} and Ce-6\emph{s} character.
LDA bands are not shown, but there is an orbital-selective shift: Ce-4\emph{f} bands shift about 2.4\,eV relative to the
LDA, while the Ce-5\emph{d} bands shift only 0.8\,eV.  (There is a smaller but orbital-selective shift of the opposite
sign as ladders are added to \qsgw, as happens for NiO, Fig.~\ref{fig:niobands}.)  The LDA gap (1.8\,eV) is not so far
removed from the optical gap ($\sim$3.2\,eV), but this is largely fortuitous, for several reasons.

\begin{itemize}[leftmargin=*]

\item The LDA badly understimates the shift in empty Ce-4\emph{f} states

\item There is a large renormalization of the Hartree part of the hamiltonian, which reduces the gap substantially
  (Table~\ref{tab:iteration0}), and this partially cancels the first error.

\item The optical gap and fundamental gap apparently differ by $\sim$0.6\,eV.  Note the absorption spectra in the right
  panel of Fig.~\ref{fig:ceo2}.  In that figure, the BSE optical spectra were red-shifted by 0.5\,eV to align them with
  the absorption data.  Thus the \qsgwl\ optical gap is $\sim$3.6\,eV, about 0.6\,eV less than the fundamental gap.
  \qsgwl\ still overestimates the optical gap by $\sim$0.5\,eV (Fig.~\ref{fig:ceo2}); however, some portion of this
  difference can be attributed to the electron-phonon interaction, as explained below.

\end{itemize}

The global valence band maximum falls on the $\Sigma$ line, about 2/3 between $\Gamma$ and K (see Fig.~\ref{fig:ceo2}),
though other local maxima are nearly degenerate with it.

Comparing \epsi\ to experiment is not straightforward because of the wide dispersion in reported experimental data, as
well as preparation conditions~\cite{Guo95} and the crystallinity of the material.  Reported values vary from
4.7~\cite{Marabelli87} to a range between 5.8 and 6.6~\cite{Inoue90} on single crystals.  A measurement of highly
oriented crystalline films yields \epsi=6.1~\cite{Guo95}.  \epsi\ from \qsgwl+BSE is found to be 5.8, which fits
comfortably within the range of reported experimental values.

An estimate of the zero-point motion can be made using Eq.~(\ref{eq:phon}).  For this equation, $a_P$ is needed
separately for the conduction band and valence band; however, the effective mass approximation is not meaningful for the
almost flat conduction band, and we consider only the valence band here.  Various experiments put $\omega_\mathrm{LO}{=}30{\pm}5$
meV~\cite{SHI2009830}; we compute the \qsgwl\ hole masses to be (0.86, 1.3, 1.8)$m_0$, for an average mass of 1.27.
Using these values, we obtain $a_P{=}18.9a_0$.  The static dielectric constant is roughly ${\sim}25$~\cite{Pradhani18}.
Using $\epsilon_\infty{=}6$, Eq.~(\ref{eq:phon}) predicts the valence band contribution to gap reduction to be 46\,meV.
Effective mass theory cannot be applied to the nearly dispersionless conduction band, but it is reasonable to expect its
contribution to the total gap reduction to be several times larger.  A factor of three larger conduction band
contribution would give a total gap correction of $\sim$0.2\,eV.  This accounts in part, but it would seem not entirely,
for the apparent $\sim$0.3-0.5\,eV overestimate of the optical gap predicted by \qsgwl.  It would seem a discrepancy
of order 0.3\,eV remains, but a better determination of the electron-phonon interaction is needed to know the
discrepancy reliably.  Assuming that for whatever reason, the \qsgwl\ gap is too large by 0.5\,eV, the fundamental
gap should be about 3.75\,eV.

% see end of this document for estimate of e-ph in CeO2

The RPA dielectric function calculated from the LDA\cite{SHI2009830} also shows a peak around 3\,eV, but this is an
artifact of error cancellation: RPA omits strong excitonic effects and the LDA gap is too small.

\subsubsection{TiO\textsubscript{2}}\label{ss:tio2}

\begin{figure}[h!]
\includegraphics[width=0.20\textwidth,clip=true,trim=0.0cm 0.0cm 0.0cm 0.0cm]{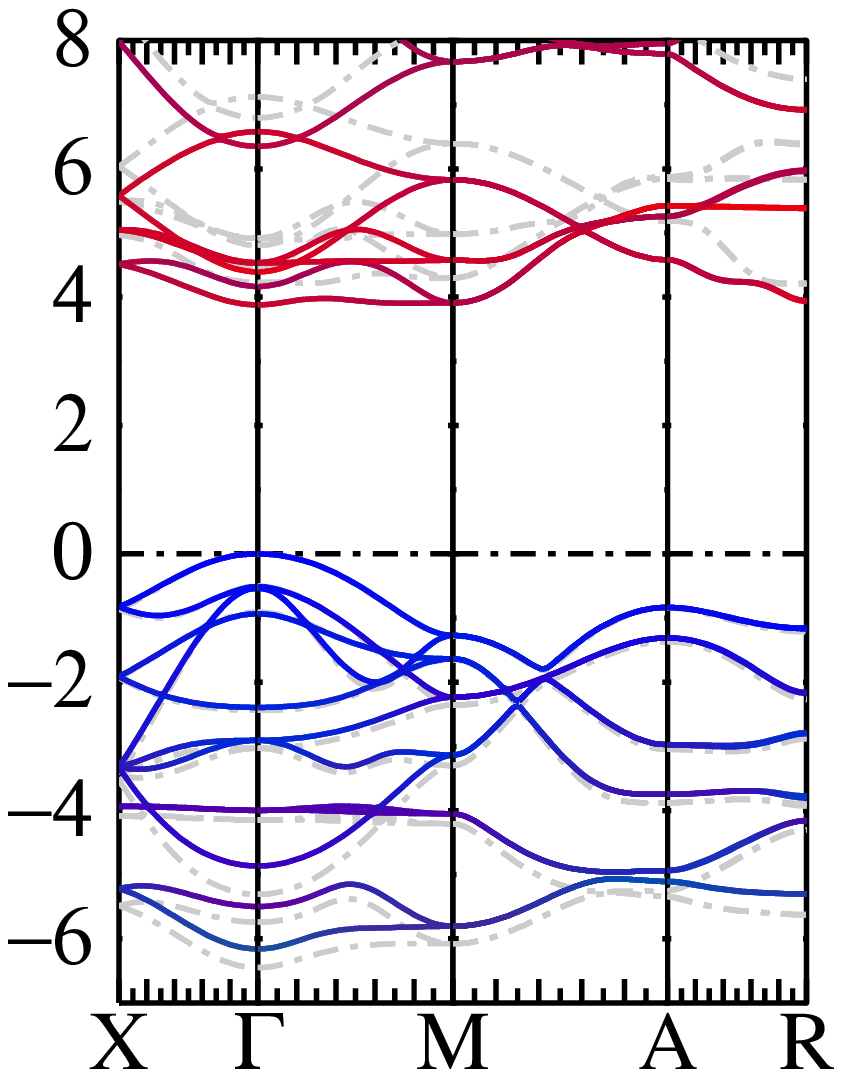}
\includegraphics[width=0.26\textwidth,clip=true,trim=0.0cm 0.0cm 0.0cm 0.0cm]{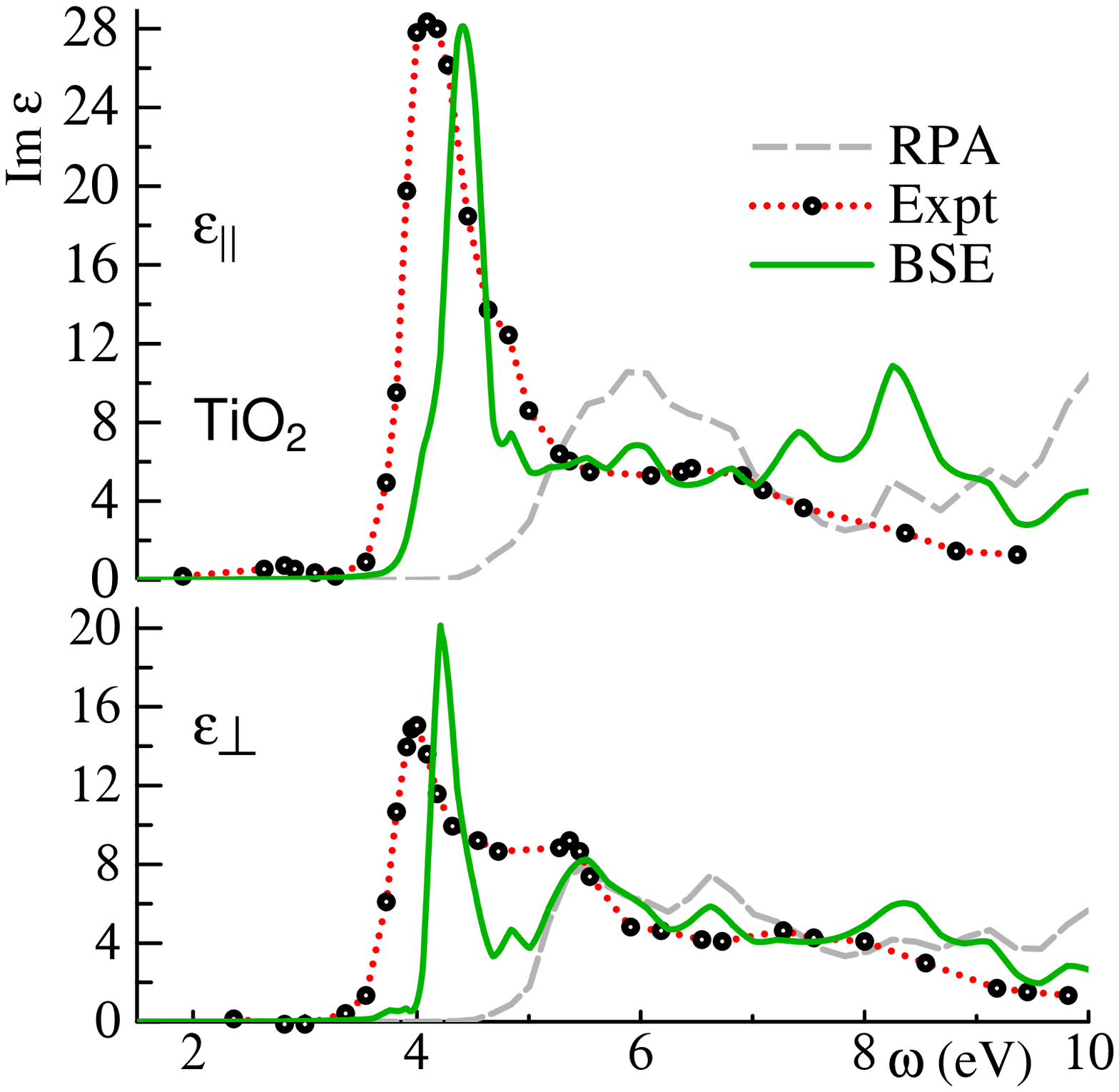}
\caption{(Left) energy bands for TiO\textsubscript{2}.  Colored bands are taken from \qsgwl\ calculations, with red and
  blue showing projections onto Ti \emph{d} and O, respectively.  Dashed gray lines show corresponding \qsgw\ results.
  (Right) dielectric function for polarization parallel and perpendicular to the \emph{z} axis.  Circles connected by
  dotted red lines are data digitized from Ref.~\onlinecite{Cardona67}.  BSE $\mathrm{Im}\,\epsilon_\perp(\omega)$ and
  $\mathrm{Im}\,\epsilon_{||}(\omega)$, computed from \qsgwl, are shown in green.  Light dashed lines compare the
  RPA results generated from the same \qsgwl\ hamiltonian.}
\label{fig:tio2}
\end{figure}

TiO\textsubscript{2}, also known as titania, is a widely used commercial compound, as food coloring (E number E171) and
as a pigment in paint for example.  We consider here only the rutile phase.  TiO\textsubscript{2} is a $d^{0}$ system
with valence band essentially O 2\emph{p} character and conduction Ti \emph{d} character (Fig.~\ref{fig:tio2}).  The
optical absorption edge was measured to be $\sim$3.3\,eV by Cardona and Harbeke from reflection
measurements~\cite{Cardona67}, and as 3.03~eV from optical transmission~\cite{Tang95}.  This is well below the calculated
\qsgwl\ fundamental gap of 3.88\,eV.  Part of the difference may be attributed to the electron-phonon interaction, which
was calculated in Ref.~\onlinecite{Miglio20} to reduce the gap by 0.34\,eV.  Reducing the gap by this amount puts it in
line with a PES/BIS study, which reported a fundamental gap of 3.3$\pm$0.5\,eV\cite{Shin94}.

The first peak in $\mathrm{Im}\,\epsilon(\omega)$ is blue shifted about 0.2\,eV compared to experiment~\cite{Cardona67},
and correspondingly \epsi\ is about 5\% smaller than experiment (Table~\ref{tab:neph}). The corresponding
\qsgwl\ birefringence is also slightly underestimated (Table~\ref{tab:birefringence}): $\Delta{n}^\mathrm{BSE}=0.22$
compared to $\Delta{n}^\mathrm{expt}=0.26$.  The corresponding RPA values from \qsgw\ are about 80\% of experiment, as
is typical, with $\Delta{n}^\mathrm{RPA}=0.16$.  Excitonic effects are strong in TiO\textsubscript{2}: compare
$\mathrm{Im}^\mathrm{RPA}\epsilon$ to $\mathrm{Im}^\mathrm{BSE}\epsilon$ in Fig.~\ref{fig:tio2} (both were generated
from the same hamiltonian).  The BSE redshifts the peak in $\mathrm{Im}\epsilon$ and significantly changes the
shape.

Both $\mathrm{Im}^\mathrm{BSE}_{||}\epsilon(\omega)$ and $\mathrm{Im}^\mathrm{BSE}_{\perp}\epsilon(\omega)$ show
reasonable resemblance to the experiment below 6\,eV: peak at 4\,eV, shoulder at 5\,eV in both
$\mathrm{Im}\,\varepsilon_{||}$ and $\mathrm{Im}\,\varepsilon_{\perp}$.  Three bound, weakly active excitons are found
in the region ($E_{c}{-}0.45$, $E_{c}{-}0.17$\,eV) below the fundamental gap, and several bright ones between
$E_{c}{-}0.13$ and $E_{c}$.  Here $E_{c}$ is the conduction band minimum.  Thus $\mathrm{Im}^\mathrm{BSE}\epsilon$ shows
a strong peak close to the fundamental gap, with a tail extending below.

The electron-phonon interaction missing in \qsgwl\ well accounts for the blue shift in leading shoulders in
$\mathrm{Im}^\mathrm{BSE}_{||}\epsilon(\omega)$ and $\mathrm{Im}^\mathrm{BSE}_{\perp}\epsilon(\omega)$ relative to
experiment.  As a proxy to account for it, we repeat the calculation with a hybrid $G_{0}$, consisting of 90\%
\qsgwl\ and 10\% LDA.  This reduces the gap by the amount calculated in Ref.~\onlinecite{Miglio20}.  With this shift,
$\epsilon_\infty$ and the birefringence both align closely to available experiments (Table~\ref{tab:neph}).

Thus, discrepancies in \qsgwl\ fundamental gap and experimental optical gap are fully explained in terms of a combination
of excitonic effects, and the electron-phonon interaction.  Adding the latter to the \qsgwl fundamental gap, we conclude
that its true value is 3.5$\pm$0.1\,eV, significantly larger than the widely accepted value of $\sim$3\,eV.

\subsubsection{SrTiO\textsubscript{3}}\label{ss:srtio3}

\begin{figure}[h!]
\includegraphics[width=0.20\textwidth,clip=true,trim=0.0cm 0.0cm 0.0cm 0.0cm]{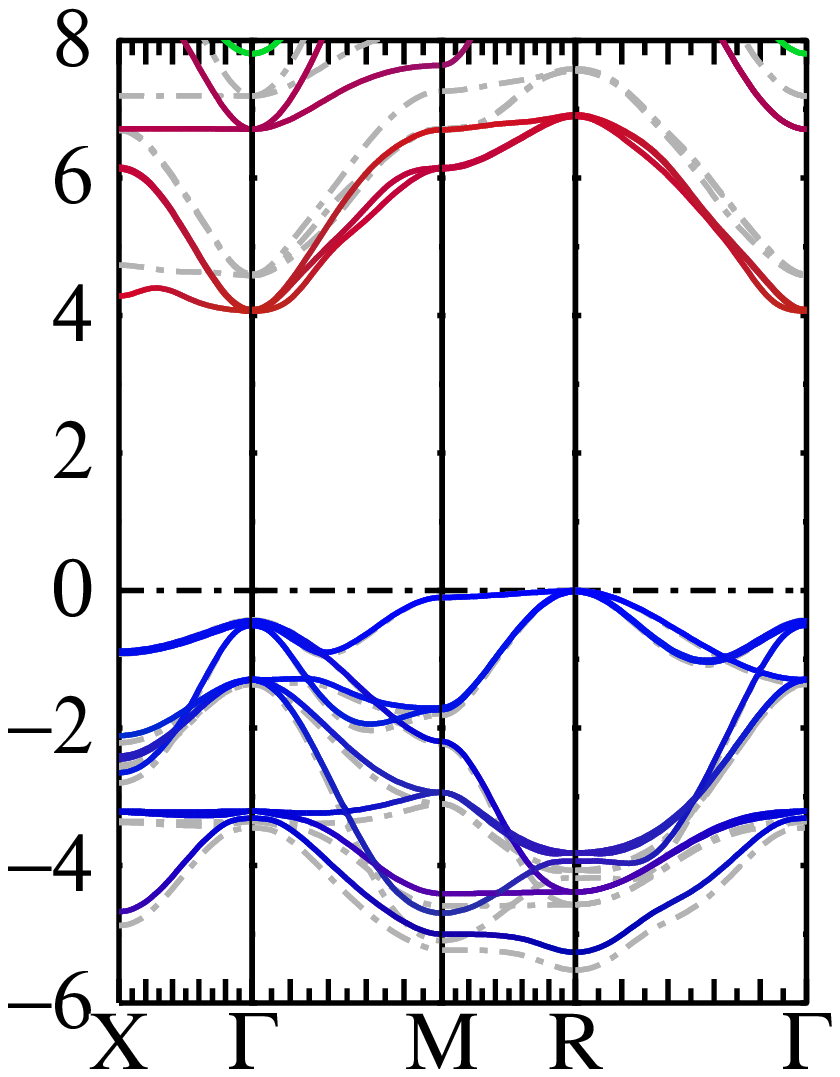}
\includegraphics[width=0.26\textwidth,clip=true,trim=0.0cm 0.0cm 0.0cm 0.0cm]{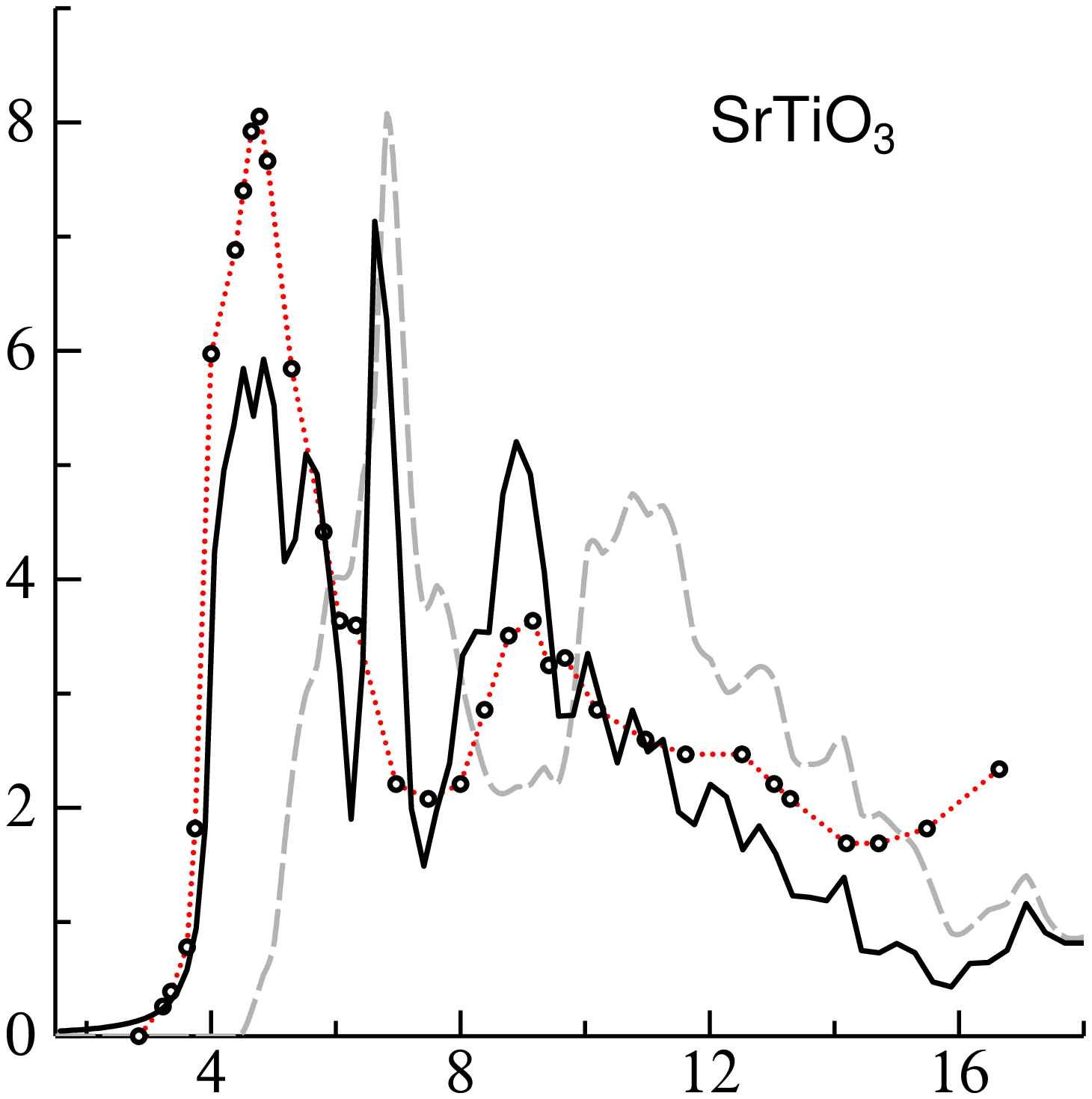}
\caption{(Left) energy bands for SrTiO\textsubscript{3}.  Colored bands taken from \qsgwl\ calculations, with red and
  blue showing projections onto Ti 4\emph{d} and O 2\emph{p}, respectively.  Gray lines show corresponding \qsgw\ results.
  (Right) dielectric function.
  Circles connected by dotted red lines are data digitized from Ref.~\onlinecite{Cardona67}.
  The black line shows the $\mathrm{Im}\,\epsilon^\mathrm{BSE}(\omega)$ generated from a \qsgwl\ hamiltonian on a $10{\times}10{\times}10$
  \emph{k}-mesh (the waviness is an artifact of incomplete $k$ convergence).
  The grey line shows $\mathrm{Im}\,\epsilon^\mathrm{RPA}(\omega)$ generated from the same hamiltonian.
  $\mathrm{Im}\,\epsilon^\mathrm{RPA}(\omega)$ vanishes at the fundamental direct gap.
}
\label{fig:srtio3}
\end{figure}

SrTiO\textsubscript{3} is a perovskite material that may exist in the usual different perovskite phases: cubic,
tetragonal and orthorhombic.  Like TiO\textsubscript{2}, it is a $d^{0}$ compound with valence band essentially O
2\emph{p} character and conduction Ti \emph{d} character (Fig.~\ref{fig:srtio3}).  In this respect it is very similar to
TiO\textsubscript{2}, and care must be taken in interpreting the experiments to determine the fundamental bandgap.  The
experimental band gap is reported to be 3.25\,eV (indirect) and 3.75\,eV (direct)~\cite{Benthem01} and according to
Bhandari et al.~\cite{PhysRevMaterials.2.013807} it is almost independent of the structure.  According to \qsgwl, the
system has a indirect gap, of 4.06\,eV, with the valence band maximum at R and conduction band minimum at $\Gamma$.  The
direct gap at $\Gamma$ is 4.51\,eV, larger than the indirect one by 0.45\,eV; however, the \qsgwl\ fundamental gap and
optical direct gap differ by about 0.75\,eV (Fig.~\ref{fig:srtio3}).  Note the BSE code has no electron-phonon coupling
and cannot detect indirect transitions.

Peak positions at 5, 9, and 12 eV correspond well to ellipsometry data, though peak amplitudes are different, especially
at 6 and 7 eV.  There are three reported experimental values for \epsi: the low frequency index of
refraction~\cite{AIPHandbook} extrapolated to 0 (\S\ref{ss:oscillators}) yields \epsi=4.71. $n$ between 2.2 and 2.3
(reported in Ref.~\cite{Benthem01}) yields \epsi\ ranging between 4.8 and 5.3; and finally a classic ellipsometry
measurement by Cardona~\cite{Cardona65} reported \epsi=5.3.  Thus it is likely $\epsilon_{\infty}{=}5.0{\pm}0.3$.  We
find from \qsgwl\ $\epsilon_{\infty}{=}4.84$ slightly lower than the average, consistent with the bandgap being slightly
overestimated.  According to Ref.~\cite{Miglio20}, the electron-phonon interaction should reduce the gap by 0.33\,eV.

\begin{table}[h]
\caption{Experimental and \qsgwl\ gaps and $\epsilon_{\infty}$ in SrTiO\textsubscript{3}.}
\begin{tabular}{@{\hspace{0.1em}}l|c|c|c|c}\hline
                 & $E_{G}$      & $E_{G}$ (dir) & $\Delta{E}$ (eph) & $\varepsilon_\infty$ \\
Expt\footnote{Ref.~\cite{Benthem01}.}
                 & 3.25           & 3.75   &          &  5.0$\pm$0.3\footnote{Average of Refs.~\cite{AIPHandbook,Benthem01,Cardona65}} \cr
\qsgwl           & 4.05           & 4.42   &  -0.33\footnote{Approximately the value in Ref.~\cite{Miglio20}}
                                                      &  4.84 \cr
0.9\,\qsgwl\,+\,0.1LDA & 3.72           &        &          &  5.11 \cr
\hline
\end{tabular}
\label{tab:srtio3}
\end{table}

The shoulder in \qsgwl\ dielectric function in Fig.~\ref{fig:srtio3} lies about 0.1\,eV above the ellipsomentry
measurement of Ref.~\cite{Benthem01}.  This suggests that the band structure is close to the true one.
$\epsilon_{\infty}$ is slightly lower than the average experimental value (Table~\ref{tab:srtio3}), which suggests that
the uncorrected \qsgwl\ gap is slightly too large.  Using a hybrid functional to reduce the gap, $\epsilon_{\infty}$
moves close to the average experimental value for $\epsilon_{\infty}$ (Table~\ref{tab:srtio3}), but the shoulder in
$\varepsilon(\omega)$ is slightly redshifted relative to the Benthem data.  Thus there is a slight inconsistency.  This
excludes a precise determination of the fundamental gap, but we conclude it is 3.75$\pm$0.1\,eV, which is about 0.5\,eV
larger than the reported optical gap.

\subsubsection{CuAlO\textsubscript{2}}\label{ss:cualo2}

CuAlO\textsubscript{2} has received a great deal of attention because it is a \emph{p}-type transparent conducting
oxide (TCO); indeed, it seems to be the only known TCO that can be doped \emph{p}-type.

\begin{figure}[h!]
\includegraphics[width=0.18\textwidth,clip=true,trim=0.0cm 0.0cm 0.0cm 0.0cm]{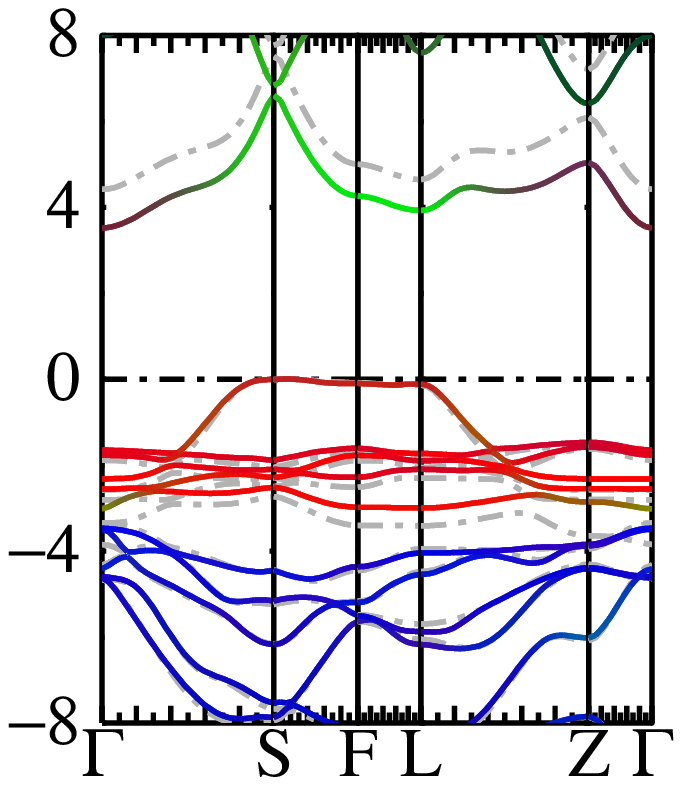}
\includegraphics[width=0.28\textwidth,clip=true,trim=0.0cm 0.0cm 0.0cm 0.0cm]{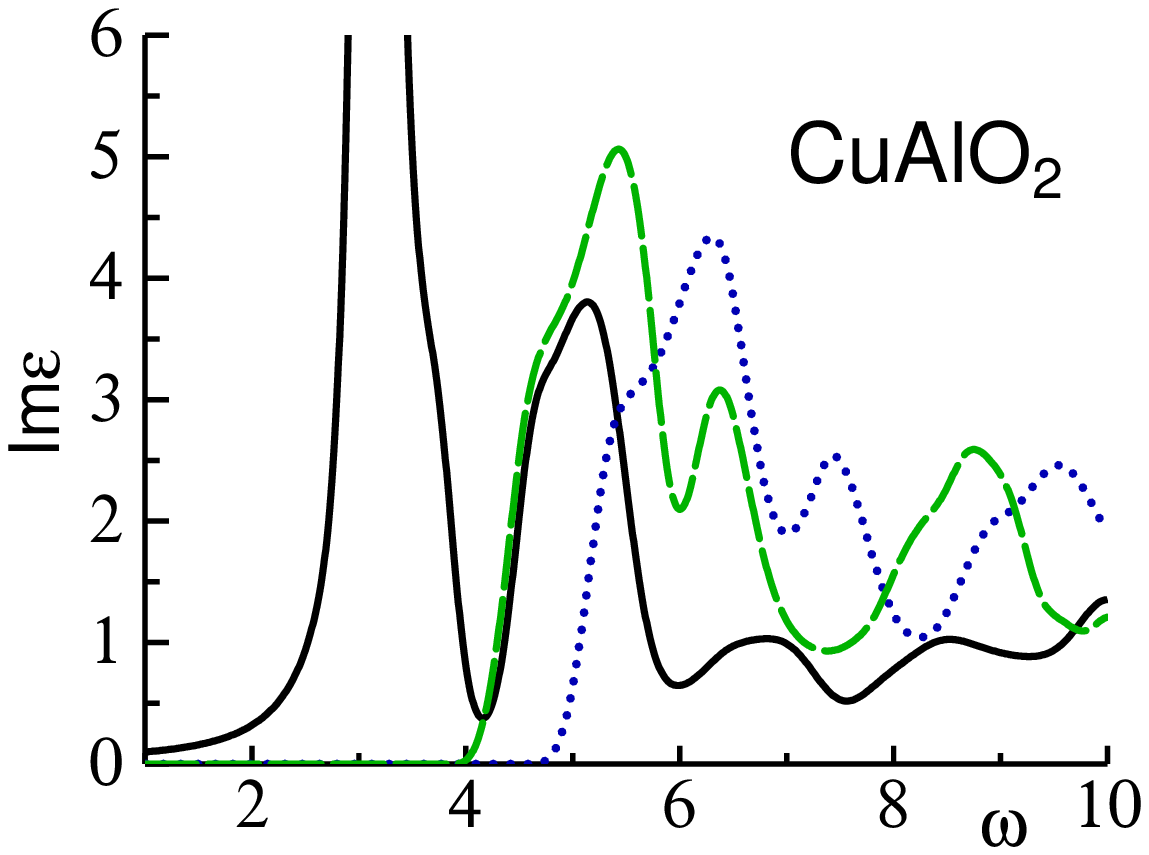}
\caption{(Left) energy bands for CuAlO\textsubscript{2}.  Colored bands are taken from \qsgwl\ calculations, with red,
  blue, and green showing projections onto Cu \emph{d}, O, and Cu \emph{sp}, respectively.  Dashed gray lines show
  corresponding \qsgw\ results.  Points S, F, and L correspond to (2/3,1/3,0), (1/2,1/2,0), and (0,0,-1/2),
  respectively, as multiples of the reciprocal lattice vectors.  (Right) dielectric function (average of \emph{x} and
  \emph{z} directions) as function of frequency $\omega$ (eV).  Solid line shows $\mathrm{Im}\,\epsilon^\mathrm{BSE}(\omega)$,
  generated from \qsgwl.  The green dashed line  shows $\mathrm{Im}\,\epsilon^\mathrm{RPA}$
  also generated from the \qsgwl\ hamiltonian.  Note $\mathrm{Im}\,\epsilon^\mathrm{BSE}$ and
  $\mathrm{Im}\,\epsilon^\mathrm{RPA}$ approach 0 near the fundamental direct gap at 4\,eV; however,
  $\mathrm{Im}\,\epsilon^\mathrm{BSE}$ has additional subgap peaks from strongly bound excitons.  The blue dotted line
   shows $\mathrm{Im}\,\epsilon^\mathrm{RPA}(\omega)$ generated from \qsgw.  The RPA functions look similar except
  for a 1\,eV shift, because of the difference in bandgaps.
}
\label{fig:cualo2}
\end{figure}

{CuAlO\textsubscript{2}} was not included in Fig.~\ref{fig:gapsandepsinfty} because reports of its magnitude vary
widely.  Reports of the lowest (indirect) gap ranges between 1.65~\cite{Yanagi00,Banerjee05} and
2.99\,eV~\cite{Pellicer06,Tate09}, and a direct gap ranging between 3.3 and 4.2\,eV~\cite{Kawazoe97,Pellicer06,Kim07}.
\qsgwl\ finds that the valence band is almost exclusively Cu 3\emph{d} character, with its maximum at a low-symmetry
point near S in Fig.~\ref{fig:cualo2}.  The conduction band minumum at $\Gamma$ is mixed Cu-\emph{sd} and O-\emph{p}
character.  The fundamental (indirect) gap is found to be 3.5\,eV, and the lowest direct gap at L is 4.0\,eV.  There is
a large region of \emph{k} space where the highest valence band is nearly dispersionless.  A prior calculation, using a
hybrid functional, found similar gaps but not the nearly dispersionless highest valence band~\cite{Scanlon10}.

The principal axes for hole mass are along low-symmetry directions, with a moderate mass (1.3) on one principal axis and
large masses (2.7, $\sim$10) on the other two.  The conduction band at $\Gamma$, by contrast, has much smaller masses
and they are along the Cartesian axes (1.3 in \emph{xy}, 0.41 along \emph{z}).  It has been argued that there should be
a large gap renormalization from the electron-phonon interaction~\cite{Vidal10b}; however the measured difference
\epsi\ (5.1) and $\epsilon_0$ (7.7) ~\cite{Pellicer09} is fairly small, and the eigenstates at the band edges are mostly
Cu-like instead of O (as it is in MgO), both of which reduce $\Sigma^\mathrm{e-ph}$ in the Fr\"olich model
(\S\ref{ss:frolich}).\footnote{It would seem that Ref.~\cite{Vidal10b} did not properly take into account the volume
confinement of \emph{W} in \textbf{q} space~\cite{phonons_walt}.}

The fundamental gap is in line with the photoemission study of Ref.~\cite{Yanagi00}, which measures the DOS, a
one-particle property.  However, when two-particle properties are considered, \qsgwl\ predicts the situation to be more
complicated.  The right panel of Fig.~\ref{fig:cualo2} compares the BSE and RPA dielectric functions.  Both approach 0
at the fundamental direct gap.  However, the BSE shows strong peaks below the fundamental gap, around 3.2\,eV; there are
also several excitons between 3.7\,eV and the fundamental gap at 4\,eV.  Such deep excitons are not typical in \emph{sp}
semiconductors, but it can be understood as an artifact of the nearly dispersionless Cu-like valence band, as well
as a relatively small dielectric constant of 5.1.  \qsgwl\ predicts \epsi\ rather well.  If the strong correlation between the reliability of
\epsi\ and the bandgap \S\ref{ss:consistency} applies equally to CuAlO\textsubscript{2}, the gap should be close to the
\qsgwl\ prediction of 3.5\,eV.

$\epsilon(\omega)$ was computed without an electron-phonon contribution, so it can only measure direct transitions.
Presumably there will be other excitons for bound electron-hole pairs coupling $\Gamma$ and states in the valence band
as well; thus the optical response will show some intensity in a spread below the peak at 3.2\,eV, larger than what is
shown in Fig.~\ref{fig:cualo2}, possibly as much as the difference between the direct and indirect gap.  Since most
determinations of the gap are performed with optical measurements, much of the confusion in the literature likely
originates from these deep excitons.  These excitons cannot explain a gap as low as 1.8\,eV, however; such a gap
likely originates from a defect band, which explains why it is not always seen.  Indeed, recent work \cite{Li18} shows
that the optical absorption edge is strongly dependent on preparation and post-annealing conditions.  Defects apparently
play an important role, which adds to the confusion about experimental reports on this materials system.

\subsubsection{VO\textsubscript{2}}\label{ss:vo2}

In the low-temperature monoclinic ({M\textsubscript{1}}) phase, {VO\textsubscript{2}} has a gap approximately
0.7\,eV~\cite{Shin90,Bermudez92}.  {M\textsubscript{1}} is a deformation of the high-symmetry rutile phase.  The unit
cell, consisting of four V atoms all equal in the rutile phase, dimerize into two pairs with short bond lengths.  It is
generally agreed that the V dimerization is what is responsible for the gap, splitting the V \emph{d} manifold into a
single occupied \emph{d} bond per dimer, and a corresponding antibond (Peierls transition).  Wentzcovitch et
al.~\cite{Wentzcovitch94} calculated the energy band structure and suggested that despite the LDA yielding no gap, that
the origin of the gap was more Peierls like than Mott-like.  This picture is further supported by the observation that
LDA augmented by single-site DMFT is also metallic~\cite{Liebsch05}, which would not happen in a simple Mott
description.  A cluster form of DMFT added to LDA does yield a gap~\cite{Biermann05}.  This indicates that the
nonlocality of the self-energy is essential, and explains why it is too small in the LDA.  Gatti et al.~\cite{Gatti07}
employed \emph{GW} to study this system, which captures the nonlocality quite well.  While they found
$G^\mathrm{LDA}W^\mathrm{LDA}$ failed to open a gap, a self-consistent \emph{GW} scheme within the COHSEX approximation
did so.  Counterbalancing this view, a DMFT work~\cite{Brito16} argued the {M\textsubscript{1}} phase should be
characterized as the Mott transition in the presence of strong intersite exchange.  In our view Gatti's work is the most
definitive, as it does not rely on the LDA, partitioning, or adjustable parameters.  It also confirms the original
Wentzcovitch conjecture: {VO\textsubscript{2}} is a simple band insulator.

\begin{figure}[h!]
\hspace{0.0cm}
\includegraphics[width=0.22\textwidth,clip=true,trim=0.0cm 0cm 0.0cm 0.0cm]{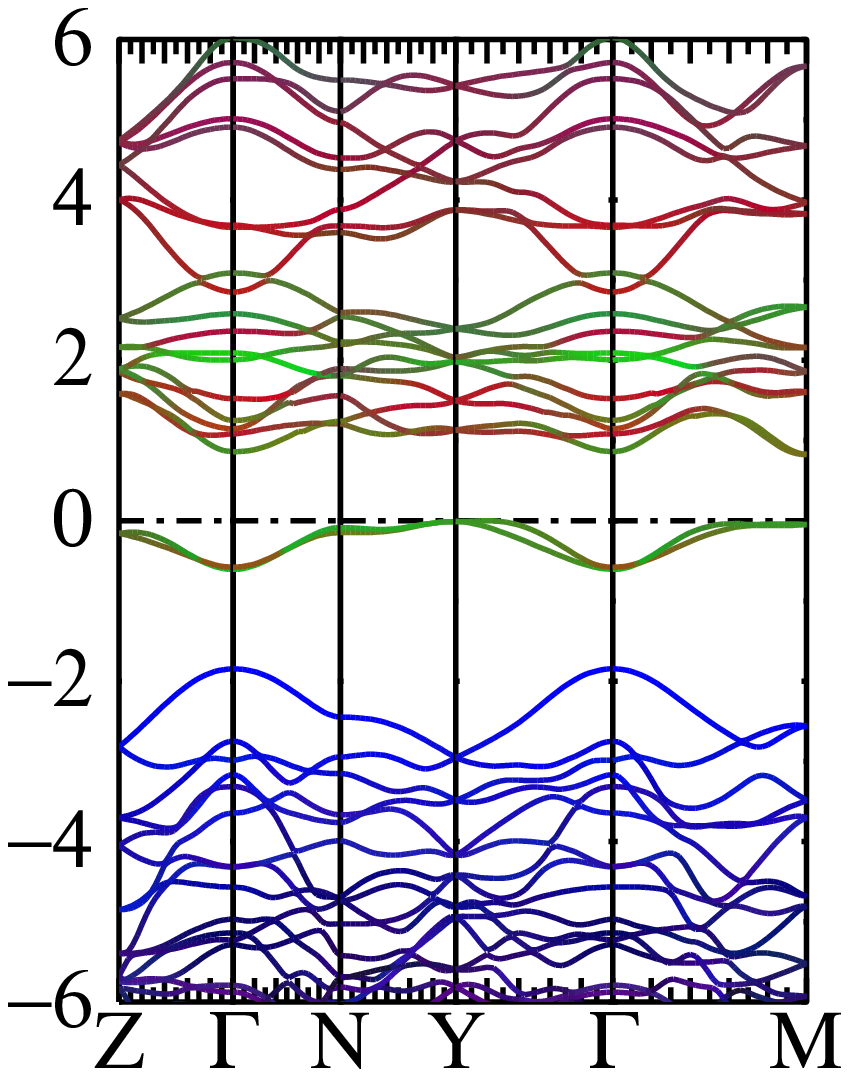}
\includegraphics[width=0.22\textwidth,clip=true,trim=0.0cm 0cm 0.0cm 0.0cm]{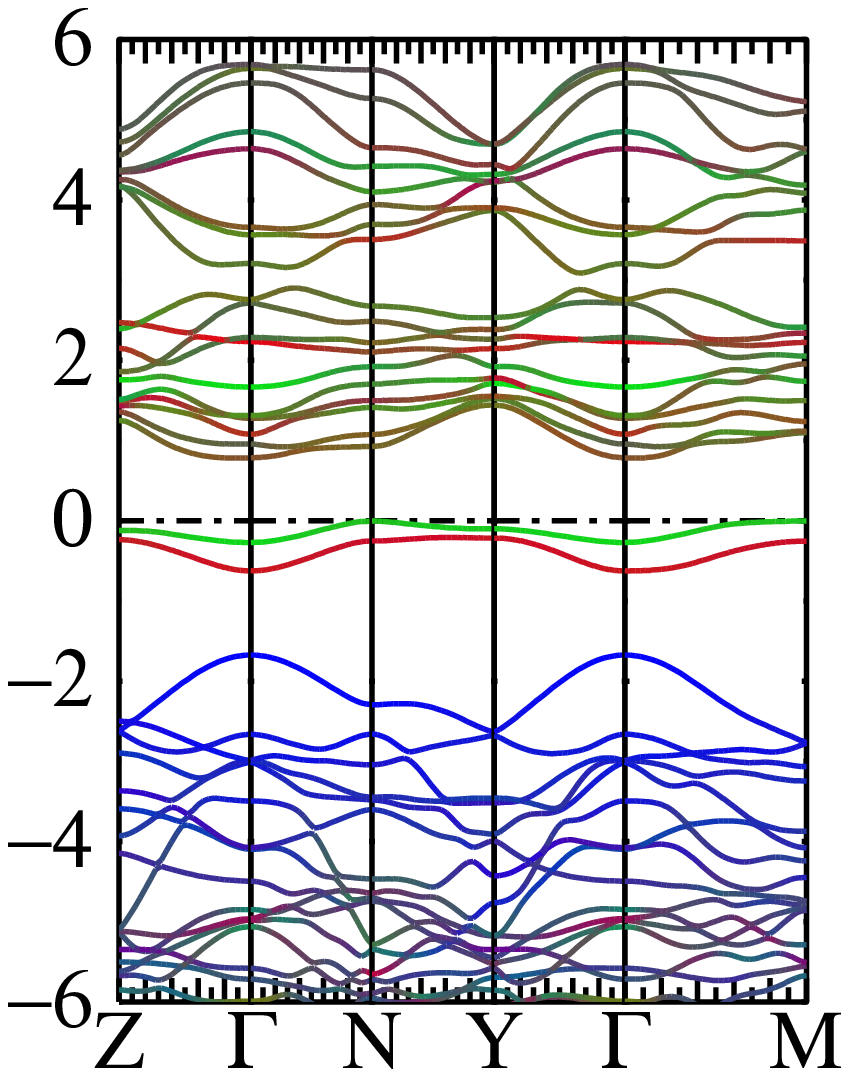}
\caption{Left: energy band structure of VO\textsubscript{2} in the {M\textsubscript{1}} phase within the
  \qsgw\ approximation.  Red, green, and blue project onto V \emph{d}($m$=$-$2,$-$1,0), V \emph{d}($m$=1,2) and O
  orbitals, respectively.  The \qsgwl\ bands (not shown) are essentially the same but with a 0.3\,eV smaller bandgap
  (see text).  Right: corresponding bands of the {M\textsubscript{2}} phase.  In this figure red and green project onto
  the dimerized and undimerized V \emph{d} orbitals, respectively.  }
\label{fig:vo2}
\end{figure}

In further support of this conjecture, magnetism appears to play no role in this phase, as we show next.  Here we
computed the electronic structure of {VO\textsubscript{2}} in the \qsgw\ and \qsgwl\ approximations.
VO\textsubscript{2} has a nearly dispersionless core-like valence band maximum (Fig.~\ref{fig:vo2}).  This makes it a
prime candidate for the gap to be underestimated, owing to the missing vertex (see discussion around
Table~\ref{tab:dvbmgaps}).  Indeed \qsgw\ predicts a rather good gap ($E_{G}$=0.76\,eV) owing to a fortuitous error
cancellation, while \qsgwl\ underestimates the gap (0.43\,eV), reminiscent of CuCl.

If Mott physics were involved, magnetism should play a role.  We find within \qsgw, that magnetism is totally suppressed
in the {M\textsubscript{1}} phase: attempts to find a magnetic solution always reverted to a nonmagnetic one with
self-consistency.  The situation is very different in the metastable {M\textsubscript{2}} phase, where half of the V
pairs dimerize and the other half do not.  Nonmagnetic \qsgw\ predicts a metallic phase.  An insulating phase forms,
however, if the system is allowed to be magnetic.  To determine the magnetic structure, each of the four V atoms was
assigned an arbitrary moment and the system driven self-consistent.  We find that the magnetic moment on the dimerized
pair vanishes, while spins on the undimerized pair becomes antiferromagnetically aligned with a local moment of
0.8$\mu_B$, which opens a gap of 0.7\,eV\footnote{In the {M\textsubscript{2}} phase the magnetism is likely disordered.
However a paramagnetic state can maintain essentially the same gap as the antiferromagnetic one, as it does for many
antiferromagnetic insulators such as NiO, CoO and La\textsubscript{2}CuO\textsubscript{4}.}.  The band structure looks
remarkably similar to the {M\textsubscript{1}} phase, even though the physical basis for the gap is very different.
Strikingly, one of the two states forming the upper valence band consists almost purely of dimerized V, while the other
is almost purely undimerized V.  There is very little hybridization between them, or between V and O.

That the physical basis for gap formation differ in the {M\textsubscript{1}} and {M\textsubscript{2}} phases was already
pointed out in a comment to the Wentzcovitch paper~\cite{Rice94}.  Their argument was based on NMR and EPR evidence for
low-lying spin excitations in the {M\textsubscript{2}} phase, which is consistent with the present work.

The picture from \qsgwl\ is similar to that of the DMFT calculation of Ref.~\cite{Brito16} for the {M\textsubscript{2}}
phase, but differs for the {M\textsubscript{1}} phase.  It finds a simple Peierls distortion accounts for the known
properties, and magnetism plays essentially no role for the latter.  Ref.~\cite{Brito16} argues that the
temperature-dependence of the bandgap is electronic in origin, and uses this as support for the Mott picture; however,
\qsgw\ calculations point to phonons playing an important role in controlling the bandgap at high temperature, with
strong support from experimental data~\cite{cedricvo2}.  Fig.\,3 of that work also presented the conductivity derived
from the BSE dielectric function, with \qsgw\ as a reference hamiltonian.  Agreement with ellipsometry
data~\cite{Okazaki06} is quite satisfactory.

\subsection{Antiferromagnetic insulating oxides}\label{ss:antiferro}

The monoxide crystal structures MnO, FeO, CoO and NiO are all of rocksalt form.  The magnetic structure consists of
sheets of spins antiferromagnetically ordered, which doubles the size of the unit cell.  According to the classic paper
by Roth~\cite{Roth58}, the alternating sheets lie in the (111) plane, but the spin orientation depends on the monoxide.
MnO and NiO are predicted to be band insulators even within the LDA.  In these cases the spin orientation scarcely
affects the electronic structure, and we assume the simpler [001] orientation.  CoO and FeO are different: LDA predicts
both to be metallic.  Spins point in the $[\bar{1}\bar{1}7]$ direction in CoO, and perpendicular to the (111) plane in
FeO~\cite{Roth58}.  For these sytems we orient the spin quantization axis along these directions and also do not assume
time-reversal symmetry.

All of the rocksalt structure oxides have sizeable magnetic moments.  By contrast, CuO is monoclinic with 8 formula
units in the unit cell (\S\ref{ss:cuo}) and a small local moment (Table~\ref{tab:afminsulators}).

R{\"o}dl and Bechstedt modeled the QP band structure of the rocksalt oxides with \emph{GW} starting from a GGA+U
functional~\cite{Rodl09}, and later these authors used the BSE framework to examine the optical response, using a
reference potential generated by a $G^\mathrm{HSE03}W^\mathrm{HSE03}$ functional~\cite{Rodl12}.

In each of these systems, (and probably Fe\textsubscript{3}O\textsubscript{4} \S\ref{ss:fe3o4}), \qsgw\ significantly
overestimates the bandgap, but not the local moment (Table~\ref{tab:afminsulators}), the difference being more
pronounced than in nonmagnetic counterparts.  \qsgwl\ greatly amelioriates this overestimate, sometimes slightly
overcorrecting \qsgw\ because of the missing vertex.  Precise benchmarking is difficult owing to the large uncertainty in
experimental data, especially in the strongly correlated cases.  One measure of correlation is the $Z$ factor,
Eq.~\ref{eq:Zren}.  Table~\ref{tab:afminsulators} presents a band- and $k$- averaged $Z$ factor, namely the ratio of the
interacting to non-interacting spectral functions $A(\omega)/A^0(\omega)$ at an energy just below the Fermi level.  The
degree of correlation differs in each case so each system is dealt with individually.

\begin{table}[h]
\caption{Bandgap; $\varepsilon_\infty$; local magnetic moment; band- and $k$-averaged $Z$ factor in selected antiferromagnetic insulators.}
%\begin{tabular}{|@{\hspace{0.6em}}c@{\hspace{0.6em}}|@{\hspace{0.6em}}c@{\hspace{0.6em}}|@{\hspace{0.6em}}c@{\hspace{0.6em}}|@{\hspace{0.6em}}c@{\hspace{0.6em}}|@{\hspace{0.6em}}c@{\hspace{0.6em}}|}
% NiO : basp.GWversion16.tppc2.tpdc4
\begin{tabular}{|c|r|c|c|c|c|c|c|c|}\hline
                              &      & MnO   & NiO    & CoO    & FeO      & CuO  & LSCO  \\ \hline\hline
                              & gap  & 0.71  & 0.55   &  -     &  -       &  -   &  0.01 \\
LDA                           &\epsi & 8.81  & 32.0   &  -     &  -       &  -   &  -  \\
                              & $m$  & 4.48  & 1.21   &  -     &  -       &  -   & 0.27  \\ \hline
\qsgw                         & gap  & 3.77  & 5.03   & 4.00   & 1.9      & 2.80 & 3.09 \\
                              &\epsi & 3.72  & 4.27   & 3.87   & 4.08     & 4.86 & 4.04 \\
                              & $m$  & 4.76  & 1.71   & 2.73   & 3.65     & 0.71 & 0.64 \\ \hline
\qsgwl                        & gap  & 3.05  & 3.23   & 3.28   & 0.67     & 1.52 & 1.66 \\
                              &\epsi & 4.76  & 6.15   & 5.15   & 17.6     & 7.85 & 5.5\footnote{average of $\epsilon_{xx}$ (6.5) and $\epsilon_{zz}$ (4.5)} \\
                              & $m$  & 4.73  & 1.67   & 2.70   & 3.66     & 0.66 & 0.53 \\
                              & $Z$  & $\sim$0.75
                                             & $\sim$0.7
                                                      &$\sim$0.75
                                                               &$\sim$0.4 & $\sim$0.5\footnote{Strongly orbital dependent:
                                                                          $Z{\sim}0.65$ for the (mostly O) valence bands,  and ${\sim}$0.45 for the (mostly Cu) conduction bands}
                                                                                 & $\sim$0.5\footnote{Strongly state-dependent:
                                                                          $Z{\sim}0.65$ for the highest valence bands, and ${\sim}$0.4 for lowest conduction band} \\
\hline
Expt                          & gap  & 3-3.9 & 4.0,4.3\footnote{Refs.~\cite{Hufner84,Sawatzky84,Zimmermann99}}
                                                      & $\sim$2.6\footnote{Optical absorption edge, Ref.~\cite{SpicerNiOCoO}; BIS Refs~\cite{Elp91}.  See \S\ref{ss:coo}}%
                                                               & $<$1\footnote{Ref.~\cite{Zimmermann99}}%
                                                                          & 1.3,1.4\footnote{Optical absorption edge~\cite{Marabelli87}; PES/BIS and XPS \cite{Ghijsen84,Massolo88}}%
                                                                               & $\sim$2\footnote{Reflectivity, Ref.~\cite{Falck92}, optical conductivity, Ref.~\cite{Baldini20}} \\
                              &\epsi & 4.95\footnote{Reststrahlen spectrum, Ref.~\cite{Plendl69}.  See \S\ref{ss:mno}}
                                             & 5.43-6.0\footnote{Refs~\cite{SpicerNiOCoO,Pecharroman94,Chern92}}%
                                                      & $\sim$5\footnote{Refs~\cite{SpicerNiOCoO,Gielisse65,Rao65}.  See \S\ref{ss:coo}}
                                                               & 9.24-11.1\footnote{Refs.~\cite{Kugel77,Prevot77}}%
                                                                          & 6.5\footnote{Refs.~\cite{Ito98,Tahir12}}
                                                                               & $\sim$5\footnote{reported in Ref.~\cite{Falck92}.  Refs.~\cite{Reagor89,Chen91}
                                                                               report anomalously large index of
                                                                               refraction (so that $\epsilon_\infty{\sim}$25-50),
                                                                                 which is likely connected to excess
                                                                                 holes in nominal {La\textsubscript{2}CuO\textsubscript{4}}.} \cr
                              & $m$  & 4.79  & 1.64,1.77\footnote{values cited in Ref.~\cite{Anisimov90}, taken from Ref.~\cite{Alperin62} and Ref.~\cite{Fender68}}%
                                                      & 2.47\footnote{spin moment from Ref.~\cite{Csiszar05}. Orbital moment estimated to be $\sim$1$\mu_B$.}%
                                                               & 3.32\footnote{Ref.~\cite{Roth58}}%
                                                                          & 0.68\footnote{Refs.~\cite{Yang88,Graham91}}%
                                                                                & 0.64\footnote{A consensus value of $0.64\mu_B{\pm}10\%$ from several sources, Ref.~\cite{Kaplan91}}% \\ \hline
\end{tabular}
\label{tab:afminsulators}
\end{table}

\subsubsection{NiO}\label{ss:nio}

Fig.~\ref{fig:niobands} shows the (noninteracting) energy bands of NiO, and compares the DOS to BIS data.  On the scale
of the figure, agreement is excellent.  However, \qsgwl\ apparently slightly underestimates the bandgap, which is
apparent in both the BIS data and the optics data of Fig.~\ref{fig:niobands}.  Also, the BSE value for
$\varepsilon_\infty$, at 6.15, is outside the range of reported values (Table~\ref{tab:afminsulators}).  Replacing
$\Sigma$(\qsgwl) with a hybrid $0.9\,\Sigma$(\qsgwl)\,+\,$0.1\,\Sigma$(\qsgw) (Eq.~\ref{eq:hybrid}), the gap increases
by 0.17\,eV, and $\varepsilon_\infty$ decreases to 5.97.  This is perhaps the best characterized correlated materials
system, though even for NiO there is some spread in reported values for both the dielectric function and the fundamental
gap.  We can conclude that to within this experimental uncertainty, the close connection between gap and
$\varepsilon_\infty$ (\S\ref{ss:otheroutliers}) is affirmed.

\begin{figure}[h!]
\includegraphics[width=0.20\textwidth,clip=true,trim=0.0cm 0.0cm 0.0cm 0.0cm]{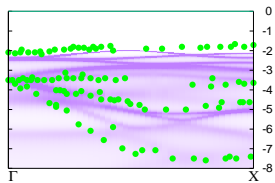}
\caption{
  Spectral function NiO along the $\Gamma$-X line, compared against ARPES measurements, Fig. 6 of
  Ref.~\cite{PhysRevB.44.3604} (green circles).
}
\label{fig:nioarpes}
\end{figure}

Some ARPES experiments on this correlated antiferromagnet have been published~\cite{PhysRevB.44.3604}.  Accordingly we
generated the fully dynamical self-energy to compute the $k$ resolved spectral function and compare to it
(Fig.~\ref{fig:nioarpes}).  The extent to which a particular band is broadened is strongly band-dependent.  Agreement
with ARPES is satisfactory, in light of the fact that that the matrix element and final state effects would have to be
included for a direct comparison.  Quite remarkably, the \qsgwl\ spectral function looks nearly identical to one
generated by LDA+DMFT, Ref.~\cite{Mandal19}.  The close similarity between these two completely different approaches
lends support to the thesis that both are characterizing the actual spectral function of NiO.

\subsubsection{CoO}\label{ss:coo}

Our \qsgwl\ gap is 3.28\,eV, and larger than a gap of 2.5\,eV measured by a combination of XPS and BIS~\cite{Elp91}.
Optical gap of similar size ($\sim$2.6\,eV) has been observed~\cite{SpicerNiOCoO}.  These two experimental findings are
consistent only if there are no excitonic effects to reduce the gap.  Our \qsgwl\ calculations show, however, that there
are a multiplicity of excitons throughout the gap at $q{=}0$.  The deeper ones (ranging between 0.4 and 1.7\,eV) are
dark, but strongly active ones at 2.8, 2.96, 3.0, and 3.1\,eV also appear.  These are likely broadened somewhat,
e.g. via some phonon-mediated transitions linking different $q$, which our calculation does not take into account.
%Close inspection of Ref.~\cite{SpicerNiOCoO}, Fig.\,11 indicates a kink in $\mathrm{Im}\,\epsilon(\omega)$ near
%$\omega{=}3$\,eV, suggesting some other process is occurring in the (2.6,3)\,eV interval.
Finally, the \qsgwl\ dielectric constant (5.15), aligns well with the mean value of various experiments
(5.43~\cite{Rao65}, 5.29~\cite{Gielisse65}, 4.75~\cite{SpicerNiOCoO}).  If the consistency between gap and
$\varepsilon_\infty$ argued in \S\ref{ss:consistency} can be relied on, it provides another indication that the
\qsgwl\ fundamental gap is close to the true one.

%Dark excitons were observed at 0.4, 0.6, 1.3, 1.4, and 1.7\,eV; these may become weakly optically active if some
%symmetry-lowering perturbation were included, e.g. spin-orbit coupling.  Ref.~\cite{Elp91} also attempts to analyze
%symmetry-forbidden optical transitions and interpret them in terms of a model cluster hamiltonian.  There is not a
%one-to-one correspondence with their work, but the excitons range between 0.7 and 2.5\,eV.  We made no attempt to
%analyze the symmetry of the excitons emerging from the BSE calculation.

% 2.33^2=5.43 K. V. Bao and A. Smakula, J. Appl. Phys. 36, 2031 (1965).
% 2.30^2 = 5.29 P. J. Gielisse et al. , J. Appl. Phys. 36, 2446 (1965)
% 2.18^2 = 4.75 SpicerNiOCoO

\begin{figure}[h!]
\includegraphics[width=0.20\textwidth,clip=true,trim=0.0cm 0.0cm 0.0cm 0.0cm]{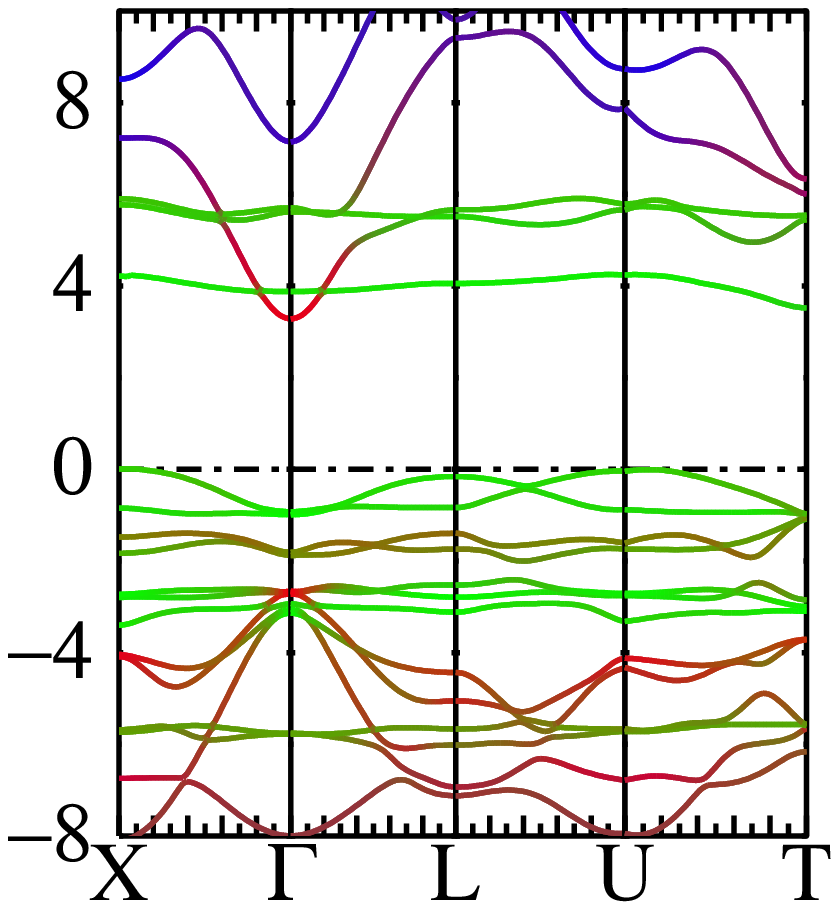}
\raisebox{-2mm}[0pt][0pt]{
\includegraphics[width=0.22\textwidth,clip=true,trim=0.0cm 0.0cm 0.0cm 0.0cm]{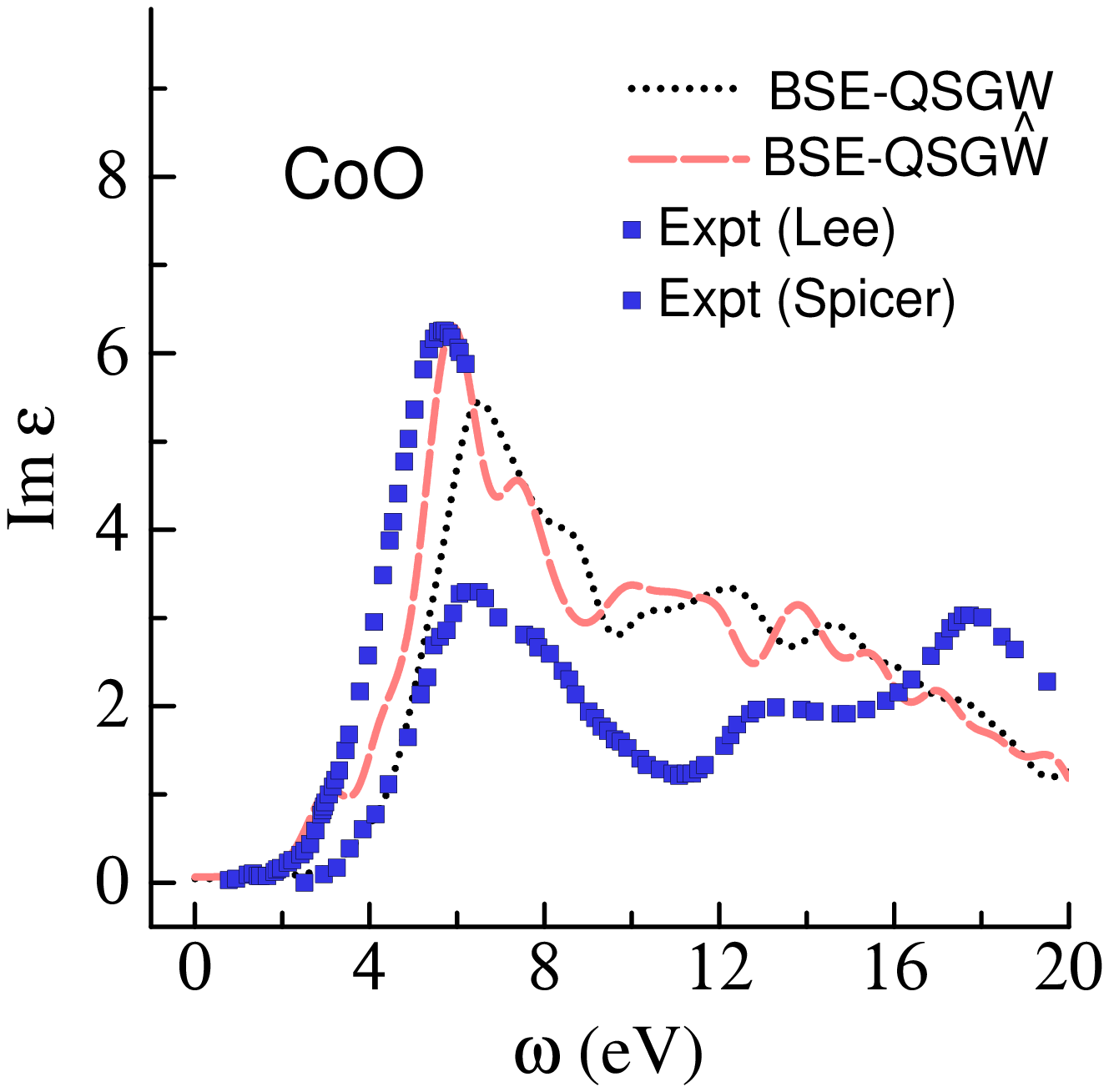}
}
\caption{Left: energy band structure of CoO.  Green and blue represents Co-centered orbitals (green: \emph{d}, blue: \emph{spf}),
  and red O-centered orbitals.  Right $\mathrm{Im}\,\epsilon(\omega)$ measured by Messick et al.~\cite{Messick72} compared
  to BSE@\qsgwl.}
\label{fig:coo}
\end{figure}

\subsubsection{MnO}\label{ss:mno}

J. van Elp et al. measured the fundamental gap of MnO by x-ray photoelectron and BIS spectroscopies, and obtained a gap
of 3.9\,eV~\cite{vanElp91}.  Three kinds of subgap transitions have been recorded by several groups, labelled as A, B, C
transitions, and identified with the following symmetries
$\mathrm{^{6}A}_{1g}{\rightarrow}\mathrm{^4T}_{1g}$ (A band),
$\mathrm{^{6}A}_{1g}{\rightarrow}\mathrm{^4T}_{2g}$ (B band),
$\mathrm{^{6}A}_{1g}{\rightarrow}\mathrm{^4A}_{1g}{+}\mathrm{^4E}_{1g}$ (C band) \cite{Pratt59,Treindl77}.
These transitions are forbidden due to spin and parity selection rules, though
significant oscillator strengths have been observed.
Huffman et. al reported two additional peaks~\cite{Huffman69}, the highest at $\sim$3.5\,eV.

\begin{figure}[h!]
\includegraphics[width=0.20\textwidth,clip=true,trim=0.0cm 0.0cm 0.0cm 0.0cm]{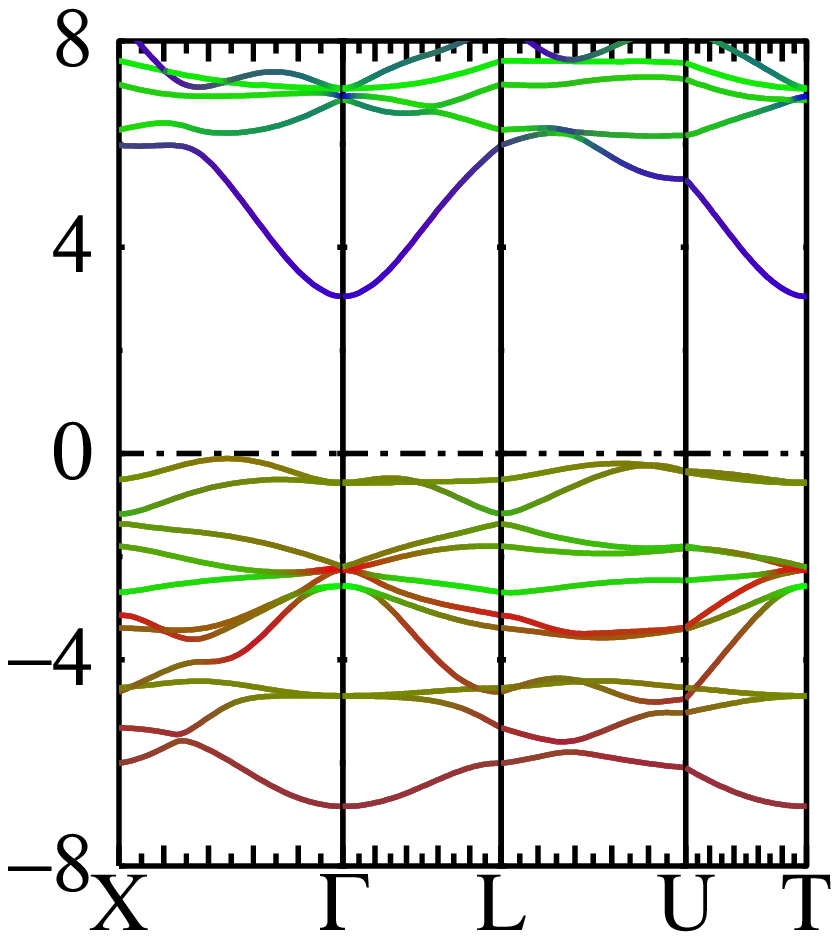}
\includegraphics[width=0.22\textwidth,clip=true,trim=0.0cm 0.0cm 0.0cm 0.0cm]{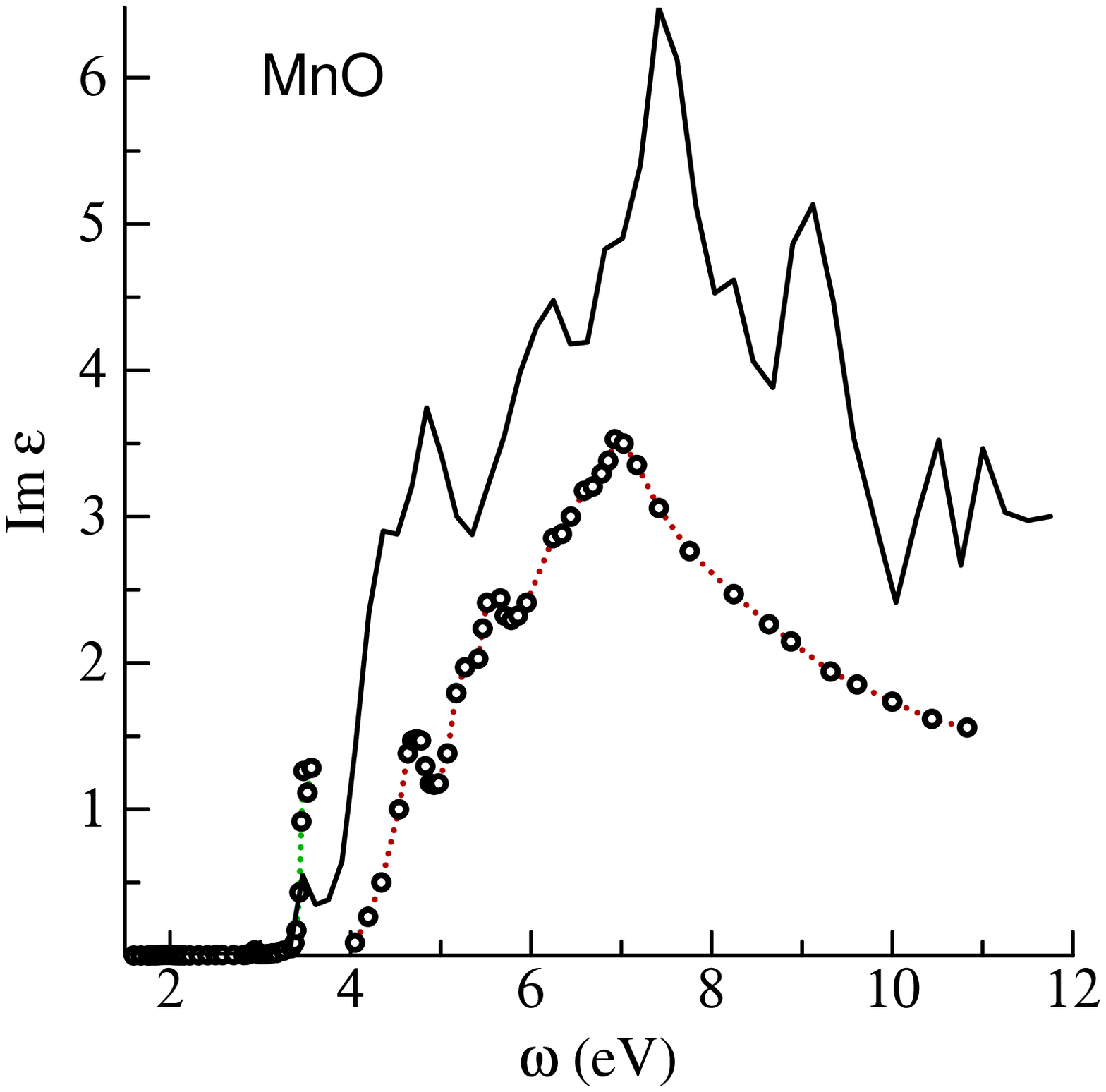}
\caption{Left: energy band structure of MnO.  Green and blue represents Mn-centered orbitals (green: \emph{d}, blue:
  \emph{spf}), and red O-centered orbitals.  Right $\mathrm{Im}\,\epsilon(\omega)$ measured by Messick et
  al.~\cite{Messick72} (circles connected by red dots) compared to BSE@\qsgwl.  Also shown at low energy (circles
  connected by green dots) is a scaled absorption, $\alpha^2/\omega^2$, taken from Ref.~\cite{Huffman69}.
}
\label{fig:mno}
\end{figure}

The band structure depicted in the left panel of Fig.~\ref{fig:mno}, is roughly similar to the one depicted by R\"odl
et al.~\cite{Rodl12}.  Unlike NiO, the conduction band is essentially pure Mn \emph{s} character, the Mn $d(t_{2g})$
appearing at 6-8\,eV.  The direct gap is at $\Gamma$, and is calculated to be 3.6\,eV, slightly smaller than the XPS/BIS
value (3.9\,eV) reported in Ref.~\cite{vanElp91}.

The BSE value for $\varepsilon^\mathrm{BSE}_\infty$ is slightly smaller than observed in a Reststrahlen experiment,
Ref.~\cite{Plendl69} (Table~\ref{tab:afminsulators}).  This suggests an inconsistency with the gap being underestimated;
however, the maximum value of $\mathrm{Im}\,\epsilon$ derived from $n$ and $k$ presented in that experiment is about an
order of magnitude too large, so it is not clear how reliable the measurement is.

We find a dark exciton at 3.07\,eV and several bright ones at $\sim$3.5\,eV, which can be seen from the shoulder in
$\mathrm{Im}\,\epsilon(\omega)$ below the fundamental gap.  These possibly correspond to the highest peaks observed by
Huffman et al..~\cite{Huffman69}.  We do not find the weak A and B excitons~\cite{Pratt59,Treindl77}; possibly these are
associated with a phonon-assisted transition and an electronic part at finite $q$.  The main shoulder in
$\mathrm{Im}\,\epsilon(\omega)$ starts to rise about 0.3\,eV earlier than the reflectance data of Ref.~\cite{Messick72}
(Fig.~\ref{fig:mno}).  This is consistent with the discrepancy in the XPS/BIS measurement of Ref.~\cite{vanElp91}.  Thus
we tentatively conclude that the \qsgwl\ gap is $\sim$0.3\,eV too low, though reliable experimental evidence is too
limited to draw strong conclusions.

% mno/M_S_Seehra_1983_J._Phys._C__Solid_State_Phys._16_L411.pdf shows how A,B,C bands shift as T approaches Tn

\subsubsection{FeO}\label{ss:feo}

\begin{figure}[h!]
\includegraphics[width=0.20\textwidth,clip=true,trim=0.0cm 0.0cm 0.0cm 0.0cm]{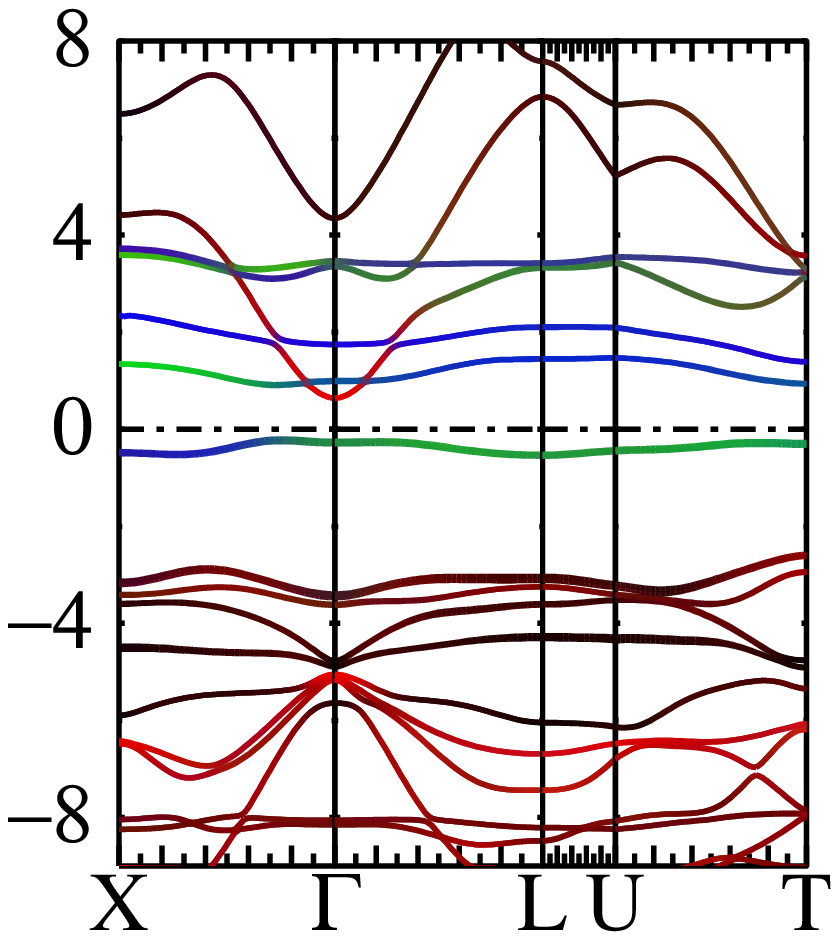}
\includegraphics[width=0.22\textwidth,clip=true,trim=0.0cm 0.0cm 0.0cm 0.0cm]{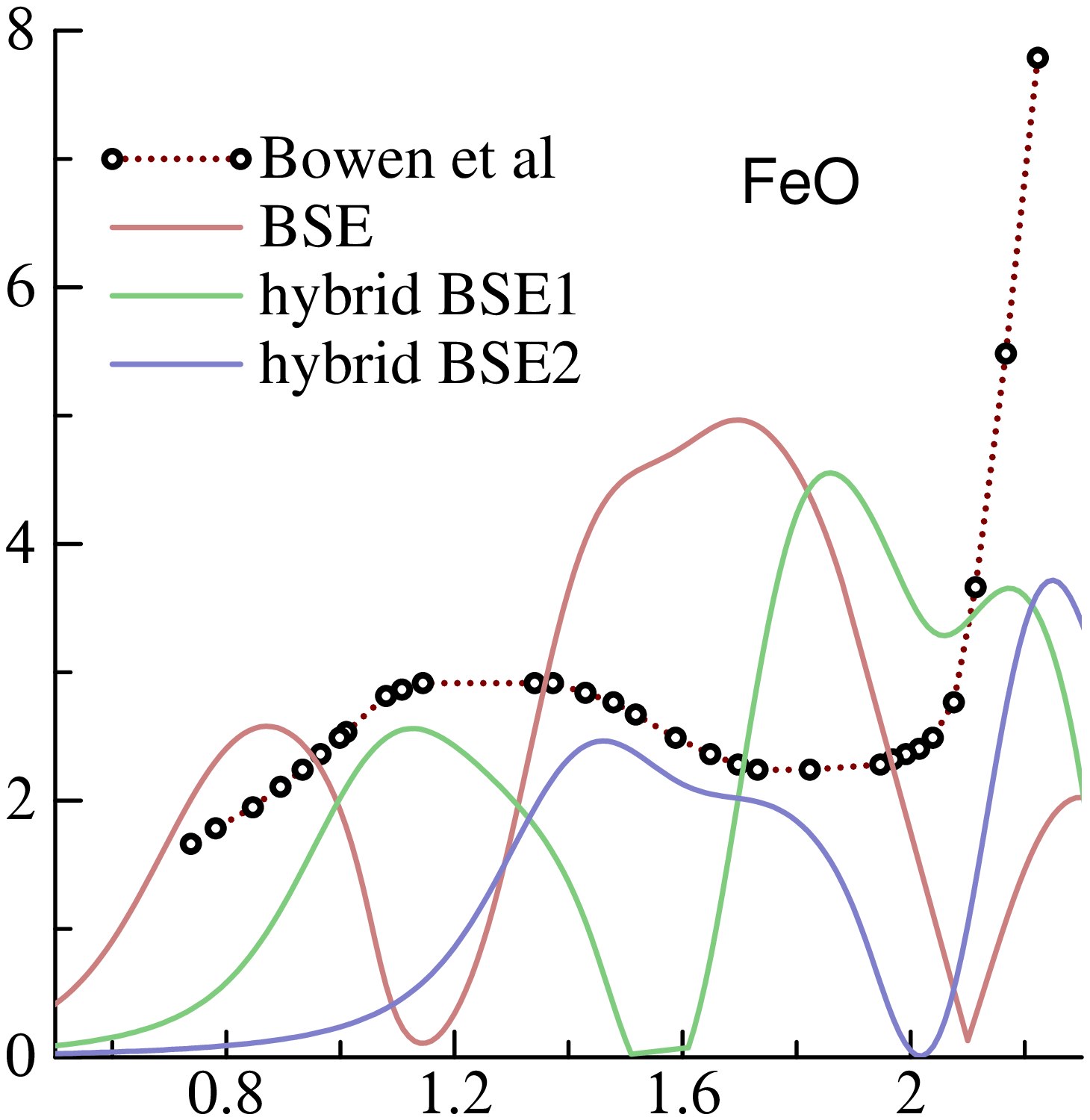}
\caption{Left: energy band structure FeO.  Green and blue represents Fe-centered \emph{d} orbitals (green: $m{=}{-}2,{-}1$; blue: $m{=}0,1,2$),
  and red O-centered orbitals.  Right: absorption $\alpha(\omega)$ measured by Bowen et al.~\cite{Bowen75} compared to
  BSE@\qsgwl\ (red) with absorption computed from Eq.~\ref{eq:defalfa}.  Also shown are BSE generated from a hybrid
  of \qsgwl\ and \qsgw\ $\Sigma$, Eq.~\ref{eq:hybrid}, with $\beta{=}0.3$ (BSE1) and $\beta{=}0.6$ (BSE2);
  see Tab.~\ref{tab:feohybrid}.
}
\label{fig:feo}
\end{figure}

FeO poses one of the most challenging benchmarks in this study.  Its highest valence state consists of a single, almost
dispersionless \emph{d} orbital whose $m$ character changes with wave number (Fig.~\ref{fig:feo}).  The small $Z$ factor
(Tab.~\ref{tab:afminsulators}) provides a clear indication that FeO is strongly correlated.

Experimental information about FeO is sparse and somewhat inconsistent.  Two values for \epsi\ have been reported:
9.24~\cite{Kugel77} and 11.9~\cite{Prevot77}.  The former may be more reliable, since the latter experiment was
performed on Fe$_{x}$O, with $x$ deviating several percent from unity.  Bowen et al. investigated the
infrared absorption~\cite{Bowen75}, which we use here to benchmark against the BSE.

Regarding the fundamental gap, there is a general expectation in both experimental and theoretical literature that it is
of order 2.5\,eV~\cite{Anisimov90,Rodl09,Wei94,Mandal19}.  R\"odl et al. associated the sharp rise in $\alpha(\omega)$
observed around 2.4\,eV in Ref.~\cite{Bowen75}, with the fundamental gap. Hiraoka et al.~\cite{Hiraoka09} also assumed
the fundamental gap was of this order but observed a peak in $\mathrm{Im}\,\epsilon$ at $\sim$1\,eV and tentatively
assigned it to a defect band.  Absorption data shows peaks at both 1.2\,eV and 2.4\,eV~\cite{Bowen75}; see
Fig.~\ref{fig:feo}.

Turning to theory, at the \qsgw\ level the fundamental gap is found to be 1.9\,eV, with the smallest direct gap 2.4\,eV.
In all the other antiferromagnetic oxides of this study, \qsgw\ overestimates the gap by $\sim$1\,eV
(Tab.~\ref{tab:afminsulators}), so these gaps are likely too high.  At the \qsgwl\ level, the fundamental gap is much
smaller, 0.64\,eV.  This leads to a puzzle: why does \qsgwl\ yield such a gap so different from the accepted values in
the literature?

Counterbalancing the experiments just mentioned, Zimmermann observed the one-particle spectral function
XPS/BIS~\cite{Zimmermann99}.  He did not attempt to extract a bandgap, but based on his Fig. 15, it would be of order
1\,eV.  The XPS/BIS data and the optical measurements both point to the lowest excitation of order 1\,eV, even though no
deep excitons were found by \qsgwl\ to explain the absorption peak there.  Thus our \qsgwl\ analysis suggests a different
interpretation, namely that the observed peak in the absorption $\alpha(\omega)$ around 1.2\,eV (Fig.~\ref{fig:feo})
corresponds to the true fundamental gap.

The \qsgwl\ prediction for $E_{G}$ is likely too small: FeO's practically dispersionless valence band strongly resembles that
of VO\textsubscript{2} (\S\ref{ss:vo2}), and we can expect the gap to be similarly underestimated in \qsgwl.  Using
materials in Table~\ref{tab:dvbmgaps} as a guide, the gap can be expected to be underestimated by $\sim$0.5\,eV.  Two
other pieces of evidence point to the gap being underestimated: $\varepsilon^\mathrm{BSE}_\infty$ is much larger than
the two experiments noted earlier (Table~\ref{tab:feohybrid}) and the peak in $\alpha(\omega)$ falls at $\sim$0.8\,eV,
well below the 1.2\,eV peak reported in Ref.~\cite{Bowen75} (Fig.~\ref{fig:feo}).

To adjust for the probable \qsgwl\ gap underestimate, we consider a hybrid of \qsgwl\ and \qsgw, Eq.~(\ref{eq:hybrid}),
and benchmark both \epsi\ and the two peaks in absorption, for different admixtures $\beta$ of \qsgw\ into
\qsgwl\ (Fig.~\ref{fig:feo}).  Table~\ref{tab:feohybrid} shows the variation of $E_{G}$ and \epsi\ with $\beta$.
Perfect alignment with the best available experimental value for \epsi\ corresponds to $E_{G}{=}1.05$\,eV.

\begin{table}[h]
\caption{Optical properties in FeO as a function of hybridization parameter $\beta$, Eq.~(\ref{eq:hybrid}).
$E_{G}$ is the fundamental gap in eV; $E_{G}$ ($\Gamma{\rightarrow}\Gamma$) is the direct gap at $\Gamma$.
Labels in the second column are used in Fig.~\ref{fig:feo}.
}
\begin{tabular}{|@{\hspace{1.0em}}cc@{\hspace{1.0em}}|@{\hspace{1.0em}}c@{\hspace{1.0em}}|c|@{\hspace{1.0em}}c@{\hspace{1.0em}}|}\hline
\vbox{\vskip 10pt}
$\beta$ &  label  & $E_{G}$ & $E_{G}$ ($\Gamma{\rightarrow}\Gamma$)
                           & \epsi \cr \hline
 0      & BSE     & 0.64   &  0.94  & 17.6  \cr
 0.3    & BSE1    & 0.90   &  1.28  & 11.2  \cr
 0.6    & BSE2    & 1.19   &  1.63  & 8.04  \cr \hline
\end{tabular}
\label{tab:feohybrid}
\end{table}

%Taken together, it seems likely that the observed peak in $\alpha$ around 1.2\,eV (Fig.~\ref{fig:feo}) corresponds to
%the true fundamental gap.  To reconcile the probable \qsgwl\ gap underestimate, we use a hybrid of \qsgwl\ and \qsgw\,
%Eq.~(\ref{eq:hybrid}), and benchmark both \epsi\ and the two peaks in absorption for different admixtures $\beta$ of
%\qsgw\ into \qsgwl\ (Fig.~\ref{fig:feo}).  Table~\ref{tab:feohybrid} shows how $E_{G}$ and \epsi\ vary with $\beta$.

Thus if FeO has a fundamental gap 1.05-1.10\,eV, a consistent picture emerges.  First the two peaks in $\alpha(\omega)$
for BSE2 and BSE1 (Fig.~\ref{fig:feo} and Table~\ref{tab:feohybrid})) bracket the two experimental peaks from above and
below.  Second, the one-particle DOS is consistent with XPS/BIS~\cite{Zimmermann99}. Finally, \epsi\ is consistent with
the best available experimental data.

\subsubsection{CuO}\label{ss:cuo}

CuO has a monoclinic lattice structure of 4 formula units~\cite{Asbrink70}, while the magnetic structure is
antiferromagnetic, and is a $\sqrt{2}{\times}1{\times}\sqrt{2}$ supercell of lattice with 8 formula
units~\cite{Yang88,Filippetti05}.  The nominal configuration Cu$^{2+}$O$^{2-}$ would imply a single unpaired \emph{d}
electron; however the magnetic moment is substantially smaller than 1$\mu_B$/atom (Table~\ref{tab:afminsulators}).

\begin{figure}[h!]
\hspace{0.0cm}
\includegraphics[width=0.22\textwidth,clip=true,trim=0.0cm 0cm 0.0cm 0.0cm]{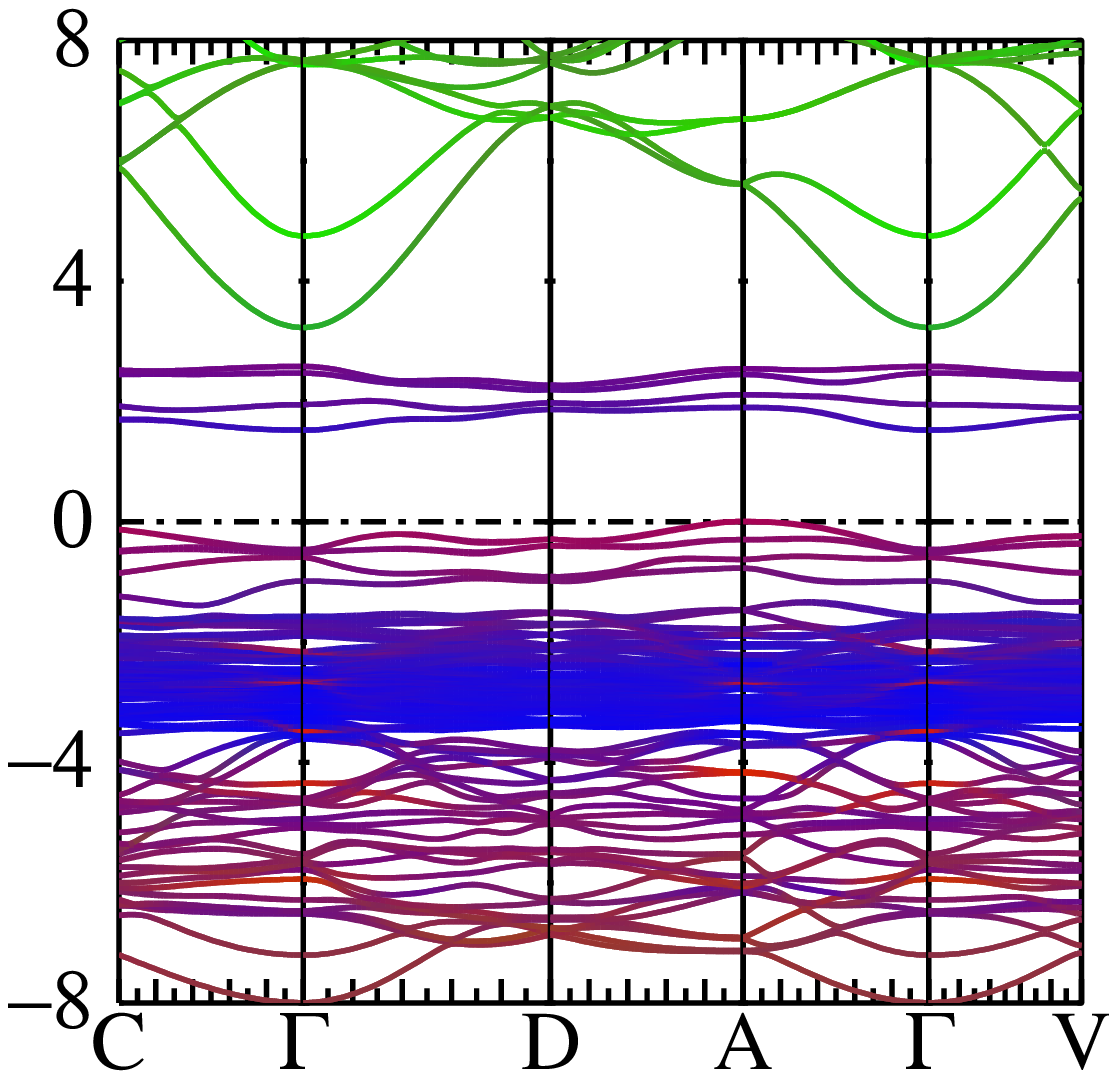}
\includegraphics[width=0.205\textwidth,clip=true,trim=0.0cm 0cm 0.0cm 0.0cm]{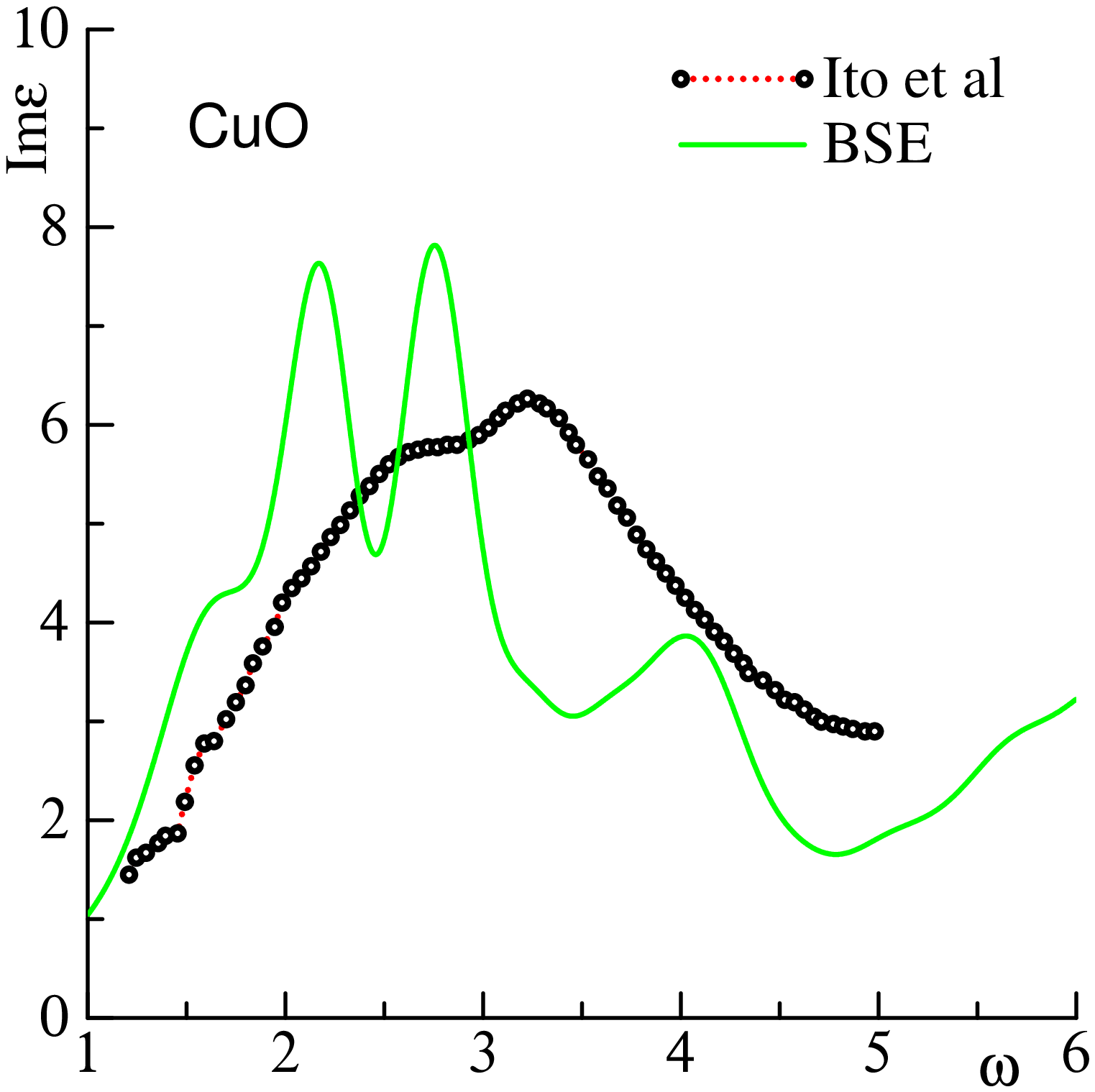}
\caption{Left: energy band structure of CuO.  Colors depict the following orbital characters:
  blue, Cu-\emph{d}; green, Cu-\emph{sp}, red, O-\emph{sp}.  Right:
  $\mathrm{Im}\,\epsilon(\omega)$ measured by Ito et al.~\cite{Ito98} compared to BSE@\qsgwl.
  Discrepancies with experiment are discussed in the text.
}
\label{fig:cuo}
\end{figure}

The \qsgwl\ energy band structure is depicted in the left panel of Fig.~\ref{fig:cuo}.  The valence band consists of approximately
2/3 O \emph{p} character, and 1/3 Cu \emph{d} character, and the conduction bands 1/3 O \emph{p} character, and 2/3 Cu
\emph{d} character.  Orbital weighting is quite different from cuprates such as La\textsubscript{2}CuO\textsubscript{4},
where both band edges are dominated by Cu.  This finding is roughly in line with the DFT calculation of Filippetti and
Fiorentini~\cite{Filippetti05}, who assigned the highest valence to O \emph{p} (the LDA puts O \emph{p} too high, so the
O character will be overestimated).

\begin{figure}[h!]
\hspace{0.0cm}
\includegraphics[width=0.17\textwidth,clip=true,trim=0.0cm 0cm 0.0cm 0.0cm]{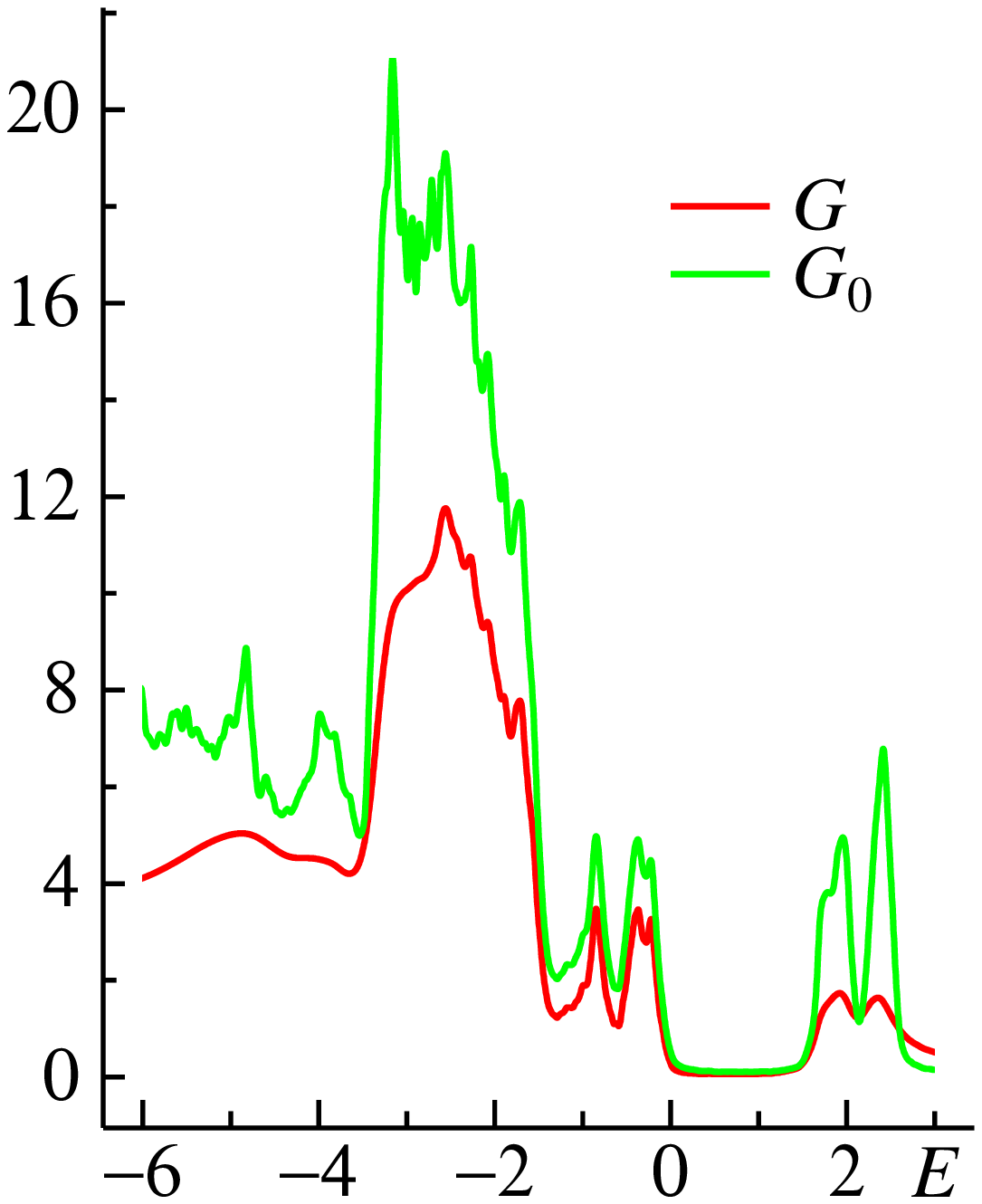}\hspace{0.4cm}
\includegraphics[width=0.25\textwidth,clip=true,trim=0.0cm 0cm 0.0cm 0.0cm]{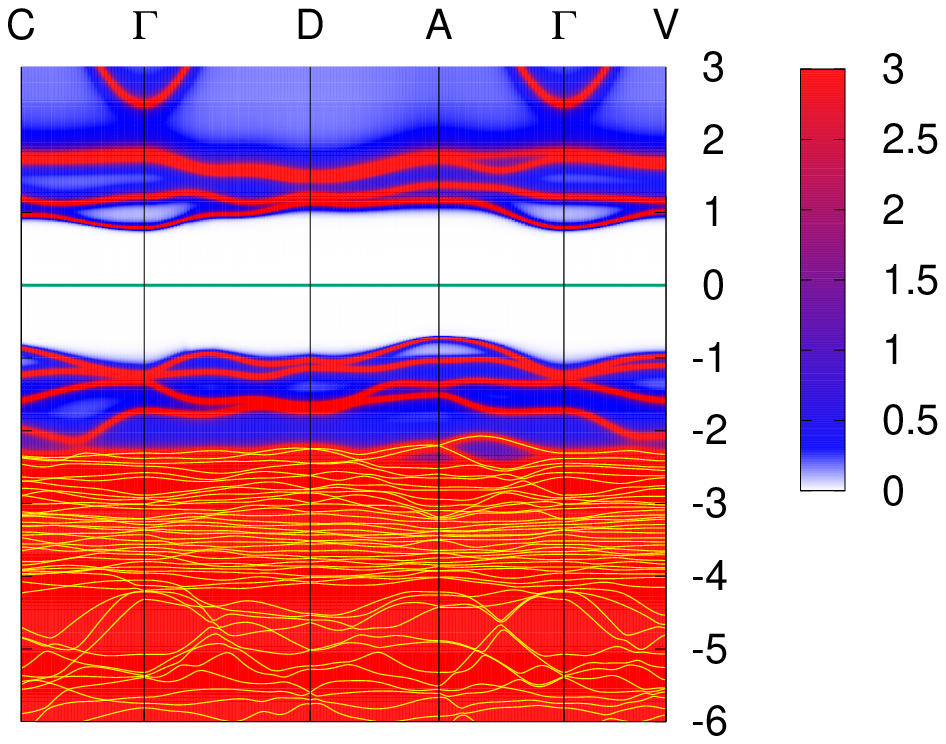}
\caption{(Left) Density-of-states generated from $G$ compared to that generated from the \qsgwl\ $G_0$.  Zero energy
  corresponds to the valence band maximum. (Right) spectral function from interacting $G$ generated by \qsgwl.
  Yellow lines below $-2$\,eV trace out the \qsgwl\ bands, and are equivalent to those in the left panel of
  Fig.~\ref{fig:cuo}.  }
\label{fig:cuodos}
\end{figure}

A band gap has been measured optically from the absorption edge~\cite{Marabelli87}, and also by PES/BIS~\cite{Ghijsen84}
and XPS~\cite{Massolo88}.  All three measurements report bandgaps in 1.3-1.4\,eV range, slightly smaller than the
1.52\,eV fundamental gap from \qsgwl\ (Table~\ref{tab:afminsulators}).  However, \qsgwl\ overestimates
\epsi\ (Table~\ref{tab:afminsulators}) by about 20\%, which if the consistency between the gap and $\varepsilon_\infty$
can be relied on (\S\ref{ss:consistency}), the \qsgwl\ fundamental gap must be too small.  Moreover, if we compare
$\epsilon(\omega)$ against ellipsometry measurements of Ito et al.~\cite{Ito98}, the peaks of
$\mathrm{Im}\,\epsilon(\omega)$ are seen to fall $\sim$0.3\,eV below the experimental data.  From this we conclude it is
likely that the fundamental gap is closer to 1.6\,eV, assuming the dielectric data of Ito et al.~\cite{Ito98} is
reliable.  This would mean the PES/BIS is underestimated.  It is perhaps not unexpected since the BIS should be
larger than the optical gap.

That experimentally $\mathrm{Im}\,\epsilon(\omega)$ is smoother than the \qsgwl\ one, can be attributed (at least in
part) to $\epsilon(\omega)$ being generated from a noninteracting $G_{0}$ (\qsgwl).  The frequency-dependence of
$\Sigma$ reduces quasiparticle weights (compare the DOS of the interacting $G$ to that of $G_{0}$, left panel of
Fig.~\ref{fig:cuodos}), and the imaginary part smears out the quasiparticle (right panel of Fig.~\ref{fig:cuodos}).  CuO is
very strongly correlated: note the sharp reduction in the DOS around 2\,eV.  These dynamical effects do not shift the
average position of the bands (owing to the \qsgw\ construction) but will smooth out transitions between occupied and
unoccupied states, and correspondingly, the imaginary part of the longitudinal dielectric function in
  the basal plane, $\mathrm{Im}\,\epsilon_{xx}$.

\subsubsection{La\textsubscript{2}CuO\textsubscript{4}}\label{ss:lsco}

\begin{figure}[h!]
\hspace{0.0cm}
\includegraphics[width=0.22\textwidth,clip=true,trim=0.0cm 0cm 0.0cm 0.0cm]{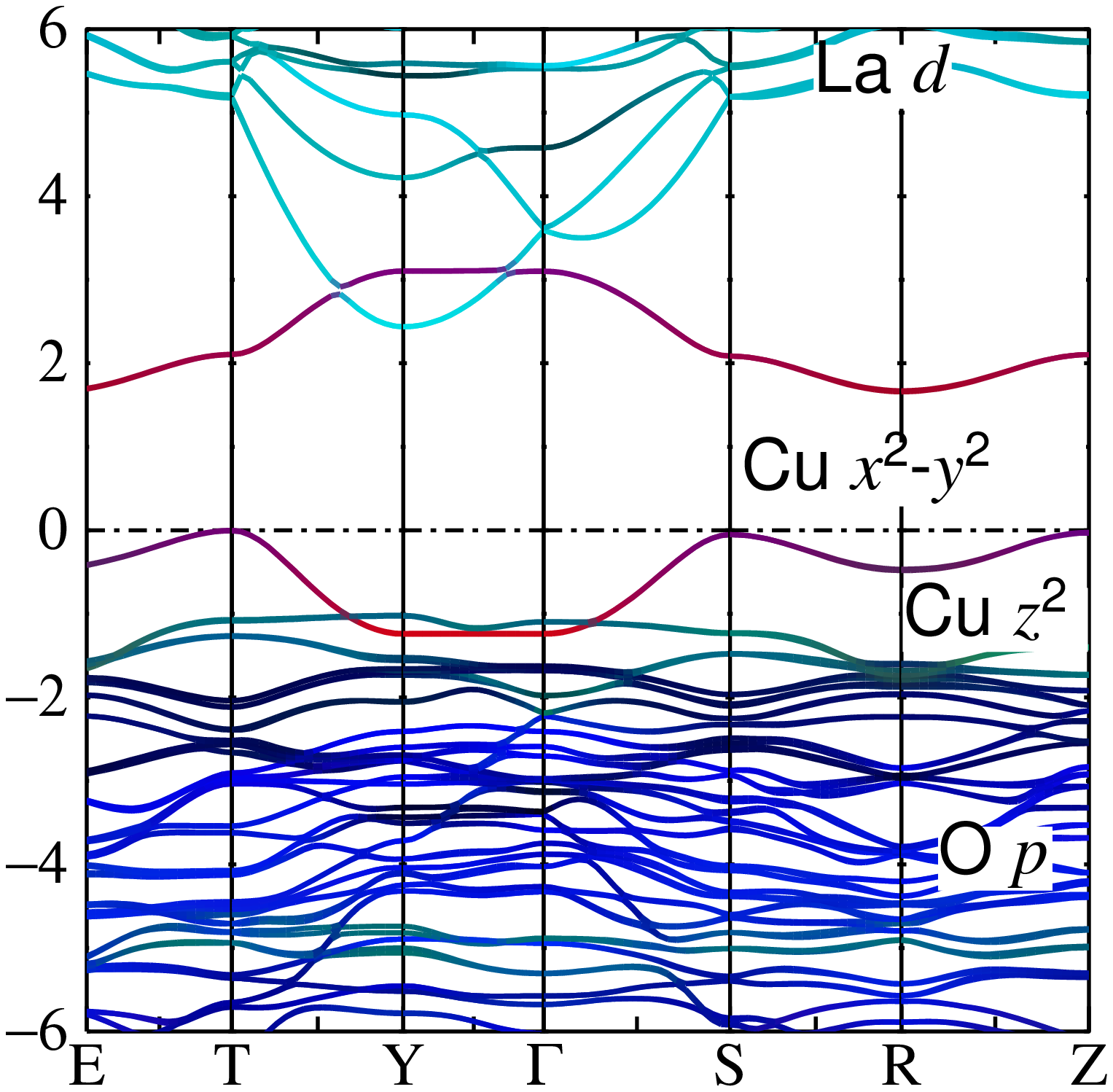}
\includegraphics[width=0.22\textwidth,clip=true,trim=0.0cm 0cm 0.0cm 0.0cm]{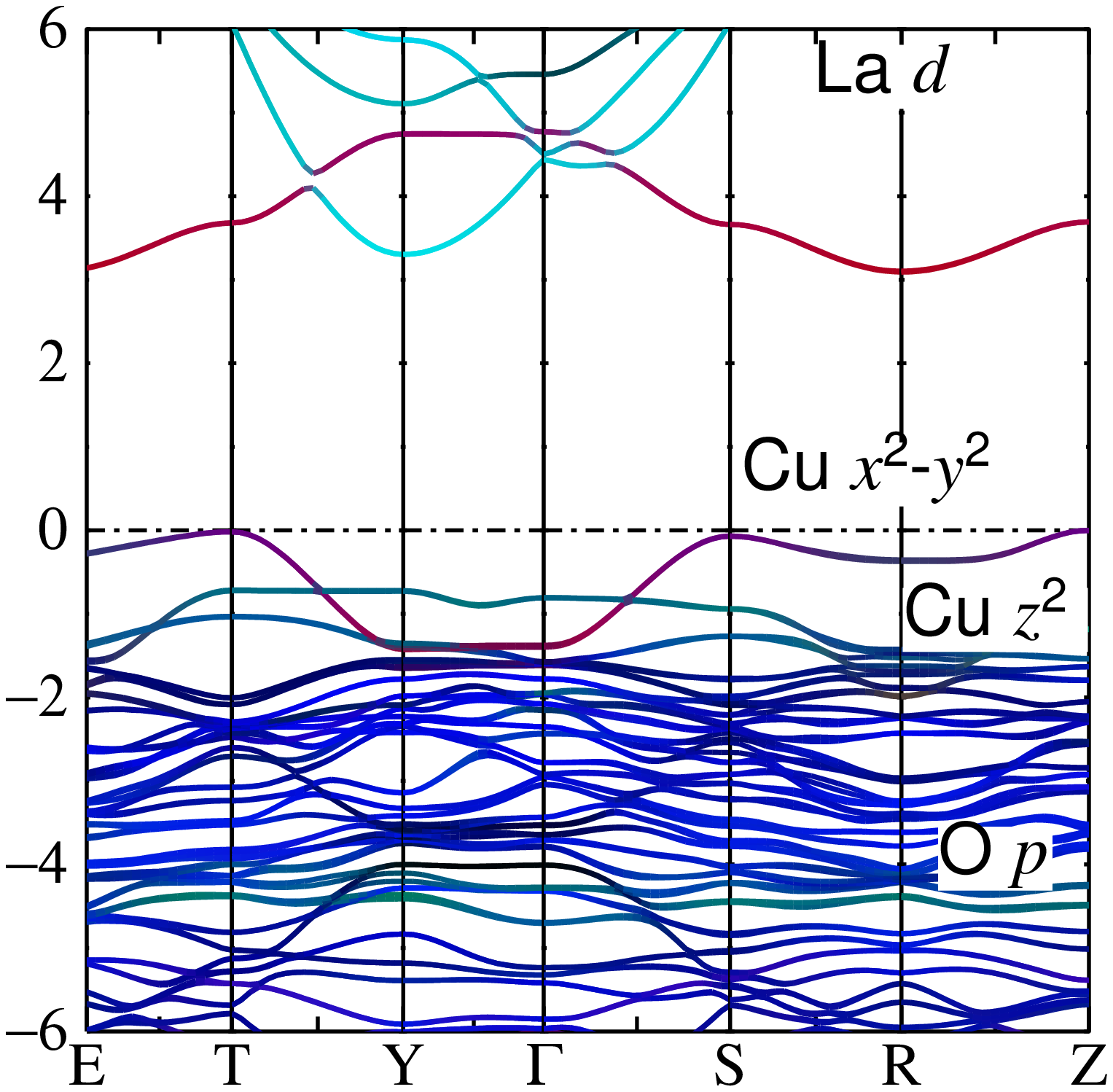}
\includegraphics[width=0.22\textwidth,clip=true,trim=0.0cm 0cm 0.0cm 0.0cm]{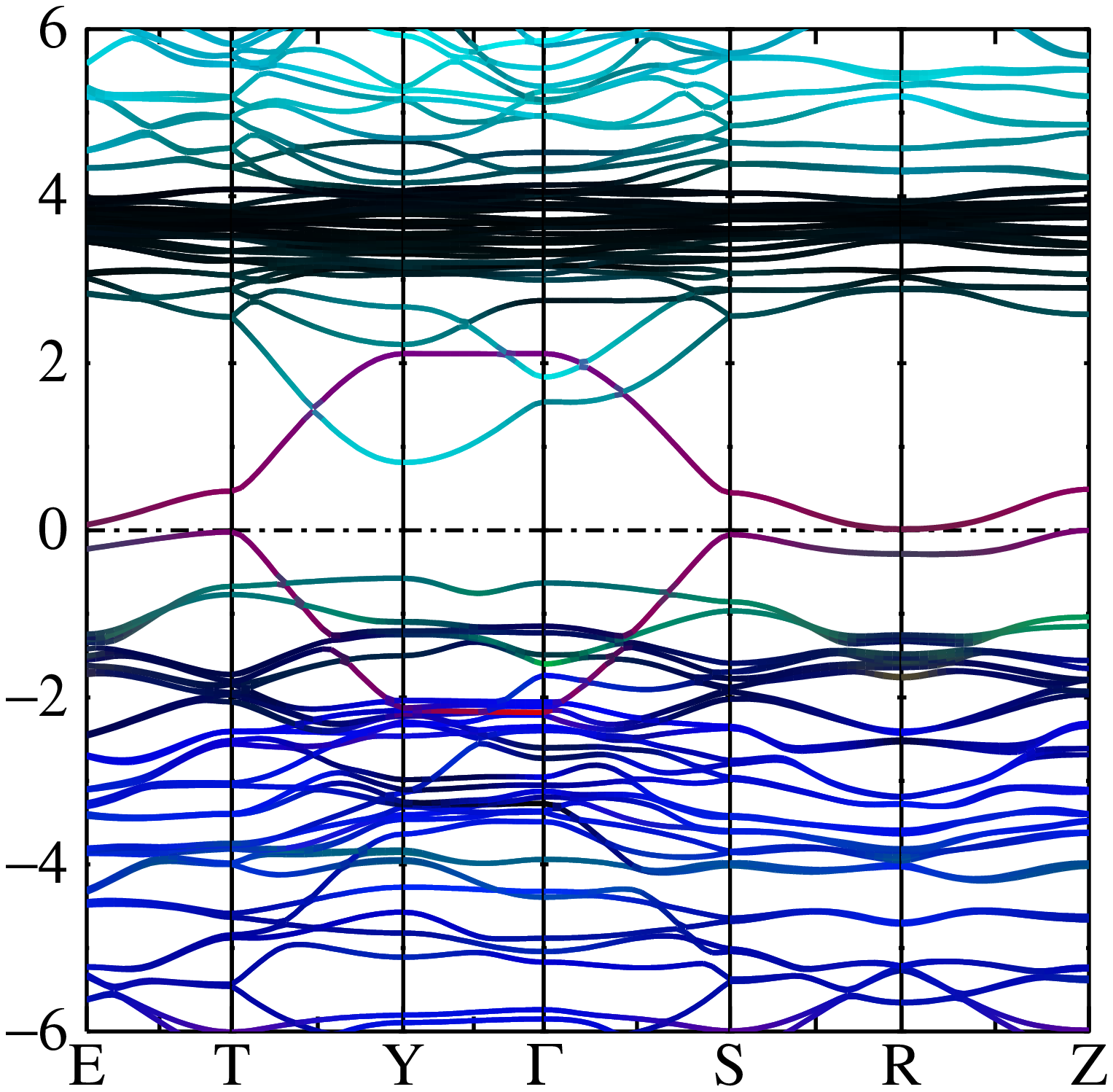}
\includegraphics[width=0.22\textwidth,clip=true,trim=0.0cm 0cm 0.0cm 0.0cm]{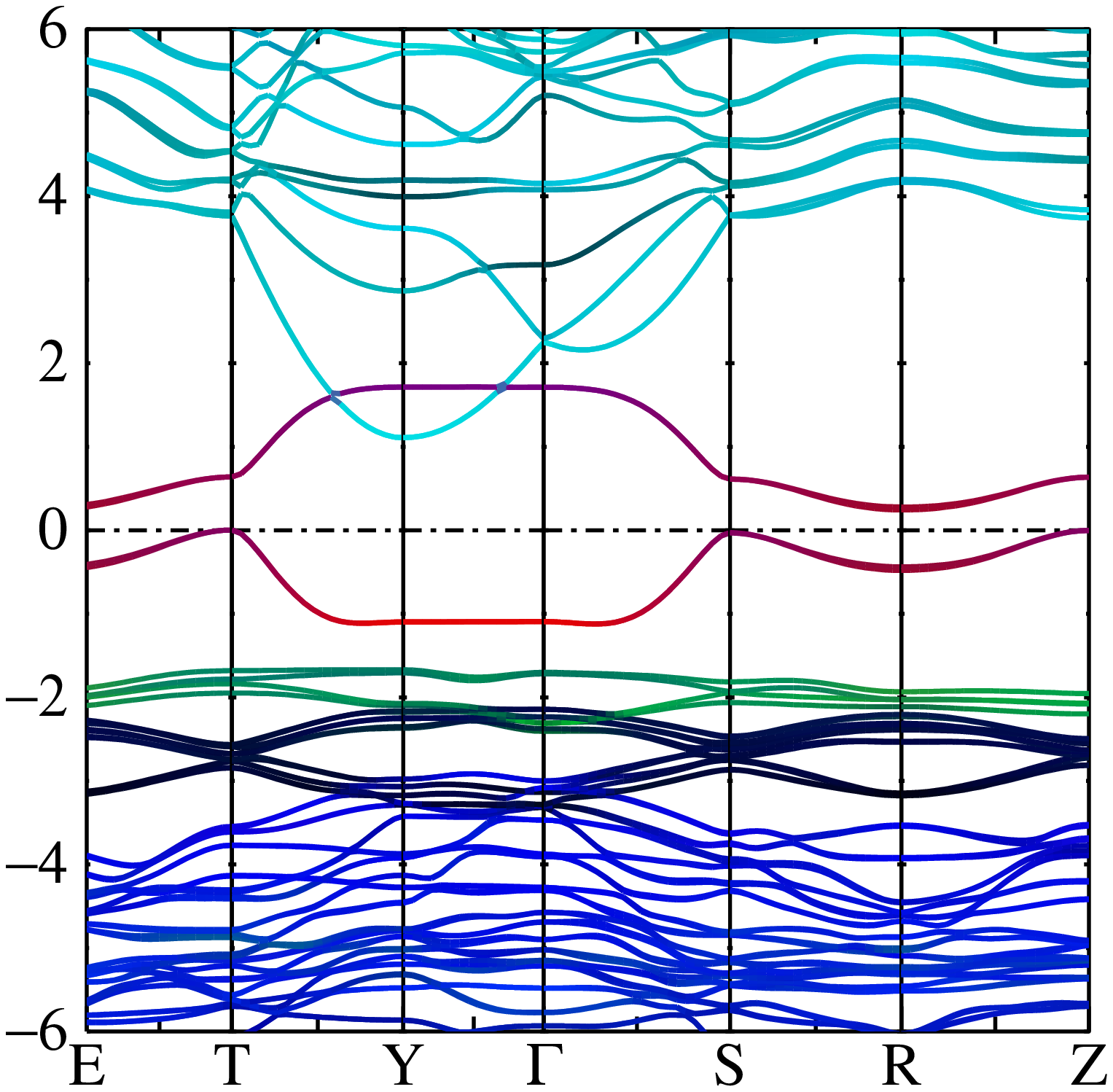}
\caption{Energy band structure of {La\textsubscript{2}CuO\textsubscript{4}}.  (a) and (b) panels apply to
  \qsgwl\ and \qsgw\ approximations, respectively.  Colors depicts the following orbital character: red: Cu $d_{x^2-y^2}$;
  green: Cu $d_{3z^2{-}1}$; blue: O $2p$; cyan: La $5d$.  Black valence bands are the remaining Cu $3d$ orbitals.
  Panel (c) is the LDA band structure, with the same color scheme.  The narrow window of black bands at 3-4
  eV are of La $4f$ character.  Panel (d) is $G^\mathrm{LDA}W^\mathrm{LDA}$ result, with $Z{=}1$ and including the off-diagonal
  parts of $\Sigma$.
}
\label{fig:lsco}
\end{figure}

{La\textsubscript{2}CuO\textsubscript{4}} is the parent compound for one of the most widely studied superconductors.  A
gap forms because Cu $d_{x^2-y^2}$ bands split into a bond-antibonding pair, owing to the formation of a local Cu moment.
\qsgw\ and \qsgwl\ energy band structures are shown in Fig.~\ref{fig:lsco}, with the Cu $d_{x^2-y^2}$ shown in red.  There are
several striking points of contrast:

\begin{itemize}[leftmargin=*]

\item \qsgwl\ reduces the Cu $d_{x^2-y^2}$ bond-antibond splitting relative to \qsgw\ by about 1.5\,eV.  The relatively
  flat $d_{x^2-y^2}$ conduction band shifts more than the La $d$ band (cyan), reminiscent of NiO.  Thus the the addition
  of ladders reduce the fundamental gap to 1.66\,eV, from the \qsgw\ gap of 3.1\,eV.

\item The occupied Cu $d_{x^2-y^2}$ band narrows relative to \qsgw.  \qsgw\ itself narrows this band substantially relative
  to LDA or LDA+U (compare Cu $d_{x^2-y^2}$ in top right to bottom left panel), but the ladders narrow it still further,
  again reminiscent of NiO.  Such mass renormalization plays a critical role in the correlations of this orbital, which
  drives superconductivity.  Spin fluctuations will narrow this band still further, but whether low-order perturbation
  theory will be sufficient to yield the true bandwidth remains an open question.

\item The O \emph{p} band is pushed down relative to \qsgw, thus reducing the hybridization of O into the Cu $d_{x^2-y^2}$
  state.  The LDA often misaligns orbitals of different character, but it is notable that ladder diagrams not only
  reduce the gap, but induce a shift to the \qsgw\ valence bands.

\item For both \qsgwl and \qsgw\ the La $4f$ states are pushed well above the Fermi level, something the LDA fails to do.

\item Self-consistency plays a very important role in this system (compare $G^\mathrm{LDA}W^\mathrm{LDA}$ to
  \qsgw\ energy bands, and see Table~\ref{tab:iteration0}).  As with NiO, the $G^\mathrm{LDA}W^\mathrm{LDA}$ bandgap is
  severely underestimated~\cite{ferdi94}.  The one-shot gap can be improved by using LDA+U or a hybrid functional
  instead of the LDA, but the resulting energy bands depend on the choice, as will other parts of the spectrum (e.g. the
  position of O $2p$ states).

\end{itemize}

\begin{figure}[h!]
\hspace{0.0cm}
\includegraphics[width=0.22\textwidth,clip=true,trim=0.0cm 0cm 0.0cm 0.0cm]{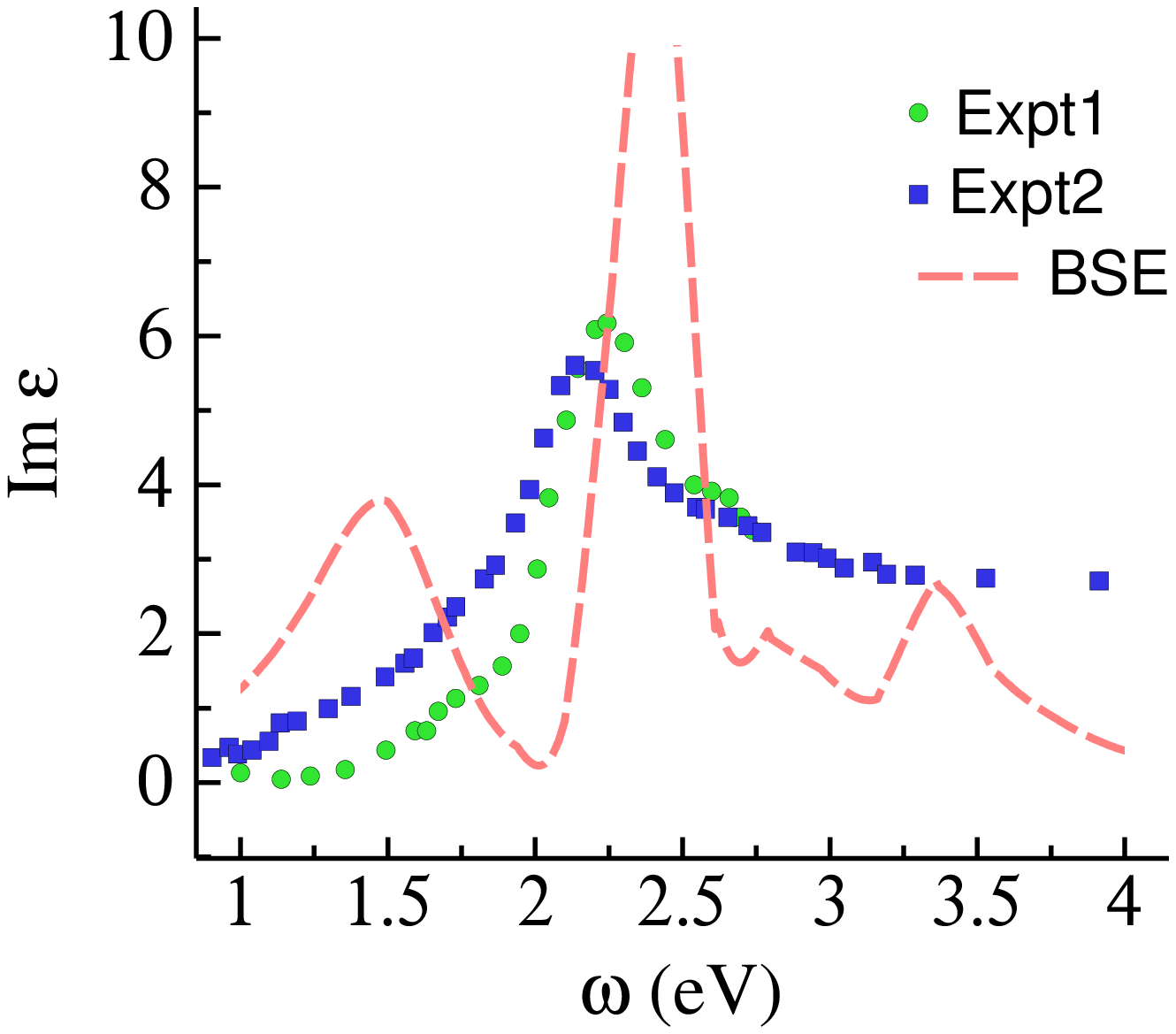}\hspace{0.5cm}
\raisebox{4mm}[0pt][0pt]{
\includegraphics[width=0.22\textwidth,clip=true,trim=0.0cm 0cm 0.0cm 0.0cm]{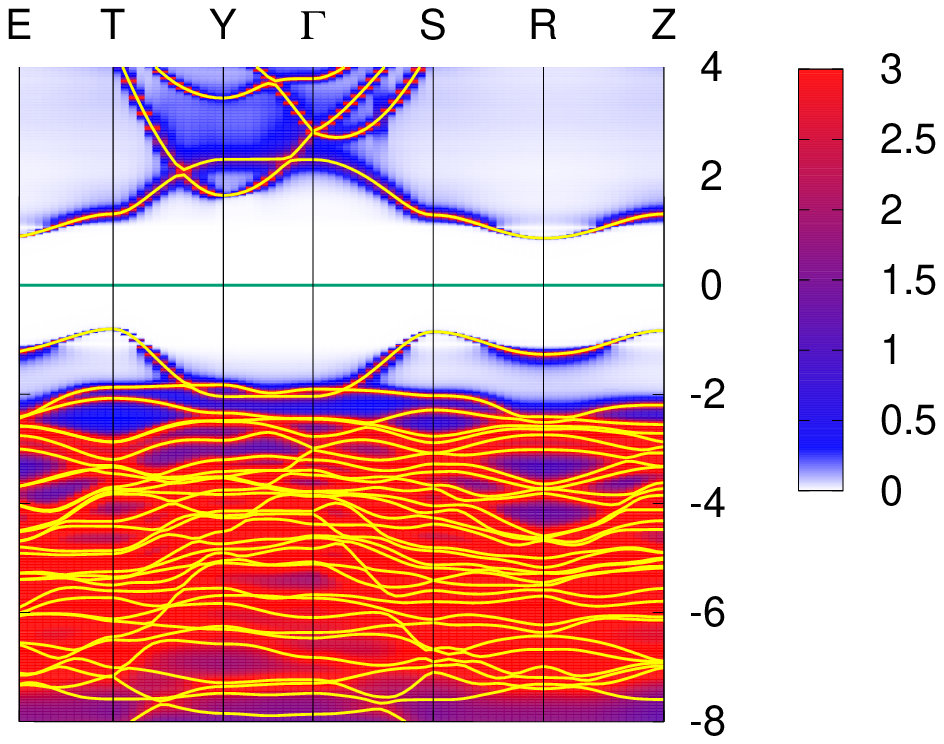}}
\caption{(Left) BSE dielectric function $\mathrm{Im}\,\epsilon_{xx}$ in {La\textsubscript{2}CuO\textsubscript{4}}, computed
  from \qsgwl, compared to reflectivity data from Ref.~\cite{Falck92} (Expt1) and conductivity data from
  Ref.~\cite{Baldini20} (Expt2). Data in Ref.~\cite{Falck92} is anomalously small.  They also report
  $\epsilon_\infty{\sim}5$, which does not seem to be compatible with their scale for $\mathrm{Im}\,\epsilon_{xx}$, from
  the Kramers Kronig relation.  Expt 1 shown in the figure scales data taken from Ref.~\cite{Falck92} by a factor of 5
  to bring it approximately in line with Ref.~\cite{Baldini20}.  (Right) Spectral function from the interacting $G$
  generated by \qsgwl.  Yellow lines are energy bands from the \qsgwl\ $G_{0}$.
 }
\label{fig:epslsco}
\end{figure}

The left panel of Fig.~\ref{fig:epslsco} shows two measurements of the dielectric function,
$\mathrm{Im}\,\epsilon_{xx}$, one inferred from reflectivity at 122\,K~\cite{Falck92}, and the other from
low-temperature optical conductivity~\cite{Baldini20}.  Ref~\cite{Falck92} also shows results from a photoconductivity
measurement, which looks similar to the blue squares in the figure but slightly blue-shifted.  \qsgwl\ results are also
shown: the peak in $\mathrm{Im}\,\epsilon_{xx}$ appears at slightly higher energy (0.1-0.2\,eV) than the experimental
data.  The \qsgwl\ result has sharper structure, in particular there appears a pronounced sub-gap peak centered at
$\sim$1.5\,eV.  A corresponding peak (albeit much weaker) is seen in the 122\,K reflectivity data, though this peak is
washed out as the temperature increases~\cite{Falck92}.  \qsgwl\ predicts a spectrum of 30 or so subgap excitons,
ranging between 1.2\,eV and the fundamental gap with widely varying oscillator strengths.  A particularly bright exciton
appears at 1.45\,eV; it is is partly responsible for the peak in $\mathrm{Im}\,\epsilon_{xx}$ there.  As for the
fundamental gap, \qsgwl\ predicts an indirect gap of 1.66\,eV, but the lowest direct gap is $\sim$2.1\,eV
(Fig.~\ref{fig:lsco}).  Ref.~\cite{Falck92} assigned a charge transfer gap of 2.1\,eV, and Ref.~\cite{Baldini20} a
similar gap (2.2\,eV), which probably corresponds to the direct gap.  The \qsgwl\ result for
$\mathrm{Im}\,\epsilon_{xx}$ shows sharper peaks than the experiment, as was shown for CuO (\S\ref{ss:cuo}).
For the same reason explained there, dynamical effects will smooth out $\mathrm{Im}\,\epsilon_{xx}$.

\subsubsection{Fe\textsubscript{3}O\textsubscript{4}}\label{ss:fe3o4}

Magnetite, or {Fe\textsubscript{3}O\textsubscript{4}}, has a cubic inverted spinel structure above the Verwey
transition at 123\,K~\cite{Verwey39}, with 6 Fe and 8 O atoms in the unit cell.  Two Fe are tetrahedrally bonded to O
(O-Fe-O bond angles 109.5\textsuperscript{o}) and four occupy octahedral sites with slightly larger bond lengths (bond
angles 90$\pm$2\textsuperscript{o} and 180\textsuperscript{o}).  It is a ferrimagnet with the spins in the tetrahedral
sites parallel, spins in the octahedral sites parallel, but the tetrahedral and octahedral sites are antiparallel.

\begin{figure}[h!]
\hspace{0.0cm}
\includegraphics[width=0.22\textwidth,clip=true,trim=0.0cm 0cm 0.0cm 0.0cm]{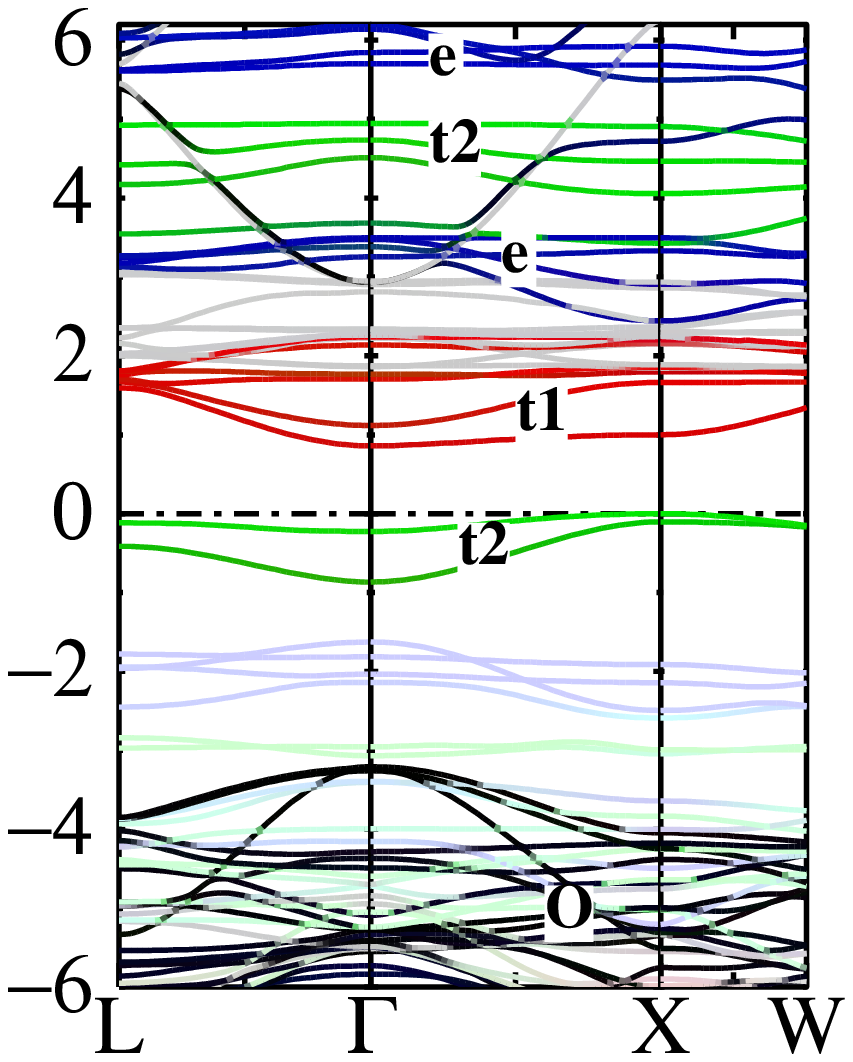}
\includegraphics[width=0.22\textwidth,clip=true,trim=0.0cm 0cm 0.0cm 0.0cm]{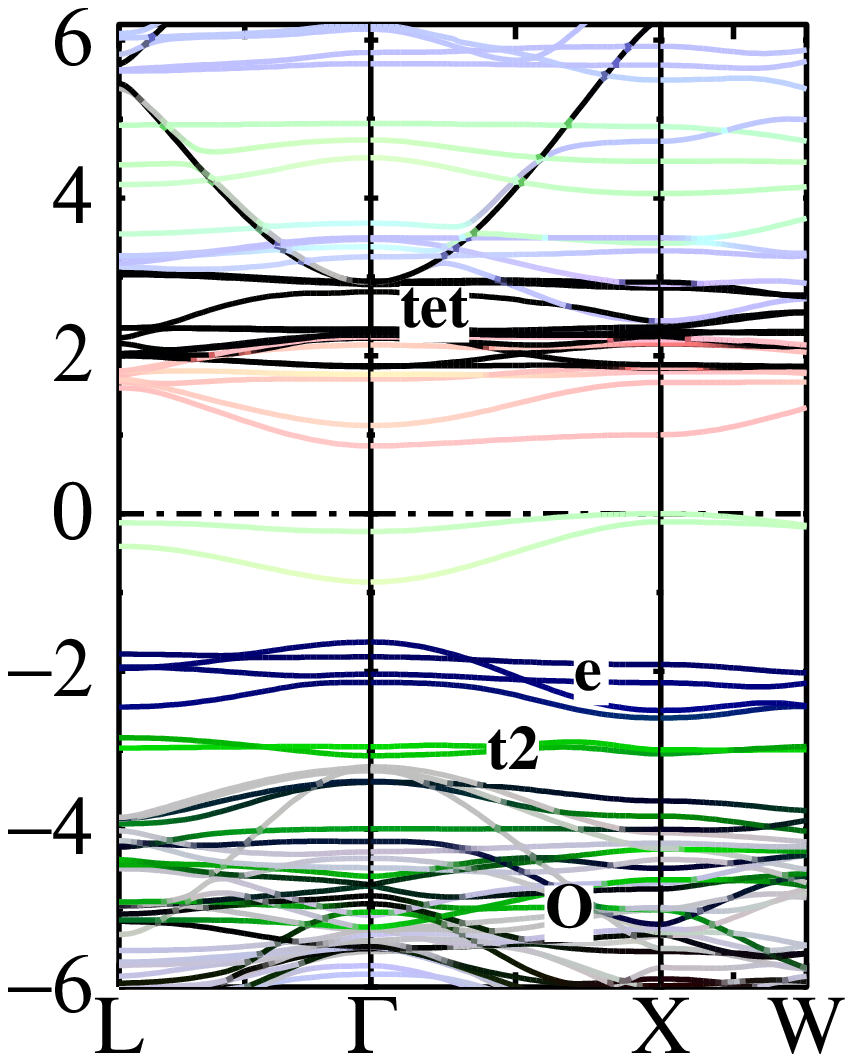}
\includegraphics[width=0.22\textwidth,clip=true,trim=0.0cm 0cm 0.0cm 0.0cm]{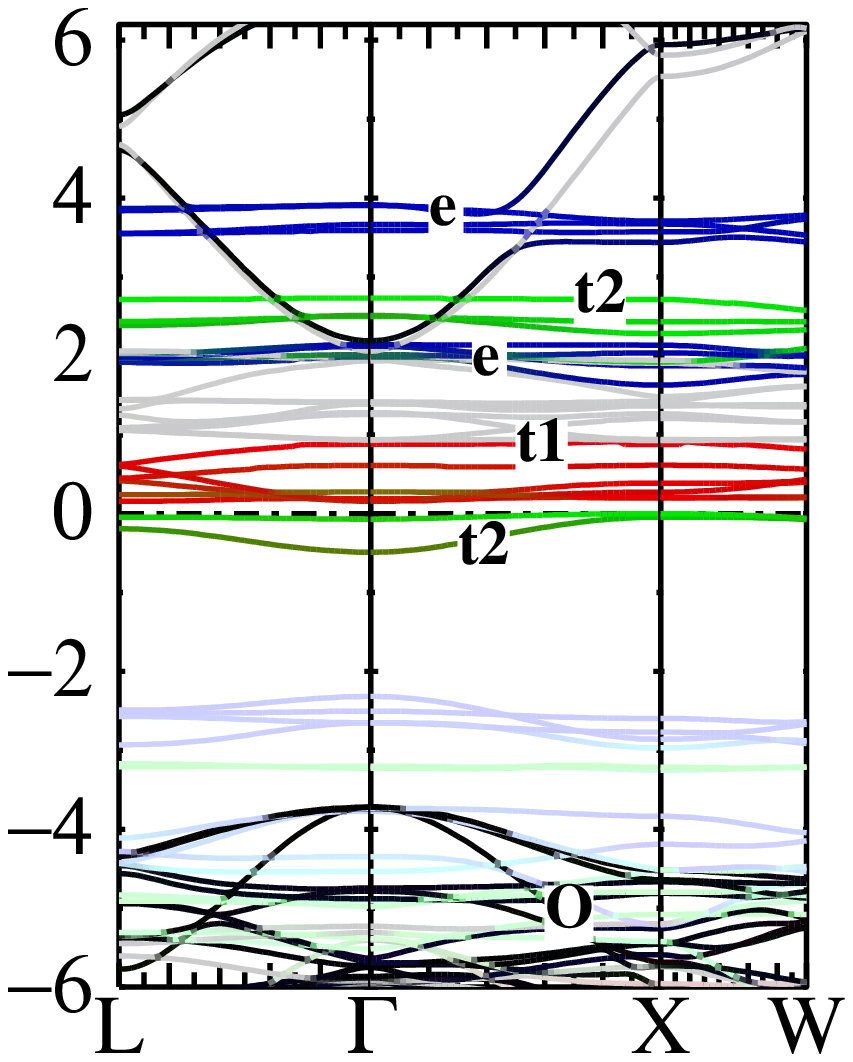}
\includegraphics[width=0.22\textwidth,clip=true,trim=0.0cm 0cm 0.0cm 0.0cm]{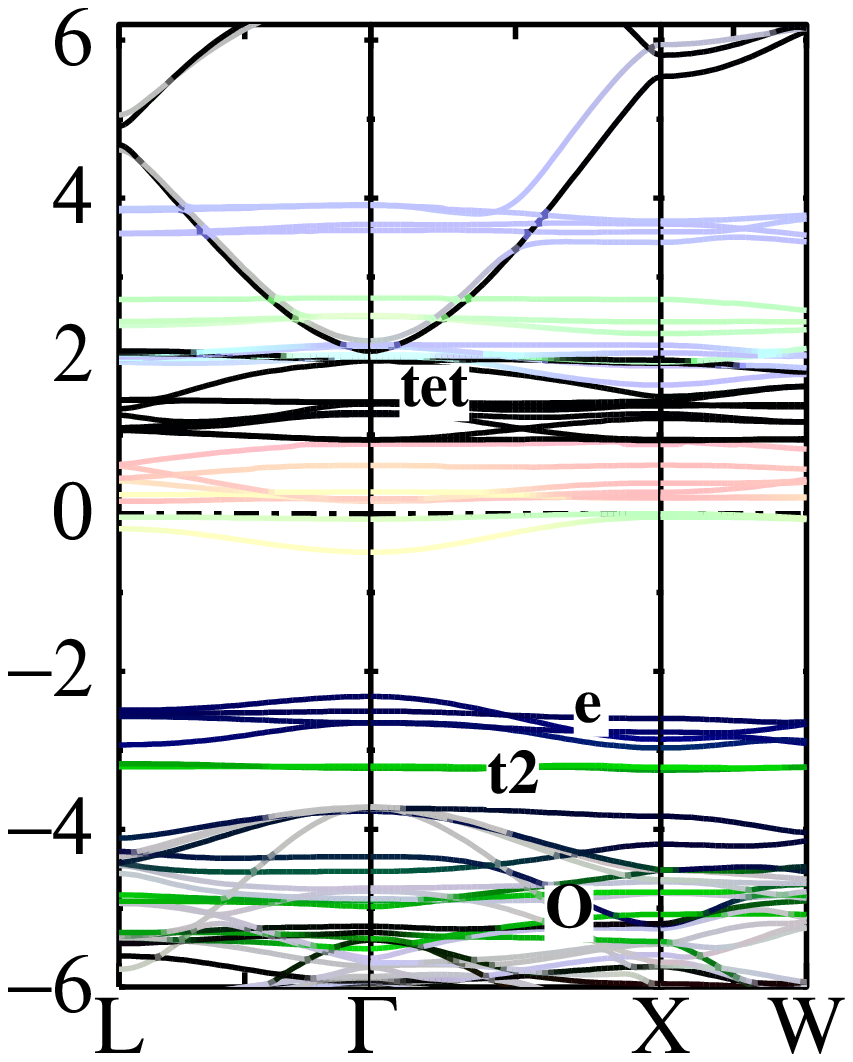}
\caption{Energy band structure {Fe\textsubscript{3}O\textsubscript{4}}, in the \qsgw\ approximation, (a) and (b) and
  the \qsgwl\ approximation, (c) and (d).  (a) and (c) show minority spin with majority spin bleached out; (b) and (d) show the reverse.  The character of relevant orbitals is labelled in the figure and the labels correspond to
  \emph{d} orbitals on the octahedral sites with the following color scheme. \textbf{t1}: $t_{2g}$, Fe\textsubscript{o1}
  (red); \textbf{t2}: $t_{2g}$ Fe\textsubscript{o2} (green); \textbf{e}: Fe\textsubscript{o1}+Fe\textsubscript{o2}
  $e_{g}$ (blue).  Labels \textbf{tet} and \textbf{O} signify bands centered mainly on Fe tetrahedral and O sites,
  respectively.
}
\label{fig:fe3o4}
\end{figure}

The conventional picture, originating from Verwey, is that the tetrahedral sites are Fe$^{3+}$ and octahedral sites
consist of equal numbers of Fe$^{2+}$ and Fe$^{3+}$.  Below the Verwey temperature magnetite is a
narrow-gap insulator with a bandgap 0.14-0.3\,eV.  It was traditionally believed that above the Verwey temperature
magnetite becomes a half-metal, in part because the conductivity increases by $\sim$100-fold across the transition, and
evidence from interpretations of PES and STS experiments suggested a finite density-of-states at $E_{F}$.  However, a
more recent high-resolution PES experiment ~\cite{Park97} found that the band gap persists above the Verwey temperature,
reduced by $\sim$50\,meV.  That a gap of order 0.2\,eV persists was confirmed by a subsequent STS
measurement~\cite{Jordan06}.  For a more detailed summary of the experimental status, see the work of Liu and Di
Valentin~\cite{Liu17}.  These authors applied various one-body techniques (LDA+U, hybrid functionals) to study magnetite
and concluded that the traditional picture of magnetic order yields a metallic ground state.  They argued that a gap
appears because the octahedral sites disproportionate into two kinds of atoms, one with a high-spin (moment
${\gtrsim}4\mu_B$) and one with a low-spin (moment ${\sim}3.5\mu_B$).

\begin{table}[h]
\caption{Spin moments and bandgap of {Fe\textsubscript{3}O\textsubscript{4}} calculated with the \qsgw\ and \qsgwl\ approximations.
Fe\textsubscript{t}, Fe\textsubscript{o1}, Fe\textsubscript{o2} indicate Fe on tetrahedral sites, and the two
inequivalent octahedral sites.
}
\begin{tabular}{|@{\hspace{1.0em}}c@{\hspace{1.0em}}|@{\hspace{1.0em}}l@{\hspace{1.0em}}|c|@{\hspace{1.0em}}c@{\hspace{1.0em}}|@{\hspace{1.0em}}c@{\hspace{1.0em}}|}\hline
           & $E_{G}$         & $\mu$ (Fe\textsubscript{t})     & $\mu$ (Fe\textsubscript{o1}) & $\mu$ (Fe\textsubscript{o2}) \cr \hline
 \qsgw     & 0.86$\pm$0.1    &  -4.05  & 4.31  & 3.65 \cr
 \qsgwl    & 0.15$\pm$0.15   &  -4.04  & 4.20  & 3.75 \cr \hline
\end{tabular}
\label{tab:fe3o4}
\end{table}

We performed \qsgw\ and \qsgwl\ calculations with symmetry suppressed, so every spin could assume an independent value.
Spins on the tetrahedral sites converged to a common value and those on the tetrahedral sites converged to two values,
one high spin and one low spin (Table~\ref{tab:fe3o4}).  The calculation was extremely difficult to stabilize and a
fully converged solution was never found, even after 100 iterations.  Local moments were stable as iterations proceeded,
but the bandgap fluctuated; for that reason error bars are given for the gaps shown in Table~\ref{tab:fe3o4}.  Here we
denote high-spin and low-spin sites as ``o1'' and ``o2.''

The Fe\textsubscript{o2} $t^\downarrow_{2g}$ manifold splits off one band (actually two, because the unit cell consists
of two {Fe\textsubscript{3}O\textsubscript{4}} formula units which weakly couple), and this split-off band forms the
valence band maximum (Fig.~\ref{fig:fe3o4}).  On the other hand, the Fe\textsubscript{o1} $t^\downarrow_{2g}$ manifold
does not split in the same way, and it forms the conduction band minimum.  Splitting of Fe\textsubscript{o2}
$t^\downarrow_{2g}$ is the main way in which cubic symmetry is broken and a gap is formed.

Ladder diagrams have two major effects: first, they sharply narrow all of the octrahedral Fe-\emph{d} bandwidths,
e.g. the $t^\downarrow_{2g}$ manifold forming the conduction band minimum is narrowed by $\sim$30\%.  Second, ladders
cause band centers to shift in a highly orbital-dependent manner.  Shifts on the occupied states are modest, but for the
unoccupied states they can be quite large.  Note for example the band center of the $t^\downarrow_{2g}$ level around
1.5\,eV is pushed down by $\sim$1\,eV, while the $t^\downarrow_{2g}$ and $e^\downarrow_{g}$ levels in the 5-6 eV range
shift by $\sim$2\,eV.  Unfortunately, no experiments are available for benchmarking.

Ordering of the Fe levels is qualitatively in agreement with the picture of Ref~\cite{Liu17}; see their Figure 5, and
the \qsgwl\ spin moments are similar to their Table 1; we thus affirm their description of magnetite.

\section{Conclusions}

We presented an extension of the \qsgw\ approximation in which we include vertex corrections to $W$, calculated at the level
of the BSE (\qsgwl).  The primary aim of this work was to establish to what extent \qsgwl\ rectifies the most severe
errors in \qsgw, with the ultimate aim to develop a high-fidelity, universally applicable theory.  If low-order diagrams
are sufficient to yield high-fidelity one- and two-particle properties, Green's function methods offer an enormous
opportunity to be both high-fidelity and relatively efficient.  \qsgwl\ supplies excitonic effects which \qsgw\ omits.
They are known to be important, as already extensively discussed in the literature (see
e.g. Ref.~\onlinecite{onida_electronic_2002}), so it is of interest to see to what extent it captures the quasiparticle
spectrum in insulators, and whether systematic errors can be discerned.

While the \qsgw\ approximation has long been known to overestimate bandgaps, the discrepancies with experiments are much
more systematic than more commonly used $G_0W_0$ approaches with \emph{G} and \emph{W} constructed, e.g. from the LDA.
Its systematic character is a consequence of self-consistency, in part because it does not rely on density-functional
theory.  We presented a few contexts where a non self-consistent approach is not important
(e.g. {Bi\textsubscript{2}Te\textsubscript{3}}), and others (e.g. TiSe\textsubscript{2},
{La\textsubscript{2}CuO\textsubscript{4}}) where it is fundamentally problematic.  Even when such an approach yields a
good bandgap, it may occur for adventitious reasons.  The relatively unsystematic nature of the errors in one-shot
approaches make it difficult to assess what diagrams are essential to realize the goal of a universally applicable,
high-fidelity theory.\footnote{The present method also employs an all-electron basis, which eliminates the dependence on
the choice of the pseudopotential.  This can---though it should not be the case---substantially influence the $GW$
results.}  Thus, self-consistency is crucial for the aims of this work.

The present work surveyed a wide range of insulators, including tetrahedrally coordinated semiconductors where
experimental information is reliable and abundant, and also a variety of other \emph{sp} systems, $d^{0}$ oxides, and
polar compounds, and a family of 3\emph{d} transition metal antiferromagnetic oxides.  Each materials system had a
distinct set of characteristics, but apart from some important exceptions critically examined in this work,
\qsgwl\ predicts with fairly high fidelity both one- particle and optical properties for all of the systems we studied.
The exceptions are important and formed a major focus of this study.  Two shortcomings clearly identified were the
omission of electron-phonon interaction, which causes gaps to be too large in wide-gap systems, and the omission of the
vertex in the exact self-energy.  This vertex pushes down nearly dispersionless core-like states, and when they form the
valence band maximum the bandgap is consistently underestimated.  By constructing hybrid self-energies, we could account
for both of these shortcomings in an approximate way, and draw the following conclusions:

\begin{enumerate}[leftmargin=*]

\item At the \qsgwl\ level, there is a very close connection between the fidelity of the fundamental gap $E_{G}$ and the
  dielectric constant \epsi.  When one is well described, so is the other, and vice-versa.  This provides a much more
  robust benchmark of a theory than benchmarking one-particle properties alone.

\item If we take the first point as an ansatz for a general principle, it can be used in cases where experimental data
  is unavailable or inconsistent.  We presented evidence for several systems (CeO$_{2}$, SrTiO$_{3}$, TiO$_{2}$, ScN,
  CuAlO$_{2}$, FeO) where the calculated results inform the experimental observations and indicate that accepted values
  of the one-particle properties need adjustment.  For FeO, the revision is rather dramatic.

\item A low-order diagrammatic theory appears to describe the dielectric response with high fidelity for all the systems
  in this study, to the extent we are able to reliably extract experimental data.  Ladder diagrams appear to be
  sufficient to capture well the main part of the optical response functions and one-particle Green's functions in most
  insulators, even strongly correlated ones.  While such an assertion is likely not universally true~\cite{noteb}, it
  appears to be the case for broad classes of materials.

\item Ladders not only shift the bandgap but further narrow the \emph{d} bandwidth in some systems (NiO,
  {La\textsubscript{2}CuO\textsubscript{4}}, {Fe\textsubscript{3}O\textsubscript{4}}).  It may be that the addition of a
  low-order \emph{GW}-like theory accounting for spin fluctuations, such as the dual-trilex formulation of Stepanov
  et. al~\cite{Stepanov19}, may adequately account for spin and charge response functions even in strongly correlated
  materials.

\end{enumerate}

These last two observations suggest the tantalizing possibility that, with some modest extensions that may be added
hierarchically, a broadly applicable, high-fidelity \emph{ab initio} approach to solving one- and two- particle
properties of the many-body problem is within reach.

\section{Appendix: the LQSGW approximation}

Kutepov's L\qsgw\ theory~\cite{Kutepov17} is a linearized form of \qsgw.  He approximates the quasiparticlized self-energy
as a Taylor series around zero frequency.   Treating each band independently and suppressing band index for simplicity of
presentation, Kutepov replaces the interacting $G$
\begin{align*}
  G^{-1} (k,\omega ) = \omega + \mu - \epsilon - \Sigma(k,\omega)
\end{align*}
by omitting the second order and higher terms of an expansion of $\Sigma$ in $\omega$:
\begin{align}
  \Sigma (k,\omega ) = \Sigma(k,0) + \omega\,\Sigma'(k,0) + \frac{1}{2}\omega^2\,\Sigma''(k,0) + ...
 \label{eq:siglin}
\end{align}
$G^{-1}$ simplifies to a linear function of $\omega$
\begin{align*}
  G^{-1} (k,\omega ) = \bar{Z}^{-1}\, \omega + \mu - \epsilon - \Sigma(k,0)
\end{align*}
and thus reduces to a linear algebraic eigenvalue problem. The bar over the $Z$ factor indicates that
is not equivalent to Eq.~\ref{eq:Zren}, since it is defined at zero frequency
$\omega{=}0$:
\begin{align*}
    1 - 1/{\bar{Z}^j} = \Sigma'(k,0)
\end{align*}

Evidently $\epsilon - \mu + \Sigma(k,0)$ is the eigenvalue of a hamiltonian defined as the one-body part of $G^{-1}$,
but including the static part of $\Sigma$.  The (linearized) energy-dependence of $\Sigma$ modifies this eigenvalue
to read
\begin{align*}
E-\mu = \bar{Z}[\epsilon-\mu + \Sigma(k,0)]
\end{align*}

$E$ is identical to the \qsgw\ quasiparticle energy if $\Sigma$ is a linear function of $\omega$.

Now let us retain the quadratic term in $\Sigma$ and determine the shift in $E$ to estimate the difference between
L\qsgw\  and \qsgw.  Let us denote the L\qsgw\ eigenvalue $E-\mu$ as $E_0$.  Expanding $G^{-1}$ to
second order we obtain, to lowest order in $\Sigma''(k,0)$ :
\begin{align}
  G^{-1} \approx \omega - (E_0 + \frac{\bar{Z}}{2}E_0^2\,\Sigma''(k,0) )
 \label{eq:lqsgwE}
\end{align}
The lowest-order difference between L\qsgw\ and \qsgw\ QP levels is the second term in parenthesis.

\begin{acknowledgments}
  The authors would like to thank all those involved in the {\it CCP flagship project: Quasiparticle Self-Consistent
    $GW$ for Next-Generation Electronic Structure}, especially Scott Mckechnie for his help. We are grateful for support
  from the Engineering and Physical Sciences Research Council, under grant EP/M011631/1. MvS and DP were supported the
  Computational Chemical Sciences program within the Office of Basic Energy Sciences, U.S. Department of Energy under
  Contract No. DE-AC36-08GO28308.  We are grateful to the UK Materials and Molecular Modelling Hub for computational
  resources, which is partially funded by EPSRC (EP/P020194/1). The research was performed using computational resources
  sponsored by the Department of Energy's Office of Energy Efficiency and Renewable Energy and located at the National
  Renewable Energy Laboratory.  This research also used resources of the National Energy Research Scientific Computing
  Center (NERSC), a U.S. Department of Energy Office of Science User Facility located at Lawrence Berkeley National
  Laboratory, operated under Contract No. DE-AC02-05CH11231. This research used resources of the National
Energy Research Scientific Computing Center, a DOE Office of Science User Facility supported by the Office of Science of
the U.S. Department of Energy under Contract No. DE-AC02-05CH11231 using NERSC award BES-ERCAP0021783.
\end{acknowledgments}

\vfil\eject

\section*{Supplemental Material:  Computational Details}

\begin{table*}[ht!]
\centering
\label{tb:comp_detailsx}
\begin{tabular}{r@{\hspace{0.3em}}|ccccccccccc}\hline
                &$a$(a.u.) &$G_\mathrm{cut}(\psi, M)$
                                               &$\Sigma_\mathrm{cut}$(Ry)
                                                       &$N_k$ &$N_{1p}$
                                                                   &$N_{v}$  &$N_{c}$
                                                                     &$N_{\rm flt}$
                                                                              &$\varphi_{z}$\\\hline\hline
C               &6.740             &4.2  3.5   &3.0     &6    &104   &4       &8   &2    &$\bar{3}s\bar{3}p\bar{4}d$\,C\\
Si              &10.24             &3.0  2.5   &3.0     &6    &104   &4       &8   &2    &$\bar{4}s\bar{4}p\bar{4}d$\,Si\\
Ge              &10.68             &3.0  2.5   &3.0     &6    &104   &4       &8   &2    &$\bar{5}s\bar{5}p    {3}d$\,Ge\\
TiO$_2$         &8.681 (0.6441)    &3.5  2.8   &3.0     &4,6  &330   &12      &10  &8    &$\bar{5}s    {3}p\bar{4}d$\,Ti;\ $\bar{3}s\bar{3}p\bar{4}d$\,O\\
SrTiO$_3$       &7.354             &3.5  2.8   &3.0     &6    &215   &9       &11  &-
    &\begin{tabular}{c}${4}s{4}p\bar{5}d$\,Sr;\ $\bar{5}s\bar{3}p\bar{4}d$\,Ti;\\ $\bar{3}s{3}p\bar{4}d$\,O\end{tabular}\\
CuAlO$_2$       &3.121 (3.422)     &4.1  3.3   &3.5     &6    &177   &8       &8   &4    &$\bar{5}s\bar{4}d$\,Cu;\ $\bar{4}s$\,Al;\ $\bar{3}s\bar{3}p\bar{4}d$\,O\\
LiF             &7.597             &4.2  3.5   &3.0     &6    &99    &3       &8   &2    &$    {1}s\bar{3}p\bar{4}d$\,Li;\ $\bar{3}s\bar{3}p\bar{4}d$\,F\\
LiCl            &9.600             &3.6  3.0   &3.0     &6    &104   &3       &8   &2    &$    {1}s\bar{3}p\bar{4}d$\,Li;\ $\bar{4}s\bar{4}\hat{p}\bar{4}d$\,Cl\\
NaCl            &10.62             &3.2  2.6   &3.0     &6    &104   &3       &8   &2    &$    {2}s    {2}p        $\,Na\\
CuCl            &10.23             &3.2  2.6   &3.0     &6    &104   &8       &8   &2    &$\bar{5}s\bar{5}\hat{p}\bar{4}d$\,Cu;\ $\bar{4}s\bar{4}\hat{p}\bar{4}d$\,Cl\\
Cu$_{2}$O       &8.069             &3.5  2.8   &3.0     &4    &330   &26      &9   &10   &$\bar{5}s\bar{5}\hat{p}\bar{4}d$\,Cu;\ $\bar{3}s\bar{3}p\bar{4}d$\,O\\
MgO             &7.933             &4.0  3.3   &3.0     &6    &104   &3       &8   &2    &$    {2}s    {2}p\bar{4}d$\,Mg;\ $\bar{3}s\bar{3}p\bar{4}d$\,O\\
CaO             &9.077             &3.5  2.9   &3.0     &6    &104   &3       &8   &2    &$    {3}s    {3}p\bar{4}d$\,Ca;\ $\bar{3}s\bar{3}p\bar{4}d$\,O\\
SrO             &9.751             &4.0  3.3   &3.0     &6    &104   &3       &8   &2    &$    {4}s    {4}p\bar{5}d$\,Sr;\ $\bar{3}s\bar{3}p\bar{4}d$\,O\\
BaO             &10.43             &3.5  2.8   &3.0     &6    &85    &3       &4   &2    &$    {5}s    {5}p\bar{6}d$\,Ba;\ $\bar{3}s\bar{3}p\bar{4}d$\,O\\
CdO             &8.874             &3.9  3.2   &3.0     &6    &104   &4       &4   &2    &$\bar{6}s\bar{6}\hat{p}\bar{5}d$\,Cd;\ $\bar{3}s\bar{3}p\bar{4}d$\,O\\
ZnO             &6.138 (1.602)     &3.7  3.0   &3.5     &6,4  &204   &6       &8   &2    &$\bar{5}s\bar{5}\hat{p}\bar{4}d$\,Zn;\ $\bar{3}s\bar{3}p\bar{4}d$\,O\\
ZnS             &10.23             &3.4  2.8   &3.0     &6    &104   &3       &8   &2    &$\bar{5}s\bar{5}\hat{p}\bar{4}d$\,Zn;\ $\bar{4}s\bar{4}\hat{p}\bar{4}d$\,S\\
ZnSe            &10.69             &3.0  2.5   &3.0     &6    &104   &3       &8   &2    &$\bar{5}s\bar{5}\hat{p}\bar{4}d$\,Zn;\ $\bar{5}s\bar{5}\hat{p}3d$\,Se\\
ZnTe            &11.53             &3.0  2.5   &3.0     &6    &104   &3       &8   &2    &$\bar{5}s\bar{5}\hat{p}\bar{4}d$\,Zn;\ $\bar{6}s\bar{6}\hat{p}4d$\,Te\\
\emph{w}CdS     &7.861 (1.620)     &2.7  2.2   &3.5     &6,4  &204   &6       &8   &2    &$\bar{6}s\bar{6}\hat{p}\bar{5}d$\,Cd;\ $\bar{4}s\bar{4}\hat{p}\bar{4}d$\,S\\
CdSe            &11.43             &2.7  2.2   &3.0     &6    &104   &3       &8   &2    &$\bar{6}s\bar{6}\hat{p}\bar{5}d$\,Cd;\ $\bar{5}s\bar{5}\hat{p}3d$\,Se\\
CdTe            &12.24             &2.8  2.3   &3.0     &6    &104   &3       &8   &2    &$\bar{6}s\bar{6}\hat{p}\bar{5}d$\,Cd;\ $\bar{6}s\bar{6}\hat{p}4d$\,Te\\
\emph{h}BN      &4.732             &3.8  3.0   &4.0     &6,3  &124   &8       &8   &2    &$\bar{3}s\bar{3}p$\,B;\ $\bar{3}s\bar{3}p$\,N\\
AlN             &5.879 (1.601)     &3.8  3.1   &3.5     &6,4  &204   &6       &8   &2    &$\bar{4}s\bar{4}p\bar{4}d$\,Al;\ $\bar{3}s\bar{3}p\bar{4}d$\,N\\
AlP             &10.32             &3.0  2.5   &3.0     &6    &104   &4       &8   &2    &$\bar{4}s\bar{4}p\bar{4}d$\,Al;\ $\bar{4}s\bar{4}p\bar{4}d$\,P\\
AlAs            &10.70             &3.3  2.7   &3.0     &6    &104   &4       &8   &2    &$\bar{4}s\bar{4}p\bar{4}d$\,Al;\ $\bar{5}s\bar{5}\hat{p}    {3}d$\,As\\
AlSb            &11.59             &3.0  2.5   &3.0     &6    &104   &4       &8   &2    &$\bar{4}s\bar{4}p\bar{4}d$\,Al;\ $\bar{6}s\bar{6}\hat{p}    {4}d$\,Sb\\
GaN             &6.027 (1.626)     &3.6  2.9   &3.5     &6,4  &204   &6       &8   &2    &$\bar{5}s\bar{5}\hat{p}    {3}d$\,Ga;\ $\bar{3}s\bar{3}p\bar{4}d$\,N\\
GaP             &10.29             &3.0  2.5   &3.0     &6    &104   &4       &8   &2    &$\bar{5}s\bar{5}\hat{p}    {3}d$\,Ga;\ $\bar{4}s\bar{4}\hat{p}\bar{4}d$\,P\\
GaAs            &10.66             &2.7  2.4   &3.0     &6    &104   &4       &8   &2    &$\bar{5}s\bar{5}\hat{p}    {3}d$\,Ga;\ $\bar{5}s\bar{5}\hat{p}    {3}d$\,As\\
GaSb            &11.50             &2.7  2.3   &3.0     &6    &104   &3       &8   &2    &$\bar{5}s\bar{5}\hat{p}    {3}d$\,Ga;\ $\bar{6}s\bar{6}\hat{p}    {4}d$\,Sb\\
InN             &6.679 (1.624)     &3.4  2.7   &3.5     &6,4  &204   &6       &8   &2    &$\bar{6}s\bar{6}\hat{p}\bar{5}d$\,In;\ $\bar{3}s\bar{3}p\bar{4}d$\,N\\
InP             &11.09             &2.9  2.4   &3.0     &6    &104   &4       &8   &2    &$\bar{6}s\bar{6}\hat{p}\bar{5}d$\,In;\ $\bar{4}s\bar{4}\hat{p}\bar{4}d$\,P\\
InAs            &11.43             &2.7  2.3   &3.0     &6    &104   &3       &8   &2    &$\bar{6}s\bar{6}\hat{p}\bar{5}d$\,In;\ $\bar{5}s\bar{5}\hat{p}    {3}d$\,As\\
InSb            &12.24             &2.7  2.3   &3.0     &6    &104   &3       &8   &2    &$\bar{6}s\bar{6}\hat{p}\bar{5}d$\,In;\ $\bar{6}s\bar{6}\hat{p}    {4}d$\,Sb\\
ScN             &8.504             &3.5  2.9   &3.0     &6    &104   &3       &8   &2    &$    {3}s    {3}p\bar{4}d$\,Sc;\ \ $\bar{3}s\bar{3}p\bar{4}d$\,N\\
PbTe            &12.15             &2.8  2.3   &3.0     &6    &104   &3       &8   &2    &$\bar{7}s\bar{7}\hat{p}\bar{6}d$\,Pb;\ $\bar{6}s\bar{6}\hat{p}    {4}d$\,Te\\
TiSe$_2$        &6.689 (1.697)     &2.8  2.2   &3.0     &3,2  &664   &16      &16  &16   &$            {3}p\bar{4}d$\,Ti\\
FeS$_{2}$       &10.22             &2.7  2.2   &3.0     &4    &720   &32      &12  &24   &$\bar{5}s        \bar{4}d$\,Fe;\ $\bar{4}s\bar{4}p\bar{4}d$\,S\\
VO\textsubscript{2}&8.536          &3.4  2.8   &2.0     &4    &332   &20      &8   &-    &            ${3}p\bar{4}d$\,V\\
CeO$_{2}$       &10.23             &3.5  3.0   &2.5     &6    &124   &6       &16  &1    &$5s5p\bar{6}d\bar{5}f$\,Ce;\ $\bar{3}s\bar{3}p$\,O\\
Bi$_{2}$Te$_{3}$&4.783 (4.015)     &3.0  2.5   &3.0     &6    &260   &18      &12  &5    &$\bar{7}s\bar{7}\hat{p}    {5}d$\,Bi;\ $\bar{6}s\bar{6}\hat{p}     {4}d$\,Te\\
\hline
MnO             &8.398             &3.5  3.0   &2.5     &4    &178   &16      &16  &4    &$\bar{5}s    {3}p\bar{4}d$\,Mn;\ $\bar{3}s\bar{3}p$\,O\\
FeO             &8.088             &3.5  2.9   &3.0     &4    &172   &9       &6   &-    &$\bar{5}s    {3}p\bar{4}d$\,Fe;\ $\bar{3}s\bar{3}p\bar{4}d$\,O\\
CoO             &8.050             &3.5  2.9   &3.0     &4    &172   &12      &12  &-    &$\bar{5}s    {3}p\bar{4}d$\,Co;\ $\bar{3}s\bar{3}p\bar{4}d$\,O\\
NiO             &7.880             &3.1  2.5   &2.5     &4    &110   &16      &16  &-    &                $\bar{4}d$\,Ni\\
CuO             &9.558$^{\dag}$     &3.5  2.8   &3.0     &3    &616   &48      &12  &-    &$\bar{5}s\bar{5}\hat{p}\bar{4}d$\,Cu;\ $\bar{3}s\bar{3}p\bar{4}d$\,O\\
MnTe            &7.823 (1.621)     &2.7  2.3   &2.2     &6,4  &170   &8       &8   &2    &                $\bar{4}d$\,Mn;\ $        \bar{6}\hat{p}4d$\,Te\\
Fe$_{3}$O$_{4}$ &15.87              &3.5  2.9   &3.0     &4    &746   &12      &12  &16   &$\bar{5}s    {3}p\bar{4}d$\,Fe;\ $\bar{3}s\bar{3}p\bar{4}d$\,O\\
{La\textsubscript{2}CuO\textsubscript{4}}
                &9.942 (1.245)     &3.1  2.5   &2.5     &3    &524   &30      &30  &-    &$5p\bar{6}d$\,La;\ $3p\bar{4}d$\,Cu;\ $\bar{3}s\bar{3}p\bar{4}d$\,O\\
\hline
\end{tabular}
\caption{Materials parameters: $a$ is lattice constant (quantity in parenthesis is $c/a$ where applicable).  $G_{\rm
    cut}(\psi)$ and $G_\mathrm{cut}(M)$ are plane-wave cutoffs for the interstitial part of the one-particle and
  two-particle basis sets, in units of $2\pi/a$; $\Sigma_\mathrm{cut}$ the energy cutoff for above which $\Sigma$ is
  restricted to a diagonal part, as described in the text.  $N_k$ is the number of divisions along each reciprocal
  lattice vector defining the $k$ mesh. When two numbers appear, the $c$ axis is assigned a mesh different than the
  basal plane. The latter number is selected to make the spacing between $k$ points as similar as possible along the
  three directions.  $N_{1p}$ is the total number of basis functions in the unit cell.  $N_{v}$ and $N_{c}$ are the
  number of occupied and unoccupied eigenstates included in the construction of the vertex.  $N_{\rm flt}$ is the number
  of points in the interstitial where envelope functions were added to increase the basis completeness.  $\varphi_{z}$
  lists the local orbitals (LO) for each element: $n-s$, $n-p$, $n-d$, where $n$ is the principal quantum number of the LO.
  $n$ without an overbar indicates the LO covers a core-like state, well below the linearization energy with a principal
  quantum number one less than that of the valence.  $\bar{n}$ indicates the LO energy is far above the linearization
  energy, and is included to better treat unoccupied states well above the Fermi energy.  Both kinds of LO are discussed
  in Ref.~\cite{questaal_paper}.  Partial waves marked as $\hat{p}$ replace the $l{=}1$ partial wave with the
  corresponding $p_{1/2}$ partial wave computed from the Dirac equation, as discussed in Ref.~\cite{questaal_paper}.
  This is a small effect but it improves the matrix elements for spin-orbit coupling.}
\label{tb:comp_details}
\end{table*}

Implementation of \emph{GW} requires both a 1-body framework and a two-body framework.  Both are described in detail in Questaal's methods paper,
Ref.~\cite{questaal_paper}, and the paper describing Questaal's implementation of QS\emph{GW} theory, Ref.~\cite{QSGW_paper}, which we denote here as papers I
and II.  I places heavier focus on the one-body part, while II focuses on the \emph{GW} theory and its implementation.

Questaal is an all-electron method, with an augmented wave basis consisting of partial waves inside augmentation spheres, constructed from numerical solutions
of the radial Schrodinger equation on a logarithmic mesh (I, \S2.2). The one-body basis consists of a linear combination of smooth, atom-centered Hankel
functions as envelope functions, augmented by the partial waves.  Two partial waves are calculated at some linearization energy $\phi_\ell$ and energy
derivative $\dot{\phi}_\ell$, which provides enough freedom to match value and slope to the envelope functions (I, \S3).

\emph{One particle basis}: In a conventional LMTO basis, envelope functions consists of ordinary Hankel functions,
parameterized by energy $E$.  Questaal's smooth Hankel functions are composed of a convolution of Gaussian functions of
smoothing radius $r_{s}$, and ordinary Hankel functions (I, \S3.1); thus two parameters are needed to define the
envelope.  In the periodic solid, Bloch sums of these functions are taken (I, Appendix C).  In the present work, $E$ is
constrained to a fixed value ($-0.4$\,Ry for most systems), and $r_{s}$ determined by optimizing the total energy of the
free-atom wave function.  These are kept fixed throughout the calculation, while the partial waves and linearization
energy float as the potential evolves.  By fixing $E$ to a universal value, we are able to take advantage of the
``screening transformation'' to render the basis set short-ranged (see I, \S2.9).  This can be useful for the
interpolation of the self-energy to an arbitrary $k$ mesh, as described below.  A second envelope function of a deeper
energy is needed to make the hamiltonian reasonably complete.  The latter energy is chosen to be 0.8\,Ry deeper than the
first.  For most materials, the envelopes of orbitals $l{=}0{\dots}4$ the first energy, and $l{=}0{\dots}3$ for the
second.  At the \emph{GW} level, a few other additions are made to make the basis closer to complete.  Completeness
of the envelope functions is sometimes improved by adding ``floating orbitals'' --- points in the interstitial regions
where smooth Hankel functions are placed without an augmentation sphere (I, \S3.11), usually for $\ell$ up to 2.
$N_{\rm flt}$ in the Table indicates how many points in the unit cell where floating orbitals are added.  To expand the
hilbert space inside the augmentation spheres, a local orbital $\phi_z$ may be added (I, \S3.7.3).  $\phi_z$ is a
solution of the radial Schrodinger equation at an energy, either well below the linearization energy for deep core-like
states, or well above it to better represent the unoccupied states.  In the Table, the $\phi_z$ used in the calculations
here are listed, with a bar over the principal quantum number to indicate the high-lying states.  For heavier elements,
the $p$ local orbital is sometimes replaced by the $p_{1/2}$ component of the Dirac equation.  This has a modest effect
but improves the accuracy of the spin-orbit coupling (I, \S3.9).  The total number of orbitals in the one-particle basis
is listed in the table as $N_{1p}$.  Another parameter is the sphere augmentation radius, $r_\mathrm{MT}$.

\emph{k convergence}:
The $GW$ mesh and the one-body mesh are generally different: the latter normally needs to be somewhat finer, as the
self-energy is a relatively smooth function of $k$ while the kinetic energy is less so.  Since the cost is low, we use a
finer mesh than necessary for the one-particle part, which obviates the need to test the mesh for $k$ convergence.
Careful tests of the $GW$ mesh were made for each system.  Most of the small unit cells used a mesh of $6$ divisions
along each axis: the number used in each materials system is listed as $N_k$ in the Table.  A finer mesh,
e.g. $8{\times}8{\times}8$ divisions, changes the result only slightly (e.g. gap changes by $\sim$0.01\,eV in $sp$
semiconductors).

To enable inequivalent meshes, the self-energy must be interpolated.  To render the interpolation everywhere smooth, (I,
\S2G) eigenfunctions and self-energy are rotated to the LDA basis, and the full self-energy matrix is kept only up to a
cutoff above the Fermi level in this basis, denoted $\Sigma_\mathrm{cut}$ in the Table.  Above this cutoff, only the
diagonal part of $\Sigma$ is kept.  $\Sigma_\mathrm{cut}$ may be made arbitrarily high, but if it is too high the
interpolation is no longer smooth.  Fortunately the result depends weakly on $\Sigma_\mathrm{cut}$, and
$\Sigma_\mathrm{cut}{\sim}2$ is typically sufficient to achieve a reasonably well converged result.

A smooth Hankel function has a plane-wave representation; thus any linear combination of them, e.g., an eigenfunction,
does also.  An eigenfunction represented in this form is equivalent to a representation in an LAPW basis: it is defined
by the coefficients to the plane waves, the shape of the partial waves and their coefficients (which are constrained to
match smoothly onto the envelope functions).  The PW cutoff for the one-particle basis is listed as $G_\mathrm{cut}(\psi)$
in the Table.

\emph{Two-particle basis}: The two-particle basis is needed to represent quantities such as the bare coulomb interaction
and the polarizability.  As with the one-particle basis, it as a mixed construction with interstitial parts and
augmentation parts (II, \S{II}A): envelope function products are represented as plane waves, since product of plane
waves is another plane wave.  Thus the interstitial parts of the mixed (product) basis are plane waves, and the PW
cutoff is listed in the Table as $G_\mathrm{cut}(M)$.  Inside augmentation spheres, all possible products of partial waves
are called product functions $B_\ell$, organized by $\ell$ with a form $B_I{=}B_{\ell}(r)Y_{{\ell}m}(\mathbf{\hat{r}})$.
The set of all possible products of partial waves is somewhat overcomplete with a relatively large rank. It is reduced
by diagonalizing the overlap matrix, and retaining the subset of functions above a cutoff eigenvalue of the overlap.  It
has been found from experience that eigenfunctions with eigenvalues below $3{\times}10^{-4}$ for $\ell$=0,1 and
$10^{-3}$ for $\ell{>}1$ have essentially negligible effect on the result, and are discarded.  The product basis is
truncated at a finite $\ell$.  For most systems, $\ell_\mathrm{cut}$ was chosen to be 6 for elements with small to
moderate $r_\mathrm{MT}$ and whose $d$ orbitals are far from the Fermi level, such as O or P; 8 for elements of
intermediate size; 10 for elements with large radii; and 12 for systems with $f$ orbitals such as Ce.

\emph{Bare coulomb interaction}: To stabilize the calculation, the bare coulomb interaction, $v(q)=1/q^2$, is
approximated by a Thomas-Fermi form, $v(q)=1/(q^2+V_\mathrm{TF})$.  This is because if $V_\mathrm{TF}$ is set to zero,
the result can become unstable.  We use a small value $V_\mathrm{TF}$, typically $2{\times}10^{-5}$\,Ry, though
sometimes somewhat larger values, up to $2{\times}10^{-4}$\,Ry were used.  The dielectric constant,
$\epsilon_\infty$, can vary by a few percent over this range.  For that reason $\epsilon_\infty$ was calculated for
several values of $+V_\mathrm{TF}$, e.g. $1{\times}10^{-5}$, $1{\times}10^{-5}$, and $3{\times}10^{-5}$\,Ry, and the
reported value is the result when extrapolated to zero.

\emph{Frequency mesh}: to construct the self-energy, an energy integration on the real frequency axis is taken.  A
regular quadratic mesh of the form $\omega_i = \texttt{dw}{\times}i + \texttt{dw}^2i^2/(2\omega_c)$ is used, with $i$
spanning $\omega_i$=0 and the largest eigenstate.  Points are linearly spaced for $\texttt{dw}\ll\omega_c$, but the
spacing increases for $\texttt{dw}\gtrsim\omega_c$.  It has been found empirically that results are essentially
independent of mesh for $\texttt{dw}{<}0.08$\,Ry and $\omega_c{\gtrsim}0.1$\,Ry.  In practice we use
$\texttt{dw}{=}0.02$\,Ry and $\omega_c{=}0.2$\,Ry to obviate the need for checking convergence.  To pick up the poles of
$G$ and $W$ to make $\Sigma$, the contour is deformed to include an integration on the imaginary axis of $\omega$ (I,
\S2F).  In all the calculations used here, we used 6 points on a Legendre quadrature.  A few checks showed that the
result hardly depended on the number of points in the quadrature.

\emph{Manual vs auto-generated input}: Questaal has an automatic generator, \texttt{blm}, to construct input files from
structural data.  Most input parameters are automatically generated by \texttt{blm}, such as the MT radii
$r_\mathrm{MT}$, the product basis cutoffs, and the plane wave cutoffs, the Gaussian smoothing radius defining the
envelope functions, and the placements for floating orbitals, when they are sought.  Also for the vast majority of
parameters, the code uses default values if inputs are not explicitly specified.  For a few parameters, manual
intervention is needed to monitor convergence, especially the number of $k$ points and the plane wave cutoffs
$G_\mathrm{cut}(\psi)$ and $G_\mathrm{cut}(M)$.  Hankel function energies $E$ must be manually set, but usually fixed
values as noted above are sufficient.  Occasionally interpolation continues to be an issue and can be stabilized by
making $E$ deeper, e.g. $E{=}-0.6$\,Ry was needed to stabilize SrTiO$_{3}$.  Results are largely insensive to the choice
of $E$, provided it is not pushed too deep.

%\section*{References}
%\bibliographystyle{ieeetr}
\bibliography{references,experiment}

\begin{thebibliography}{205}
\bibitem [{not({\natexlab{a}})}]{notea}%
  \BibitemOpen
  \href@noop {} {} ({\natexlab{a}}),\ \bibinfo {note} {with the caveat that
  there not be a unique solution. There can be multiple solutions for $G_{0}$,
  similar to what is found in Hartree Fock theory, or DFT. Nevertheless the
  fact that it satisfies a variational principle~\cite{Beigi17} implies that
  there are a limited, discrete set of solutions corresponding to metastable
  states as in Hartree Fock or density functional theory. Such states, e.g.
  solutions found for both an antiferromagnetic and a ferromagnetic state that
  are connected to local minima.}\BibitemShut {Stop}%
\bibitem [{not({\natexlab{b}})}]{noteb}%
  \BibitemOpen
  \href@noop {} {} ({\natexlab{b}}),\ \bibinfo {note} {{The screening is much
  larger in a metal than an insulators, so ladders in the charge channel are
  generally thought to be less important. A big exception to this rule is the
  low-density limit, e.g. when the average density is less than that of
  Na~\cite{Gatti23}.}}\BibitemShut {Stop}%
\bibitem [{exp()}]{exptcaution}%
  \BibitemOpen
  \href@noop {} {}\bibinfo {note} {{Several photoemission studies can be found
  in the literature, e.g. for
  NiO~\cite{Sawatzky84,PhysRevB.44.3604,Zimmermann99}. Different studies yield
  somewhat different results: ref.~\onlinecite{Zimmermann99} in particular
  shows that the observed spectra depend significantly on the instrument used
  and the photon energy.}}\BibitemShut {Stop}%
\end{thebibliography}
\end{document}